\documentclass[fleqn,usenatbib]{mnras}

\usepackage{newtxtext,newtxmath}
\usepackage[T1]{fontenc}

\DeclareRobustCommand{\VAN}[3]{#2}
\let\VANthebibliography\thebibliography
\def\thebibliography{\DeclareRobustCommand{\VAN}[3]{##3}\VANthebibliography}

\usepackage{graphicx}	% Including figure files
\usepackage{amsmath}	% Advanced maths commands
\usepackage{siunitx}    % Non-italic letters in equations
\usepackage{subcaption} % Makes multi-image figures simpler
\usepackage{booktabs}   % Table design
\usepackage{multirow}   % Table design
\usepackage{threeparttable} % Table design
\usepackage{makecell}   % Table design
\usepackage{orcidlink}  % Adding ORDiDs

\newcommand\Tstrut{\rule{0pt}{2.6ex}}
\graphicspath{{./Images}}
\defcitealias{2017MNRAS.466.3519R}{RR17}

\title[Properties of HBC 494's Outflows]{The Kinematic and Dynamic Properties of HBC 494's Wide-Angle Outflows}

\author[A. Fourkas et al.]{
Austen Fourkas$^{1,2}$\thanks{E-mail: afourkas@terpmail.umd.edu}\orcidlink{0000-0001-7387-3898},
Dary Ru\'{i}z-Rodr\'{i}guez$^{2}$\orcidlink{0000-0003-3573-8163},
Lee G. Mundy$^{1}$\orcidlink{0000-0002-8876-0690},
Jonathan P. Williams$^{3}$\orcidlink{0000-0001-5058-695X}
\\
$^{1}$Department of Astronomy, University of Maryland, 4296 Stadium Dr, College Park, MD 20742, USA\\
$^{2}$National Radio Astronomy Observatory, 520 Edgemont Rd, Charlottesville, VA 22903, USA\\
$^{3}$Institute for Astronomy, University of Hawai`i at M\={a}noa, 2680 Woodlawn Dr., Honolulu, HI 96822, USA}

\date{Accepted XXX. Received YYY; in original form ZZZ}

\pubyear{2023}

\begin{document}
\label{firstpage}
\pagerange{\pageref{firstpage}--\pageref{lastpage}}
\maketitle

\begin{abstract} 
We present Atacama Large Millimeter/sub-millimeter Array (ALMA) Cycle-5 observations of HBC 494, as well as calculations of the kinematic and dynamic variables which represent the object’s wide-angle bipolar outflows. HBC 494 is a binary FU Orionis type object located in the Orion A molecular cloud. We take advantage of combining the ALMA main array, Atacama Compact Array (ACA), and Total Power (TP) array in order to map HBC 494's outflows and thus, estimate their kinematic parameters with higher accuracy in comparison to prior publications. We use $^{12}$CO, $^{13}$CO, C$^{18}$O, and SO observations to describe the object’s outflows, envelope, and disc, as well as estimate the mass, momentum, and kinetic energy of the outflows. After correcting for optical opacity near systemic velocities, we estimate a mass of $3.0\times10^{-2}$ $\si{M_{\sun}}$ for the southern outflow and $2.8\times10^{-2}$ $\si{M_{\sun}}$ for northern outflow. We report the first detection of a secondary outflow cavity located approximately $15\arcsec$ north of the central binary system, which could be a remnant of a previous large-scale accretion outburst. Furthermore, we find CO spatial features in HBC 494's outflows corresponding to position angles of $\sim35\degr$ and $\sim145\degr$. This suggests that HBC 494's outflows are most likely a composite of overlapping outflows from two different sources, i.e., HBC 494a and HBC 494b, the two objects in the binary system. 
\end{abstract}


%%%%%%%%%%%%%%%%%%%%%%%%%%%%%%%%%%%%%%%%%%%%%%%%%%

%%%%%%%%%%%%%%%%% BODY OF PAPER %%%%%%%%%%%%%%%%%%

\section{Introduction}
Young stellar objects (YSOs) primarily gain mass due to accretion of material from their large stellar envelope onto a protostellar disc \citep{1996ARA&A..34..207H}. However, mass and luminosity predictions based on constant accretion models do not align with measured values for multiple YSOs \citep{1990AJ.....99..869K, 2014prpl.conf..387A}. As first proposed by \cite{1990AJ.....99..869K}, the `luminosity problem' highlights the difference between predicted and observed luminosity values. FU Orionis stars (FUors) serve as a solution to the luminosity problem, as their non-constant accretion rates are able to resolve differences in theory and observation. FUors are a subclass of Class 0 and I YSOs which display characteristic large-scale accretion outbursts resulting in changes to brightness, bolometric luminosity, and the object's spectral energy diagram (SED) \citep{1966VA......8..109H}. Specifically, accretion rates during outbursts can increase by up to 1000 times in comparison to quiescent epochs ($10^{-7} \rightarrow 10^{-4}$ $ \si{M_{\sun}..yr^{-1}}$), and the star may accrete up to 0.01 ${\si{M_{\sun}}}$ during an outburst \citep{2014prpl.conf..387A, 2005ApJ...631.1134A}. The large amount of disc material accreted onto the protostar during an outburst event can increase a FUor's optical brightness by 3-6 mag, and lead to bolometric luminosity increases of between 50 and 500 $\si{L_{\sun}}$ \citep{2014prpl.conf..387A}.\par
While luminosity and magnitude changes serve as observational evidence for the occurrence of outbursting events, the mechanisms which trigger these events are yet to be completely constrained \citep{2018MNRAS.474.4347C}. Possible outbursting mechanisms include: thermal instabilities in the disc \citep{1994ApJ...427..987B}, disc fragmentation \citep{2005ApJ...631.1134A, 2012ApJ...746..110Z}, tidal interactions between the protostellar disc and a companion object \citep{2004MNRAS.353..841L, 1992ApJ...401L..31B}, and a combination of gravitational and magneto-rotational instabilities \citep{2001MNRAS.324..705A, 2009ApJ...694.1045Z}.

As accretion from the protostellar disc onto the central protostar proceeds, streams of molecular material called outflows are created and propagate perpendicular to the disc in opposite directions. The creation and propagation of these flows are closely linked to the magnetic field generated by the central protostar and how this field interacts with the protostellar disc. The wind created by a magnetized disc, called the disc-wind, carries angular momentum away from the protostar through the bipolar outflows \citep{2019FrASS...6...54P}. While disc-winds assume that outflows stem from the surface of the disc, other models suggest that interactions between the inner edge of the disc and the magnetosphere of the central protostar are the primary driver of outflows. Magnetic field changes at these radii cause material to be ejected centrifugally through open field lines, and the wind generated in conjunction with the X-ring is called an X-wind \citep{2019FrASS...6...54P}. Along with impacting the accretion rate of material onto the central protostar, stellar outflows can also entrain and disperse envelope material away from the central object, thus impacting the object's final mass, and evolutionary pathway.\par
By observing and analysing an FU Ori type object's outflows, its large-scale accretion outburst mechanism can be better constrained along with its evolutionary pathway. Specifically, interferometric millimetre/sub-millimetre arrays with both long and short baselines can aid in observing FU Ori type objects due the presence of both large and small-scale emission in such systems. Greater baseline lengths are able resolve emission from the protostellar disc, as well as outflow cavities and bow-shocks, while shorter baselines image the envelope emission and material which fills the aforementioned cavities.\par
This study focuses on HBC 494, a binary FU Ori type object located at a distance of $414\pm7$ pc in the Orion A molecular cloud with a systemic velocity of $V_ {\si{sys}}=4.3\,\si{{km}.s^{-1}}$ \citep{2007A&A...474..515M}. HBC 494 was initially correlated with optical and infrared brightness increases in the reflection nebulae Re50 and Reipurth 50 N \citep{1985A&AS...61..319R, 1986Natur.320..336R}. \cite{2023MNRAS.tmp.1574N} were able to resolve both objects in the binary, HBC 494a and HBC 494b, with a separation of approximately 75 au. They report that HBC 494a has a dust mass of 1.4 $\si{M_{J}}$ and a gas mass of 143.5 $\si{M_{J}}$, while HBC 494b has a dust mass of 0.3 $\si{M_{J}}$ and a gas mass of 28.7 $\si{M_{J}}$. Furthermore, they report that HBC 494a is likely the primary object in the binary, as along with its mass, it is approximately 5 times brighter than HBC 494b. HBC 494 also presents extraordinarily wide bipolar outflows \citep[hereon \citetalias{2017MNRAS.466.3519R}]{2017MNRAS.466.3519R}. With an opening angle of 150 degrees, HBC 494's outflows are at least 20$\%$ wider than early interferometric observations suggest, as Class I YSO outflows were typically observed with opening angles angles between 30 and 125 degrees \citep[e.g.,][]{2006ApJ...646.1070A, 2016MNRAS.460..627K, 2018MNRAS.474.4347C}. Recent studies have revealed, however, that wide angle outflows are more common amongst Class 0 and I YSOs than previously thought \citep[e.g.][]{2023ApJ...947...25H}. Nevertheless, HBC 494's outflows present interesting structure, shape, and possible origin, all of which will be discussed in this work. Initial observations of HBC 494's outflows presented in \citetalias{2017MNRAS.466.3519R} only made use of the ALMA main array, making it difficult to accurately estimate outflow parameters. Specifically, these observations were limited by poor UV-plane coverage at short baselines, thus filtering out flux from large-scale structure and causing outflow parameters to be underestimated. In this paper we improve upon the results presented in 2017 through improvements to observations with regard to sensitivity and field of view (FOV), as well as an updated analysis and calculation procedure.\par
\begin{figure*}
    \centering
    \includegraphics[width=\textwidth]{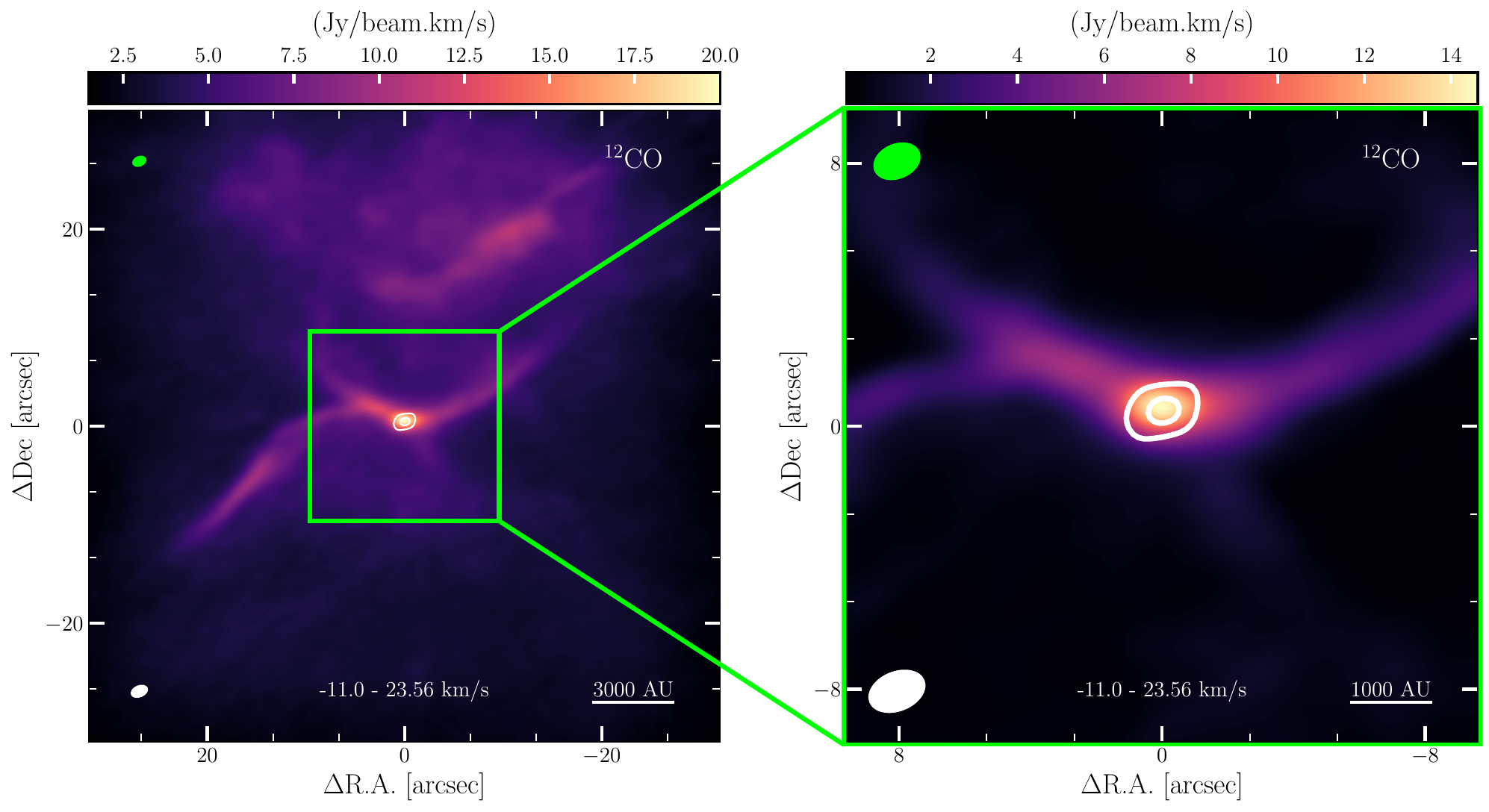}
    \caption{Moment-0 maps of $^{12}$CO integrated between LSR velocities of -11 and 23.56 $\si{{km}.s^{-1}}$. This figure illustrates the improvements made to both sensitivity and field of view by using both the combination array and mosaic pointings. The left image shows a Cycle-5 combination array observation of $^{12}$CO at a field of view of approximately one square arcminute, while the right image shows a Cycle-5 main array observation of $^{12}$CO at a field of view of approximately 400 square arcseconds, which is equal to the field of view of previously recorded ALMA Cycle-2 observations presented in \protect\citetalias{2017MNRAS.466.3519R}. The combination array synthesised beam is shown in white at the bottom left of the first image with a size of 1.85 $\times$ 1.23 arcseconds and position angle of -67.2$\degr$. The main array synthesised beam is shown in white at the bottom left of the second image with a size and position angle of 1.55 $\times$ 1.09 arcseconds and -68.8$\degr$, respectively. The white contours at the centre of each image represent continuum emission centred at 232.5 GHz, and is shown at levels of 0.25 and 0.75 times the maximum intensity of 0.11 Jy/beam. The continuum beam is shown in the top left of each frame as a green ellipse, and has an angular size of 1.50 $\times$ 1.07 arcseconds and a position angle of -66.8$\degr$.}
    \label{fig:ImageComp}
\end{figure*}

We present ALMA Cycle-5 observations of HBC 494. $^{12}$CO (J=2-1), $^{13}$CO (J=2-1), C$^{18}$O (J=2-1), and SO (J=6-5) are all used to trace different regions of the protostar and its surrounding environment. We make use of the ALMA main array (12-m array) as well as a combination of the main array, Total Power array (single-dish, 12-m diameter), and Atacama Compact Array (7-m array), which will be called the `combination array' for the remainder of this paper. The use of both long and short baselines in the combination array prevents spatial filtering normally present in ALMA 12-m array observations. The Cycle-5 observations traced a larger extent of the outflow in comparison to prior observations due to the use of X-field mosaicing instead of a single pointing for imaging (Fig. \ref{fig:ImageComp}).
Specifications of our ALMA observations are described in section \ref{sec:OBS}, while observational and computational results are presented in sections \ref{sec:OBSRES} and \ref{sec:CR}, respectively. A discussion of results and important conclusions are given in sections \ref{sec:DIS} and \ref{sec:CONC}.
\section{Observations}
\label{sec:OBS}

We obtained ALMA Cycle-5 Band 6 observations toward HBC 494 with a centre
position of $\alpha$ (J2000) = 05$^{\rm h}$ 40$^{\rm m}$ 27.45$^{\rm s}$ ; $\delta$ (J2000) = -07$^{\rm o}$27${\arcmin}$30.05${\arcsec}$. The spectral windows targeted $^{12}$CO (J = 2-1), $^{13}$CO (J = 2-1), C$^{18}$O (J = 2-1), SO (J = 6(5)-5(4)), DCN (J = 3-2), SiO (J= 5-4), $^{13}$CS (J = 5-4) and continuum at rest-frame frequencies of 230.588, 220.398, 219.560, 219.949, 217.238, 217.104, 231.220, and 232.50 GHz, respectively (see Table \ref{Table:Width}).
\begin{table*}
%\contcaption{Deconvolved radial extensions fr gom ALMA observations.}
\begin{threeparttable}
\caption{Spectral windows.}
\label{Table:Width}
\begin{tabular}{cccccccc}
    \toprule
        \multirow{3}{*}{Line} & \multirow{3}{*}{Rest Freq.} & \multicolumn{2}{c}{12-m} &  \multicolumn{2}{c}{7-m} & \multicolumn{2}{c}{TP}  \\
        \cmidrule(lr){3-4}\cmidrule(lr){5-6}\cmidrule(lr){7-8}
        &&Bandwidth&Channel&Bandwidth&Channel&Bandwidth&Channel\\ %FIX THIS
        &[GHz]&[MHz]&[$\#$]&[MHz]&[$\#$]&[MHz]&[$\#$]\\

        \hline
        
        SiO J=5-4 & 217.105 &58.5 & 480& 62.5&512& 62.5&512\Tstrut\\
        DCN J=3-2& 217.238 & 58.5 & 480 & 62.5&512&62.5&512 \\
        H$_{2}$CO J=3(2,1)-2(2,0) & 218.760 & 58.5 & 480 & 62.5&512& 62.5&512 \\
        C$^{18}$O  J=2-1  & 219.560 & 58.5 &960 & 62.5&1024 & 62.5&1024   \\
        SO J=6(5)-5(4)& 219.949 & 58.5 &960 & 62.5&1024& 62.5&1024\\
        $^{13}$CO  J=2-1  & 220.398 & 58.5 & 960 & 62.5 & 1024 & 62.5 & 1024 \\
        $^{12}$CO  J=2-1 & 230.55 & 117.2 & 960 & 62.5 & 1024 & 125 & 1024 \\
        $^{13}$CS J=5-4 & 231.221 & 117.2 & 960 & 125 & 1024 & 125 & 1024 \\
    \bottomrule
\end{tabular}
\end{threeparttable}
%\footnotesize{$^a$ CO isotopologues will be presented in Ruiz-Rodriguez et al. In prep. }\\
\end{table*}
The correlator for these spectral windows was set to have bandwidths between 58.5 MHz and 2000 MHz. In order to image HBC 494 with a full coverage from 0.5${\arcsec}$ to the map size of 30$^{\arcsec}$, we make use of the ALMA combination array. By combining the 12-m, 7-m, and TP data, the observations are sensitive to emission from disc to outflow scales. Table \ref{Table:Observations} summarizes the observational logs for this project.

 HBC 494 was observed (project ID: 2017.1.00015.S PI: J. Williams) using the compact (baselines $\sim$ 55 to 91 m) and the extended configurations, C43-2 (15 - 315 m) and C43-5 (15 - 1400 m), on 2017 October 31, 2018 Jan 21, and July 12, respectively (Table \ref{Table:Observations}). The field of view of 60$^{\arcsec}$ $\times$ 60$^{\arcsec}$ was covered by a mosaic of 27 pointings (12-m array). The shorter UV distance range was fulfilled by ACA observations using a 7 pointing mosaic (7-m array) with a maximum detectable size of 30$^{\arcsec}$, and the TP observations covered the larger-scale emission over the entire mapped region.

\begin{table*}
%\contcaption{Deconvolved radial extensions fr gom ALMA observations.}
\begin{threeparttable}
\caption{Observational details.}
\label{Table:Observations}
\begin{tabular}{cccccccc}
    \toprule
    \multirow{2}{*}{Array} & Observation  & Bandpass & Flux & Phase & Time On & Median PWV  \\
    & Date & Calibrator & Calibrator  &Calibrator& Source [mm:ss] & [mm]\\
    \hline

    12-m $\times$ 44 & 21-Jan-2018   &J0607-0834 & J0607-0834 & J0541-0541 & 17:21 & 0.98 \\
    12-m $\times$ 43 & 12-Jul-2018   & J0542-0913 & J0522-3627 & J0542-0913 & 08:48 & 2.25 \\
    7-m $\times$ 11 & 31-Oct-2017  & J0522-3627& J0607-0834 & J0607-0834 & 34:45  & 1.43 \\
    7-m $\times$ 11 & 01-Nov-2017  & J0522-3627& J0607-0834 & J0607-0834 & 34:11 & 0.58 \\
    TP & 03-May-2018   &J0542-0913&J0542-0913 &J0542-0913 & 38:01 & 0.63 \\
    TP & 04-May-2018   &J0542-0913& J0542-0913 &J0542-0913 & 38:01 & 0.71 \\
    TP & 06-May-2018  & J0542-0913 & J0542-0913 &J0542-0913 & 38:01 &  1.40 \\
    TP & 15-May-2018  &J0607-0834 & J0607-0834 &J0607-0834 & 38:01 & 0.89 \\

    \bottomrule
\end{tabular}
\end{threeparttable}
%{FHWM value obtained from Gaussian fit. For details, see Sec}
%{Radial extension obtained after deconvolving FWHM from beamsize.}
\end{table*}

\subsection{Data Calibration and Reduction}\label{Sec:Analysis}

Data were calibrated using the standard calibration script in the CASA software package (Version 4.7.2- Cycle-5). In order to concatenate all data sets, we initially converted the TP map into visibilities by using the Total Power to Visibilities (TP2VIS)\footnote{https://github.com/tp2vis/distribute} package that runs on the CASA platform \citep[and references therein]{Koda2019}. We adopt the root-mean-square (rms) noise in the TP map
%\footnote{A more detailed description of the reduction process can be found in Appendix \ref{App:C}.} 
as the optimal weight of the TP visibilities. For line imaging, we initially subtracted the continuum emission by fitting a first-order polynomial to the continuum in the UV-plane. After subtracting continuum emission determined at the emission-free channels, we then combined the 12-m, 7-m, and TP data in UV space using the CASA task \textit{concat} and deconvolved them jointly. We determined weights of the visibility data using the task \textit{statwt}. Unfortunately, the  sensitivity of the extended configuration C43-5 from Cycle-5 observations is not optimal for the combination. We attempted to combine the 12-m, 7-m, and TP data by concatenating the visibility data together using different weights; however, the resulting rms became higher when including the C43-5 array.  We opted to not include these observations in the final products. Nevertheless, for completeness, we list in Table \ref{Table:Observations} the Cycle-5 C43-5 array observations.\par

After concatenating the visibility data, the reduction process was done using the TCLEAN task with a multi-scaled deconvolver of 0, 20, 60, which applies an iterative procedure with a decreasing threshold parameter to automatically mask regions during the cleaning process \citep[e.g.][]{Kepley2020}. To achieve a good balance between sensitivity and angular resolution, the Briggs weighting parameter R was set to 0.5 for the spectral-line image cubes. The spectral data cubes were constructed on a 640 $\times$ 640 pixel grid with 0.05$^{"}$ pixel size, and with 0.18 km s$^{-1}$ velocity resolution.\par

The broadband (2 GHz baseband) 232.5 GHz visibility was imaged by using the TCLEAN algorithm and applying similar imaging parameters as mentioned above. The analysis in this work, however, is focused mainly on the line data detected towards HBC 494, see section \ref{sec:OBSRES}.

\section{Observational Results}
\label{sec:OBSRES}

\subsection{\texorpdfstring{$^{12}$CO observations}{Lg}}
\label{sec:12COObs}
$^{12}$CO emission is the primary tracer of HBC 494's outflow material, and as such, we created a data cube covering LSR velocities from -20 to 33.82 $\si{{km}.s^{-1}}$. The cube can be split into three velocity sections: blueshifted, systemic, and redshifted. Blueshifted emission is present from -20 to 3.4 $\si{{km}.s^{-1}}$; emission near systemic velocities is present between 3.58 and 5.2 $\si{{km}.s^{-1}}$, and redshifted emission is present from 5.38 to 33.82 $\si{{km}.s^{-1}}$.\par

\begin{figure*}
    \centering
    \includegraphics[width=0.31\linewidth]{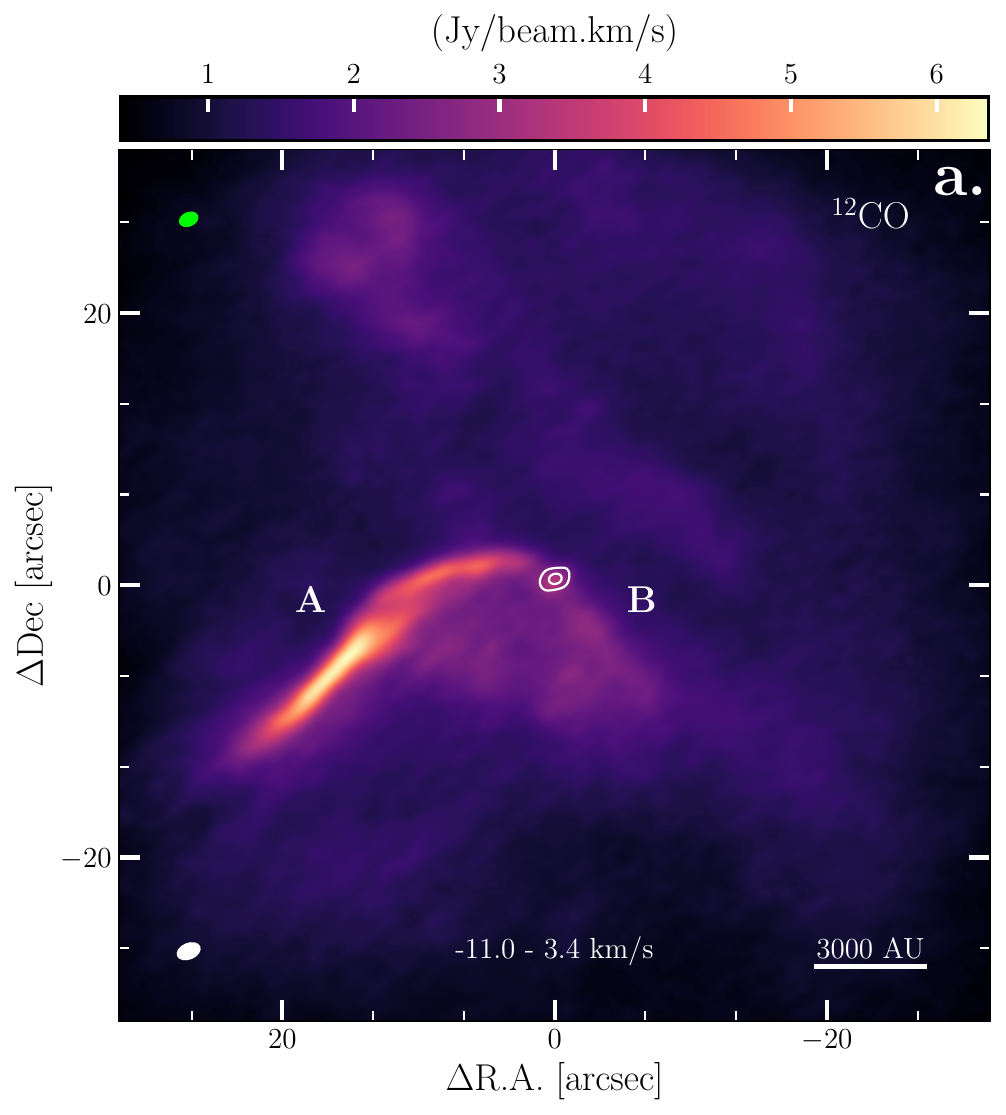}
    \includegraphics[width=0.31\linewidth]{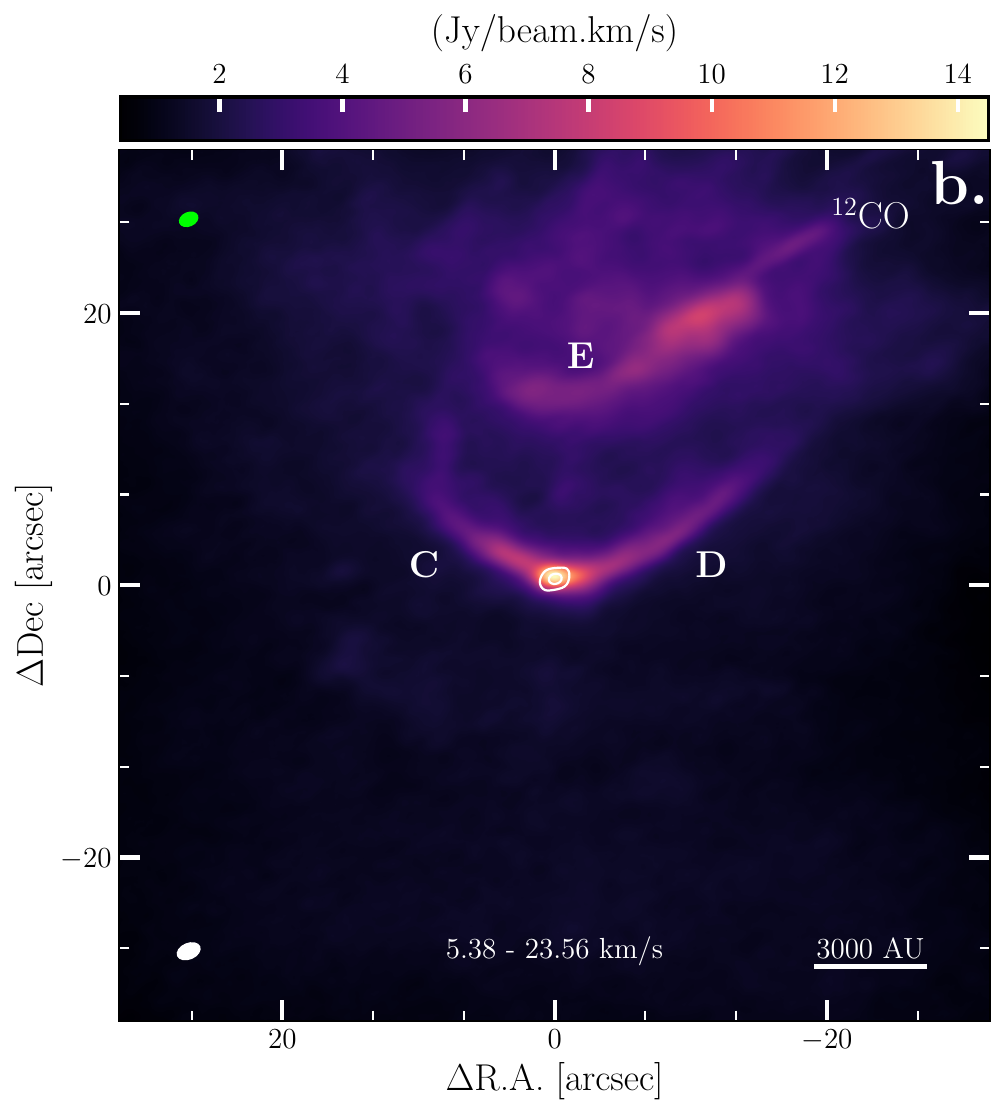}
    \includegraphics[width=0.31\linewidth]{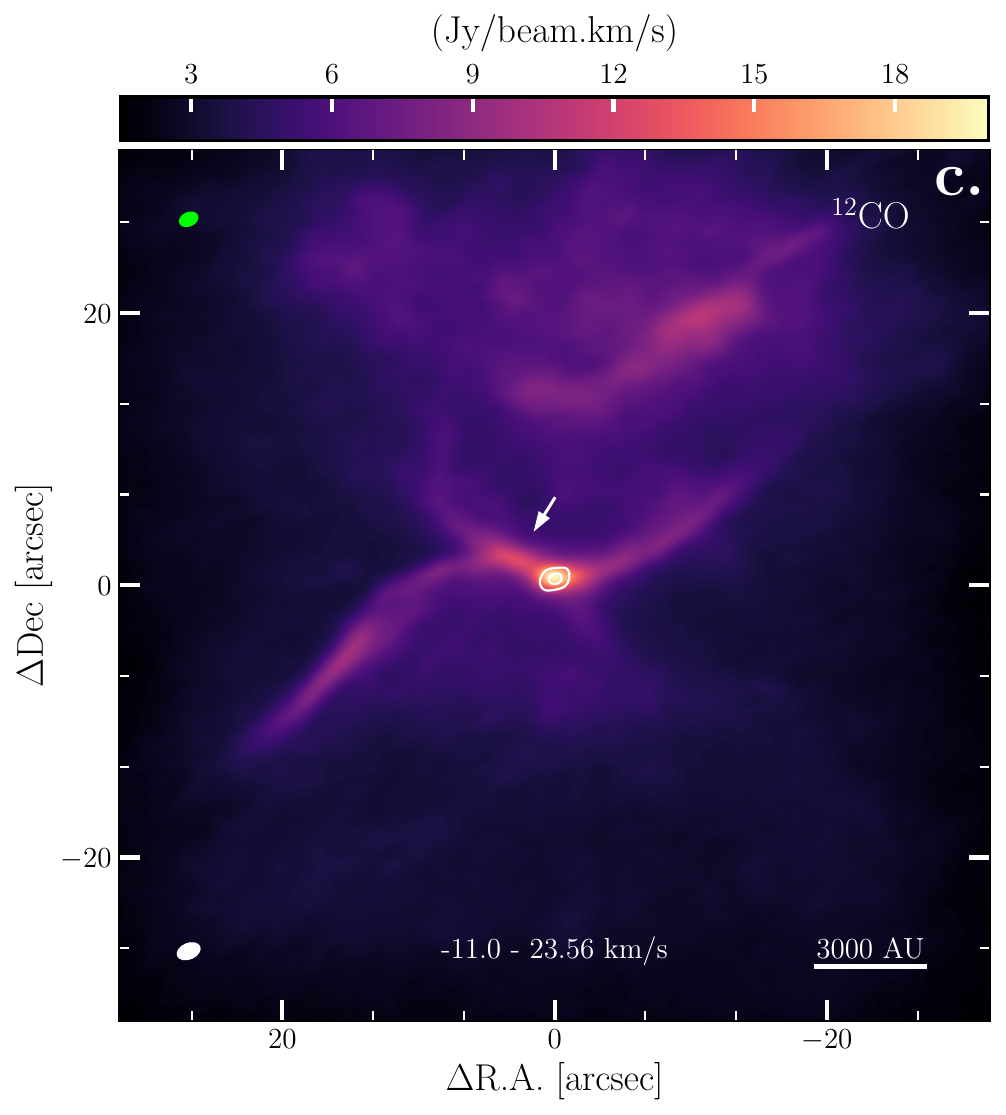}
    \caption{Moment-0 maps of blueshifted $^{12}$CO emission integrated between velocities of -11 -- 3.4 $\si{{km}.s^{-1}}$ (\ref{fig:12OutflowAB}a -- left), redshifted emission between velocities of 5.38 -- 23.56 $\si{{km}.s^{-1}}$ (\ref{fig:12OutflowAB}b -- centre), and both velocity ranges to encompass the northern and southern outflow (\ref{fig:12OutflowAB}c -- right). The letters A-E in shown in the above figures correspond to the following features: A -- eastern blueshifted wall, B -- western blueshifted wall, C -- eastern redshifted wall, D -- western redshifted wall, E -- the secondary cavity. The white contours at the centre of each image represent continuum emission centred at 232.5 GHz at levels of 0.25 and 0.75 times the maximum detected intensity of 0.11 Jy/beam. The white ellipse in the bottom left corner of each image represents the synthesised beam, which has an angular size of 1.85 $\times$ 1.23 arcseconds and a position angle of -67.2$\degr$. The continuum beam is shown in the top left of each frame as a green ellipse, and has an angular size of 1.50 $\times$ 1.07 arcseconds and a position angle of -66.8$\degr$.}
    \label{fig:12OutflowAB}
\end{figure*}

In order to reveal the large outflow structure traced by blueshifted and redshifted $^{12}$CO emission, we integrate over two velocity ranges: -11 -- 3.4 $\si{{km}.s^{-1}}$ and  5.38 -- 23.56 $\si{{km}.s^{-1}}$ for blueshifted and redshifted emission, respectively. These velocity ranges do not encompass the full extent of the aforementioned data cube, as outflow emission is not present near the minimum and maximum velocities of the cube. The results of these velocity channel integrations (moment-0 maps) are presented in Figure \ref{fig:12OutflowAB}. Figure \ref{fig:12OutflowAB}a shows that a blueshifted outflow cavity wall is detected with an extension of approximately 27${\arcsec}$ to the southeast (labelled A), and a fainter structure 13${\arcsec}$ long is detected on the southwestern side (labelled B). Figure \ref{fig:12OutflowAB}b shows the northern, redshifted outflow cavity. In comparison to the southern outflow cavity, the northern cavity shows a more uniformly parabolic shape with extension in the eastern and western arms of 11${\arcsec}$ (labelled C) and 18${\arcsec}$ (labelled D), respectively. The two prior outflows will hereon be called the primary northern and southern outflow cavities, respectively. A secondary outflow cavity is traced approximately 15${\arcsec}$ north of the central binary (labelled E). Due to its location, it is possible that this cavity is the result of a previous accretion outburst. Additionally, the misalignment between the southern cavity and the two northern cavities suggests an overlap of two sets of outflows from different origins (i.e., HBC 494a and HBC 494b), and is explained further in section \ref{sec:PV}.\par

Figure \ref{fig:12OutflowAB}c shows an integration over the full velocity range which highlights the parabolic shape of the southern and northern outflow cavities, as well as the secondary northern cavity. An asymmetry between the eastern and western features of each primary outflow is seen in the centre of the figure. A $\sim5.5{\arcsec}$ long and $\sim1{\arcsec}$  wide `bar' (indicated by an arrow) of emission extends eastwards from the central binary with a position angle of 67$\degr$, and is more than twice as intense as the more extended regions of each outflow. This bar could represent an overlap between the northern and southern outflows, as the eastern arms of both outflows appear to begin their separation at the end of this bar. It is also possible that this asymmetry is further evidence of overlapping outflows from the two objects in HBC 494's binary system, which likely have differing position angles from one another.\par
\begin{figure*}
    \centering
    \includegraphics[width=\textwidth]{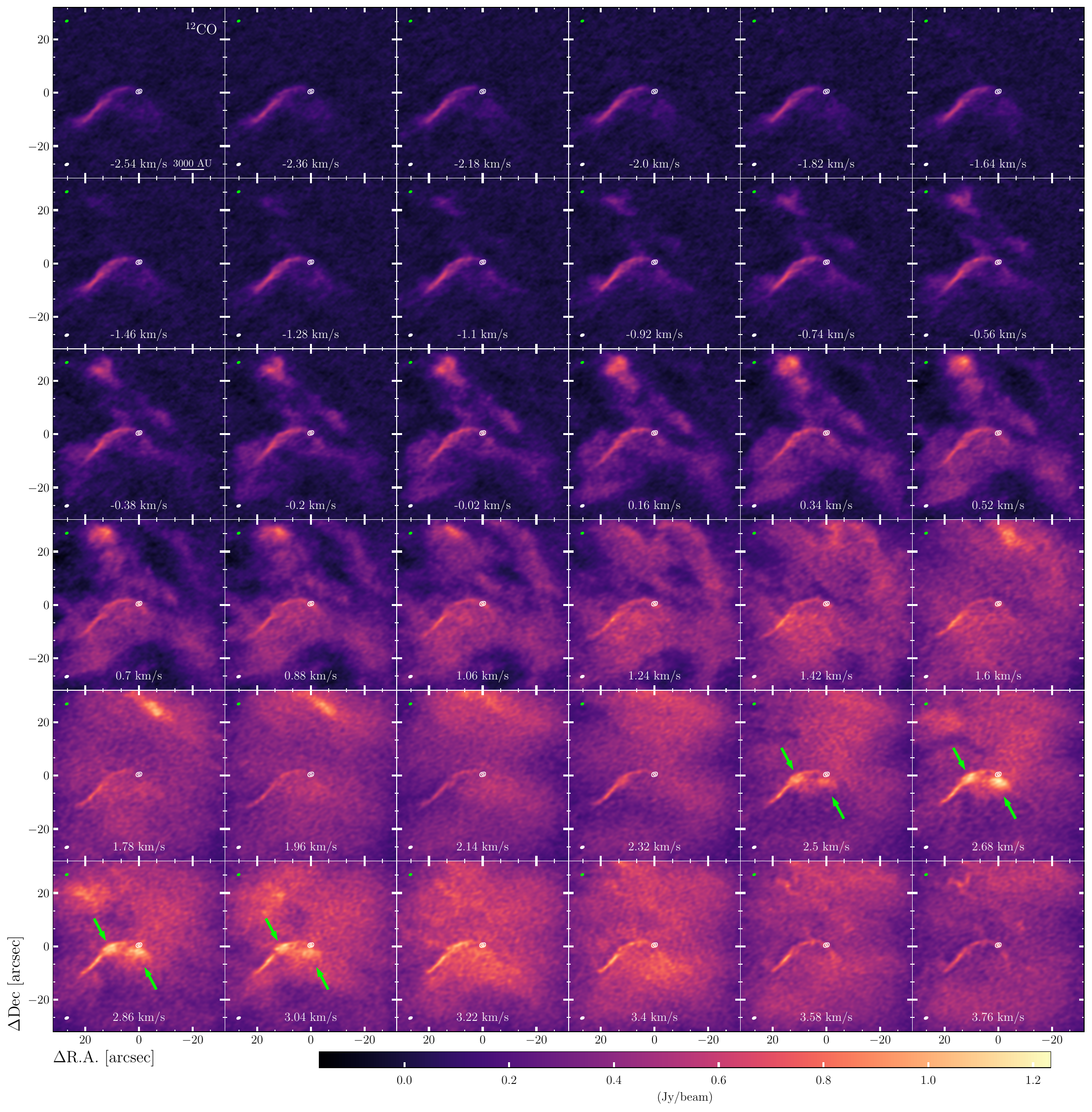}
    \caption{An individual channel map of blueshifted $^{12}$CO emission between velocities of -2.54 and 3.76 $\si{{km}.s^{-1}}$. The western arm of the southern outflow can be seen more prevalently as channel velocity increases. The green arrows shown towards the bottom of the figure highlight the lobes of emission present near the base of the outflow cavity. The central contours in each panel display continuum emission centred at 232.5 GHz at levels of 0.25 and 0.75 times the maximum recorded intensity of 0.11 Jy/beam. The white ellipse in the bottom left corner of each panel represents the synthesised beam, which has a size and position angle of 1.85 $\times$ 1.23 arcseconds and -67.2$\degr$, respectively. The green ellipse in the top left corner of each panel represents the continuum emission beam with a size of 1.50 $\times$ 1.07 arcseconds and a position angle of -66.8$\degr$.}
    \label{fig:12SOutflowInd}
\end{figure*}

The blueshifted outflow increases in both extension and intensity as channel velocities approach the systemic value, with the ambient cloud becoming prevalent at 1.6 $\si{{km}.s^{-1}}$ (Fig. \ref{fig:12SOutflow4}). The velocity evolution of the southern outflow cavity is shown in greater detail in Figure \ref{fig:12SOutflowInd}, which displays emission from the southern cavity between -2.54 and 3.76 $\si{{km}.s^{-1}}$ with a channel spacing of 0.18 $\si{{km}.s^{-1}}$. The eastern arm shows significant emission at channel velocities ranging from -11 and 3.76 $\si{{km}.s^{-1}}$; however, the western arm only displays emission from -2.54 to 3.76 $\si{{km}.s^{-1}}$. Indicated by arrows towards the bottom of the figure, emission near the base of the outflow lobe at velocities between 2.5 and 3.58 $\si{{km}.s^{-1}}$ follows the curvature the southern cavity, and separates into two distinct lobes of emission along its eastern and western walls. While the intensity of these two lobes is similar, their extension in relation to the southern outflow cavity is not. Specifically, the southwestern emission lobe extends to the end of its corresponding outflow arm, while the southeastern lobe extends approximately half way down the eastern arm. \par
\begin{figure*}
     \centering
     \includegraphics[width=\textwidth]{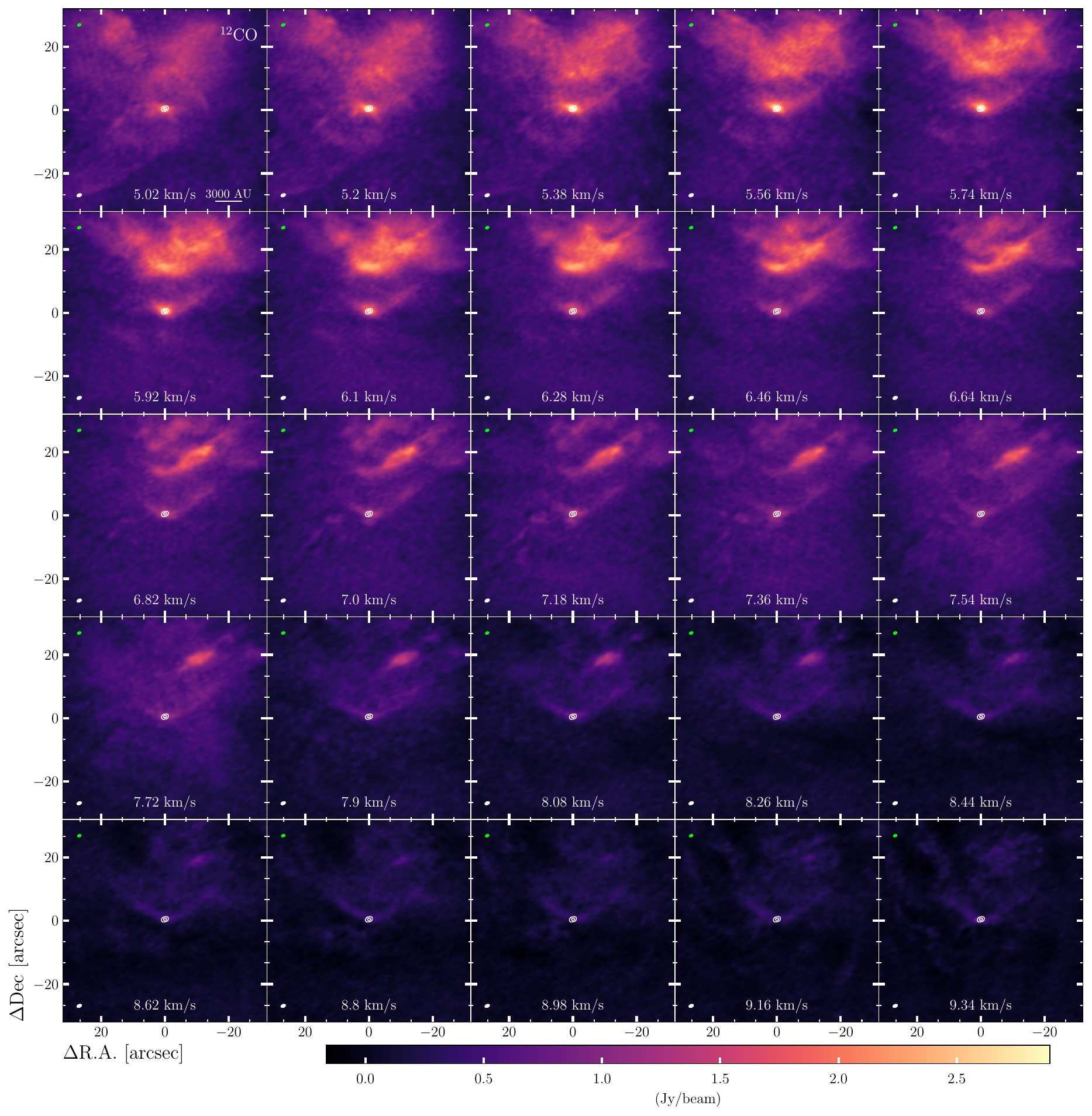}
     \caption{A channel-by-channel map of redshifted $^{12}$CO emission at velocities between 5.02 and 9.34 $\si{{km}.s^{-1}}$. The white contours visible at the centre of each panel represent continuum emission centred at 232.5 GHz at levels of 0.25 and 0.75 times the maximum detected intensity of 0.11 Jy/beam. The green and white ellipses shown at the top and bottom left of each panel, respectively, represent the $^{12}$CO continuum and synthesised beams. These beams maintain an angular size of 1.85 $\times$ 1.23 and 1.50 $\times$ 1.07 arcseconds, and position angles of -67.2$\degr$ and -66.8$\degr$, respectively.}
     \label{fig:12NOutflowInd}
\end{figure*}

As velocities increase from systemic values, both northern outflow arms extend away from the central binary towards the north east. Figure \ref{fig:12NOutflowInd} shows per-channel $^{12}$CO emission in the redshifted velocity range, specifically between 5.02 and 9.34 $\si{{km}.s^{-1}}$ with a channel spacing of 0.18 $\si{{km}.s^{-1}}$. A large cloud of material appears to be pushed upwards in the northern outflow cavity. Beginning at the front of this velocity range and becoming more prevalent as channel velocity increases, this cloud becomes separated from the central binary, continues to propagate upwards, and morphs into the secondary northern outflow cavity. At the same time, the two northern outflow arms begin to take shape; however, extension does not occur in the same velocity channels for both arms, nor does it occur at the same rate. The western arm reaches its maximum projected extension within 0.18 $\si{{km}.s^{-1}}$ of the emission cloud becoming detached from the central binary. In contrast, the eastern arm does not extend until 7 $\si{{km}.s^{-1}}$. Approaching 9.34 $\si{{km}.s^{-1}}$, emission from the secondary northern outflow cavity dissipates and the two northern outflow arms are seen clearly. \par
\begin{figure*}
     \centering
     \includegraphics[width=\textwidth]{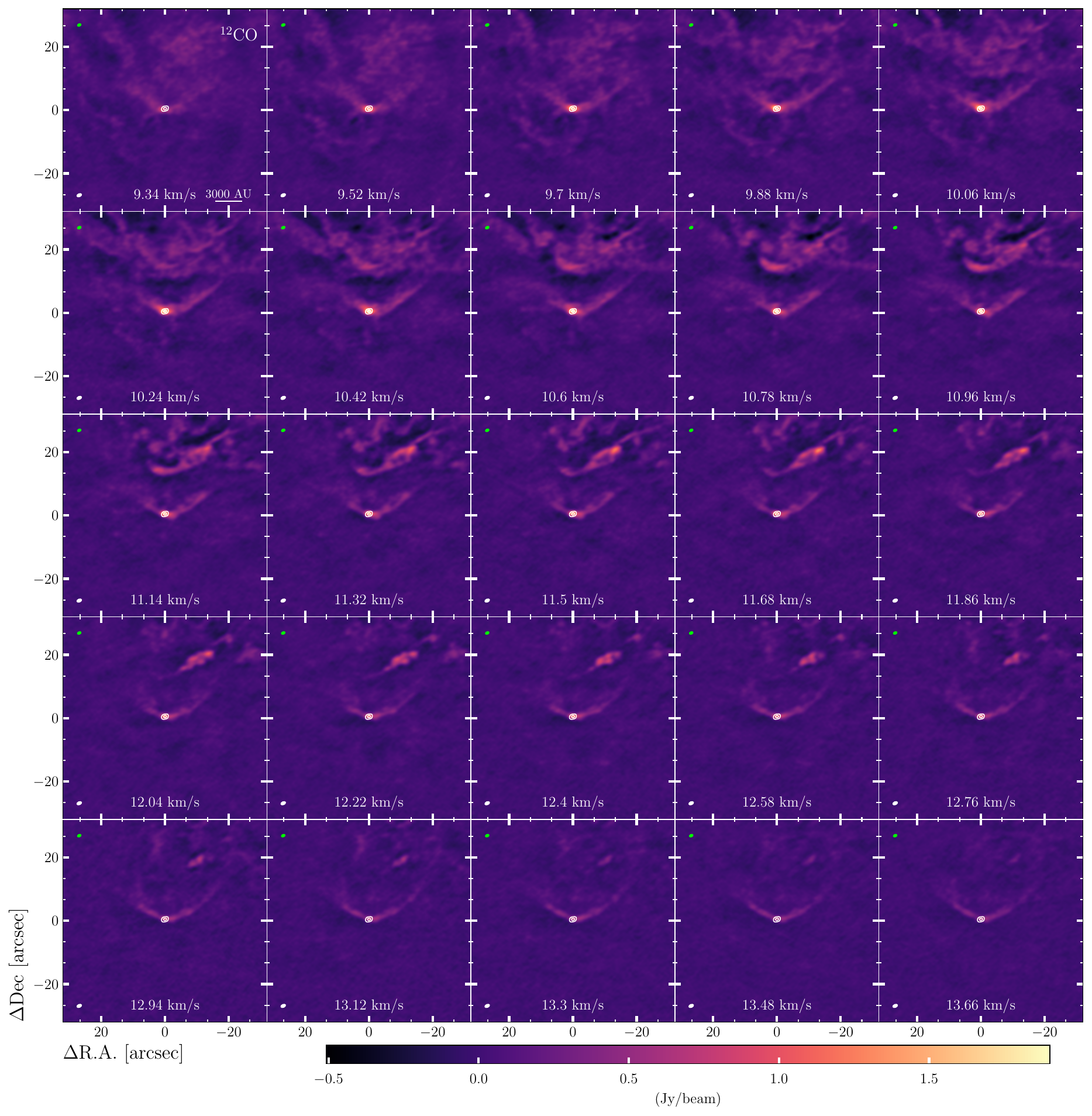}
     \caption{An individual channel map of redshifted $^{12}$CO emission at velocities between 9.34 and 13.66 $\si{{km}.s^{-1}}$. The white contours shown in the centre of each frame represent continuum emission centred at 232.5 GHz at levels of 0.25 and 0.75 times the maximum detected intensity of 0.11 Jy/beam. The green and white ellipses shown at the top and bottom left of each panel, respectively, represent the $^{12}$CO continuum and synthesised beams. These beams maintain an angular size of 1.85 $\times$ 1.23 and 1.50 $\times$ 1.07 arcseconds, and position angles of -67.2$\degr$ and -66.8$\degr$, respectively.}
     \label{fig:12NOutflowMimick}
\end{figure*}

Redshifted emission present between 5.38 and 9.16 $\si{{km}.s^{-1}}$ is mimicked between 9.34 and 13.84 $\si{{km}.s^{-1}}$ at lower intensity. In other words, the outflow/emission shapes seen between 5.38 and 9.16 $\si{{km}.s^{-1}}$ are also present between 9.34 and 13.84 $\si{{km}.s^{-1}}$ with lower overall intensity in each channel. This phenomenon is displayed in Fig. \ref{fig:12NOutflowMimick}, which shows a channel map of $^{12}$CO emission towards the tail-end of the redshifted velocity range (9.34 to 13.66 $\si{{km}.s^{-1}}$) with a channel spacing of 0.18 $\si{{km}.s^{-1}}$. As channel velocities increase towards the tail end of the data cube, the secondary northern cavity dissipates while the primary northern cavity remains prevalent. The primary cavity maintains its shape and intensity up to 23.56 $\si{{km}.s^{-1}}$, at which point it quickly disappears.

\subsection{\texorpdfstring{$^{13}$CO observations}{Lg}}
\label{sec:13CO-Obs}
\begin{figure}
    \centering
    \includegraphics[width=\columnwidth]{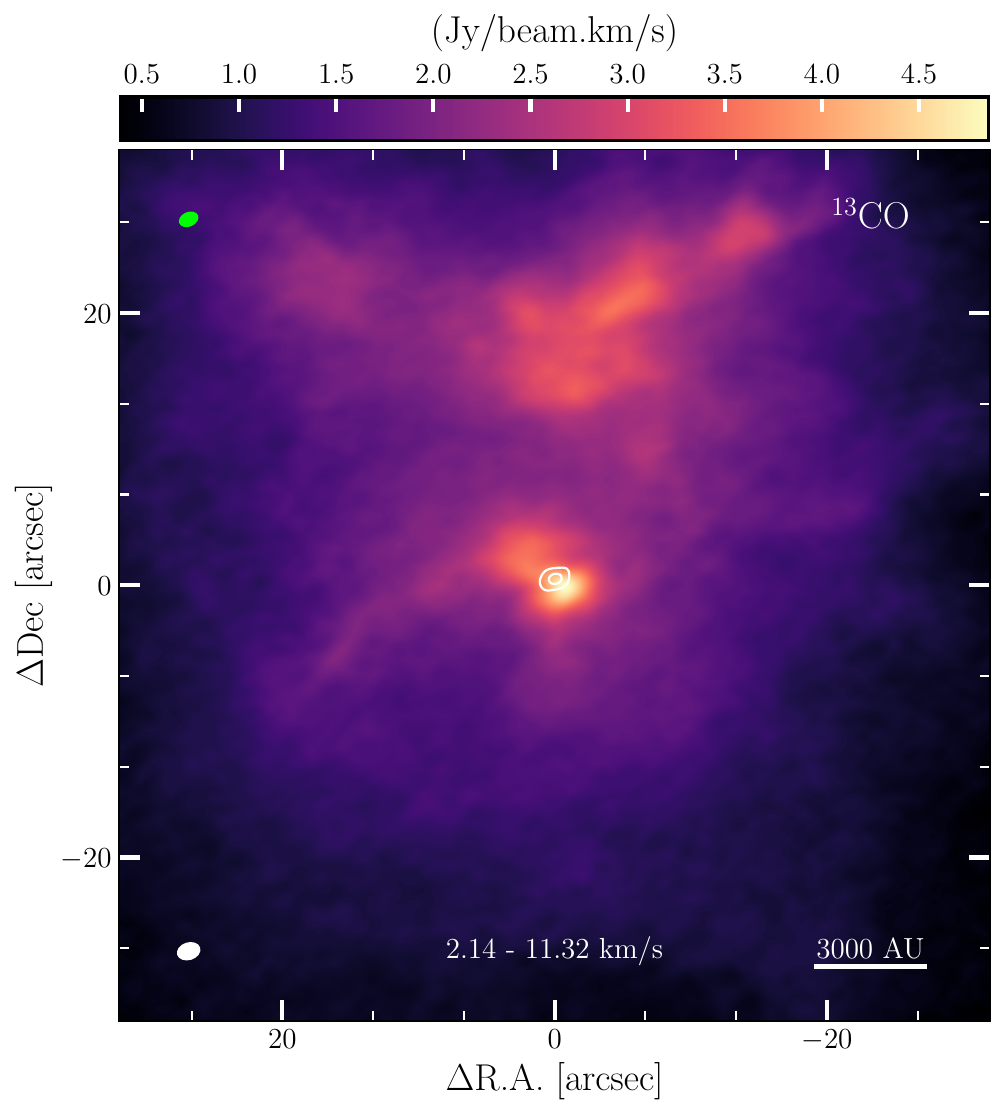}
    \caption{A moment-0 map of $^{13}$CO emission as observed by the ALMA combination array, and integrated between 2.14 and 11.32 $\si{{km}.s^{-1}}$. White contours representing 232.5 GHz continuum emission are shown at the centre of the frame at levels of 0.25 and 0.75 times the maximum intensity of 0.11 Jy/beam. The synthesized beam is shown in the bottom left of the frame with a size of 1.77 $\times$ 1.30 arcseconds, and a position angle of -72.3$\degr$. In addition, the continuum beam is shown in green in the top left corner, and maintains a size of 1.50 $\times$ 1.07 arcseconds and a position angle of -66.8$\degr$.}
    \label{fig:13EnvelopeFull}
\end{figure}

To trace the stellar envelope, as well as interactions between the envelope and HBC 494's outflows, we produced a $^{13}$CO data cube encompassing emission from -20 to 33.82 $\si{{km}.s^{-1}}$ with a channel width of 0.18 $\si{{km}.s^{-1}}$. Although $^{13}$CO detections are most prevalent near the systemic velocity ($4.3\pm0.7$ $\si{{km}.s^{-1}}$), emission is generally present between 2.14 and 11.5 $\si{{km}.s^{-1}}$. As with $^{12}$CO emission, $^{13}$CO detections can be separated into three different velocity ranges: blueshifted (2.14 -- 3.40 $\si{{km}.s^{-1}}$), systemic (3.58 -- 5.02 $\si{{km}.s^{-1}}$), and redshifted (5.2 -- 11.32 $\si{{km}.s^{-1}}$). Figure \ref{fig:13EnvelopeFull} displays a velocity integration between 2.14 and 11.32 $\si{{km}.s^{-1}}$. There are two notable features in this figure, the first being the concentration of envelope material around the central binary system, and a cloud of material north of the binary. The former feature illustrates outflow material and the stellar envelope near systemic velocities, while the latter is indicative of interactions between the stellar envelope and the secondary northern outflow cavity.\par
\begin{figure*}
    \centering
    \includegraphics[width=\textwidth]{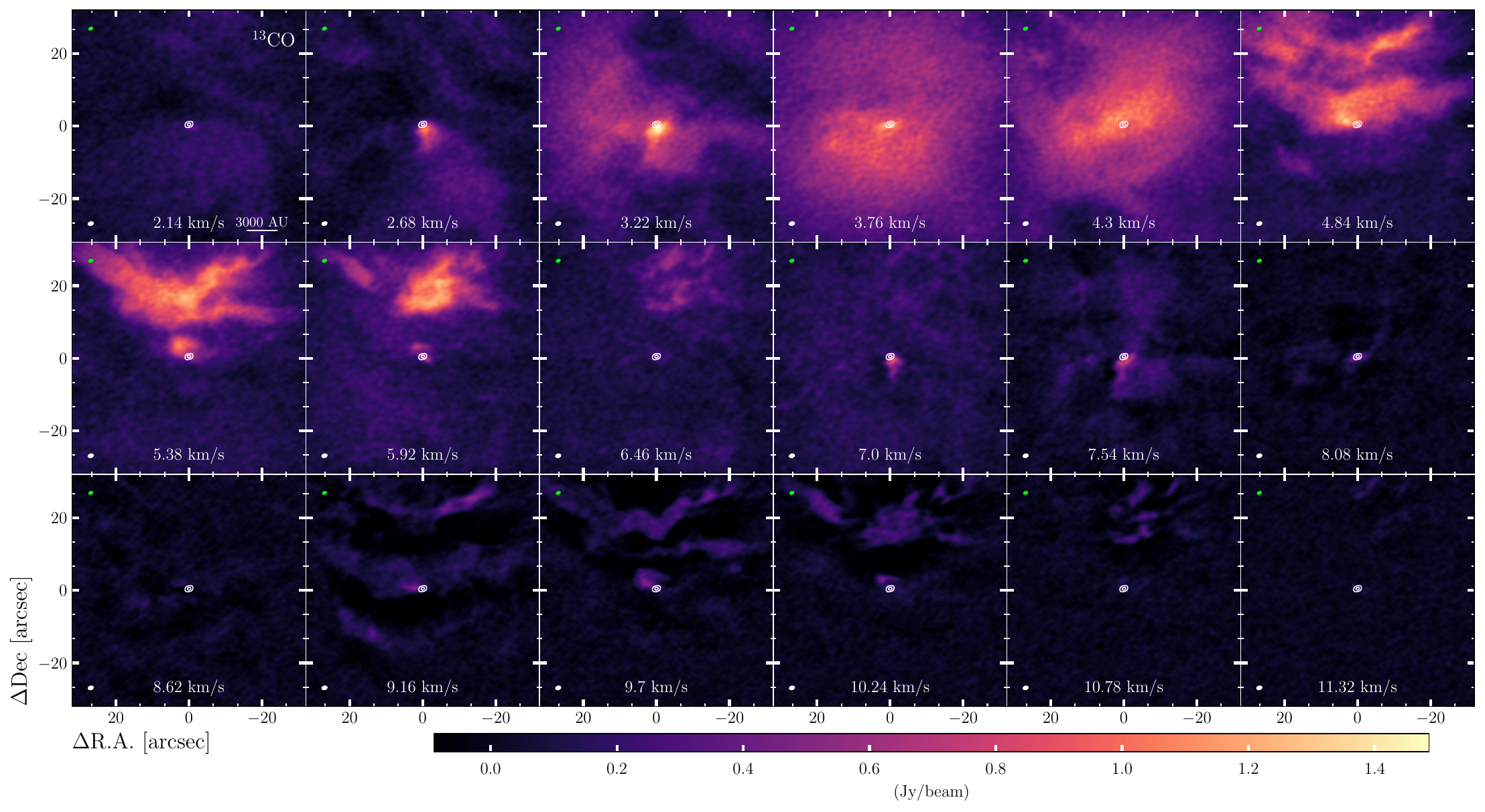}
    \caption{A channel map of $^{13}$CO emission detected between 2.14 and 11.32 $\si{{km}.s^{-1}}$, with each frame at a velocity spacing of 0.54 $\si{{km}.s^{-1}}$. The white contours at the centre of each frame represent continuum emission centered at 232.5 GHz at levels of 0.25 and 0.75 the maximum detected intensity of 0.11 Jy/beam. The synthesized beam is shown as a white ellipse in the bottom left of each frame, and has an angular size and position angle of 1.77 $\times$ 1.30 arcseconds and -72.3$\degr$, respectively. The continuum beam is shown in the top left by a green ellipse with an angular size of 1.50 $\times$ 1.07 arcseconds and a position angle of -66.8$\degr$}
    \label{fig:13EnvelopeCHMap}
\end{figure*}

In order show how $^{13}$CO detections change across a range of velocity channels, Figure \ref{fig:13EnvelopeCHMap} displays emission between 2.14 and 11.32 $\si{{km}.s^{-1}}$ with spacing of 0.54 $\si{{km}.s^{-1}}$. 
Firstly, blueshifted emission at velocities between 2.14 and 3.4 $\si{{km}.s^{-1}}$ traces an interaction between the western arm of the southern outflow and material from the stellar envelope. Emission in this region shifts from the southwest towards the northeast with increasing size as velocity approaches the systemic value. $^{13}$CO emission also takes the shape of the southern outflow as traced with $^{12}$CO at 3.04 and 3.22 $\si{{km}.s^{-1}}$, indicating that envelope material is entrained within the wide-angle wind of the outflow, a conclusion consistent with observations from \citetalias{2017MNRAS.466.3519R}. A wing of emission is also seen on HBC 494's western side at these velocities, although there is no overlap between either of the primary outflows and this wing at blueshifted, systemic, or redshifted velocities. As mentioned in section \ref{sec:12COObs}, the western arm of the southern outflow shows much less extension than its eastern counterpart, specifically at velocities where $^{13}$CO detections are prominent, meaning that the stellar envelope could be obfuscating it from view. Alternatively, a compact region of $^{13}$CO could be entrained in this arm of the outflow, thus making it appear as a clump rather than mirroring the shape of the eastern arm.\par
Near systemic velocities ($4.3\pm0.7$ $\si{{km}.s^{-1}}$), $^{13}$CO continues to trace envelope motion, extending further towards the northeast approaching redshifted velocities. Interactions between each of the primary outflows and the stellar envelope may also be accentuating this feature at the extremes of this velocity range. $^{13}$CO emission reaches its largest angular area within this velocity range at approximately $50\arcsec\times50\arcsec$. Starting at 4.48 $\si{{km}.s^{-1}}$, along with rotating envelope emission in the northwest, $^{13}$CO takes the shape of the primary cavity as observed with $^{12}$CO. This continues up to 6.82 $\si{{km}.s^{-1}}$ when $^{13}$CO emission dissipates and the envelope is no longer detected. Significant interactions between the secondary cavity and envelope material are also observed upwards of 8.8 $\si{{km}.s^{-1}}$.\par
A `wave' of emission propagates upwards at increasing velocities, which most likely represents envelope material entrained in the secondary outflow's wind. $^{13}$CO emission entrained in both the primary and secondary northern outflow cavities maintains a much larger opening angle in comparison to material entrained at blueshifted velocities, possibly due to an interaction between the outflows of HBC 494a and HBC 494b. 

\subsection{\texorpdfstring{C$^{18}$O observations}{Lg}}
\begin{figure*}
    \centering
    \includegraphics[width=\textwidth]{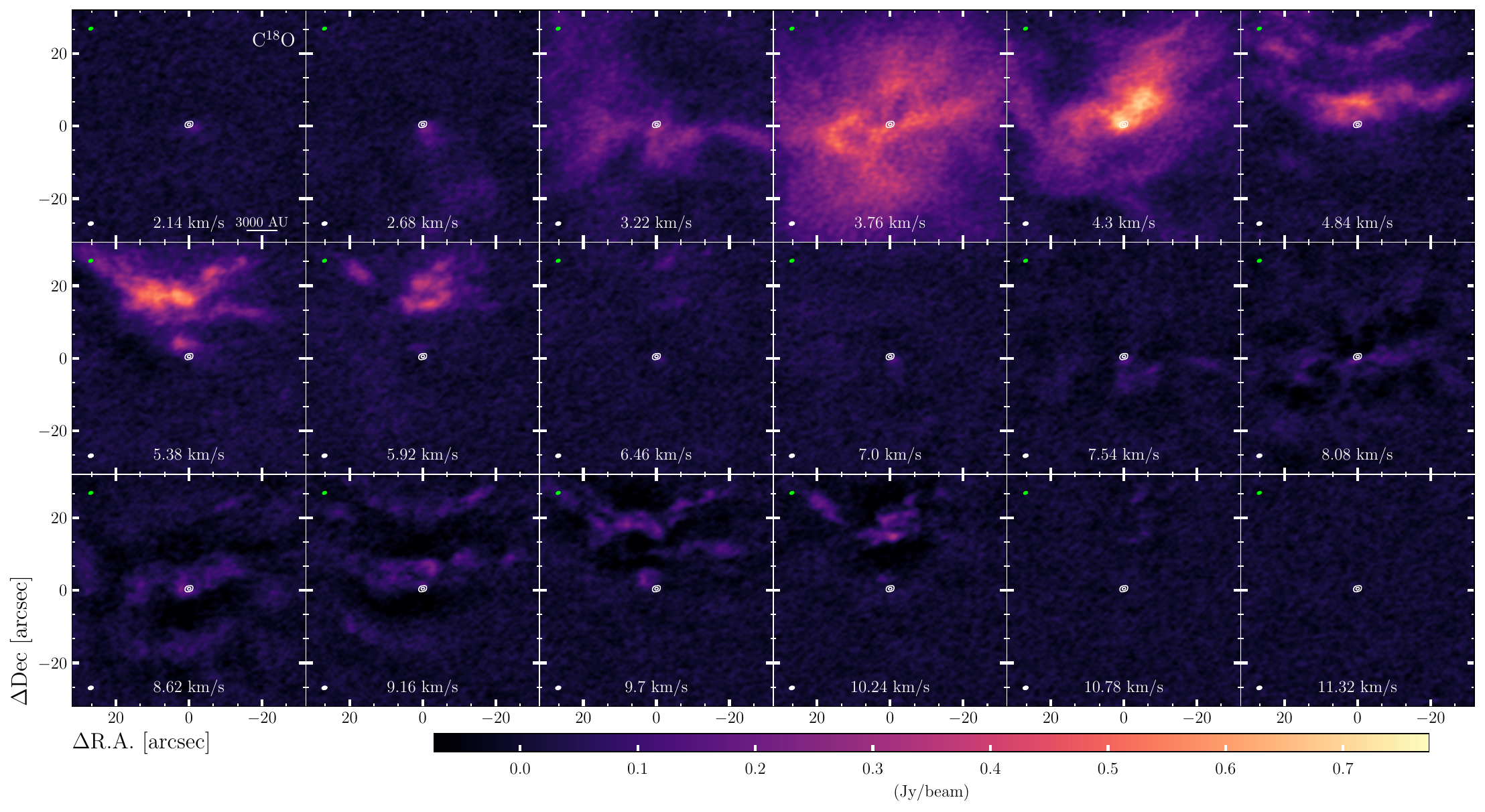}
    \caption{A channel map of C$^{18}$O emission detected between 2.14 and 11.32 $\si{{km}.s^{-1}}$, with each frame at a velocity spacing of 0.54 $\si{{km}.s^{-1}}$. The white contours at the centre of each frame represent continuum emission centred at 232.5 GHz at levels of 0.25 and 0.75 the maximum detected intensity of 0.11 Jy/beam. The synthesised beam is shown as a white ellipse in the bottom left of each frame, and has an angular size and position angle of 1.78 $\times$ 1.28 arcseconds and -74.6$\degr$, respectively. The continuum beam is shown in the top left by a green ellipse with an angular size of 1.50 $\times$ 1.07 arcseconds and a position angle of -66.8$\degr$}
    \label{fig:18Channel}
\end{figure*}

We produced a combination array C$^{18}$O data cube encompassing velocities identical to those in the $^{13}$CO data cube, again using a channel width of 0.18 $\si{{km}.s^{-1}}$. Because they both trace the stellar envelope, as well as interactions between the envelope and outflow cavities, C$^{18}$O exhibits very similar detection velocities to $^{13}$CO. Emission is detected between 2.32 and 11.14 $\si{{km}.s^{-1}}$, with the most intense and extended detections observed near systemic velocity (Fig. \ref{fig:18Channel}). 
Blueshifted emission at velocities between 2.86 and 3.4 $\si{{km}.s^{-1}}$ traces the motion of the stellar envelope envelope, as well as material entrained by the eastern arm of the primary southern outflow. Emission in this velocity range also shows the wing of material directly west of HBC 494 observed with $^{13}$CO, although with less extension and intensity -- most likely due to C$^{18}$O tracing high-density regions of the envelope. Near systemic velocities, C$^{18}$O is able to trace the shape of the stellar envelope. Specifically, channels with velocities between 4.12 and 5.02 $\si{{km}.s^{-1}}$ show the greatest extension of the envelope, as well as a strong resemblance to $^{13}$CO emission in the same velocity range.\par
\begin{figure*}
    \centering
    \includegraphics[width=\columnwidth]{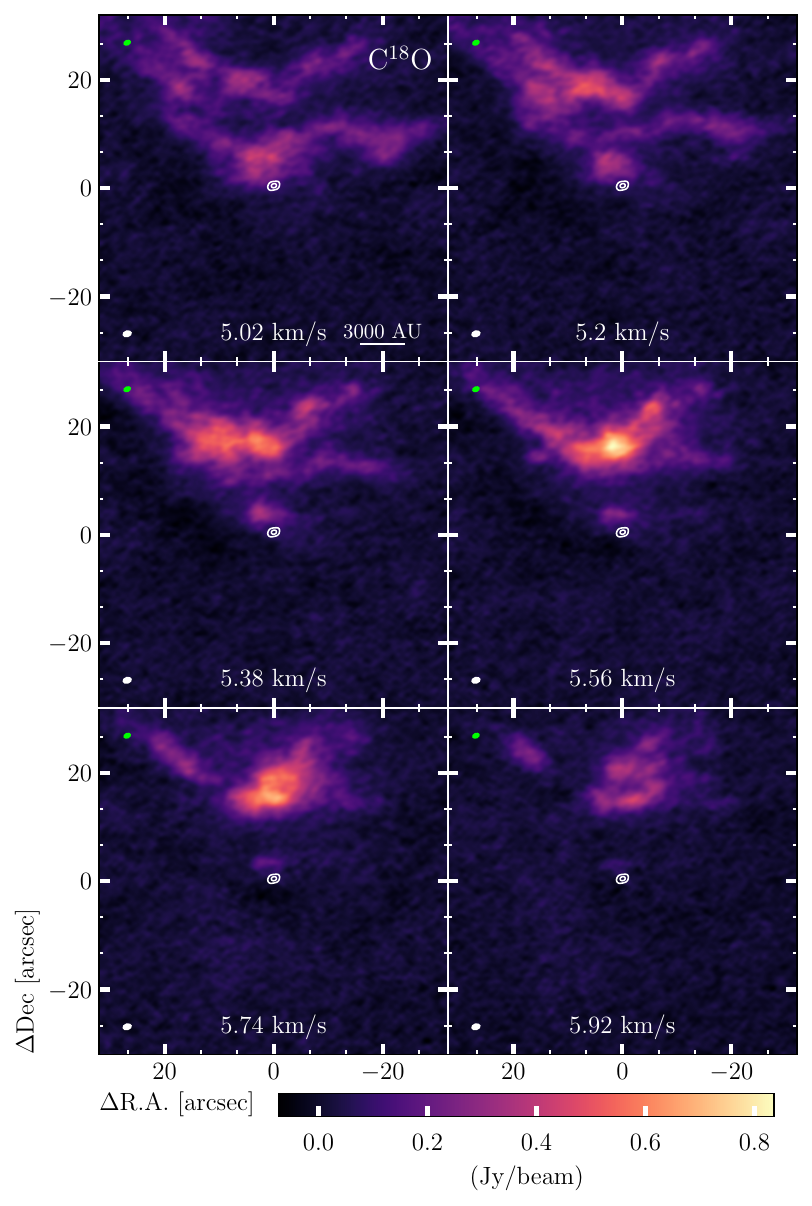}
    \includegraphics[width=\columnwidth]{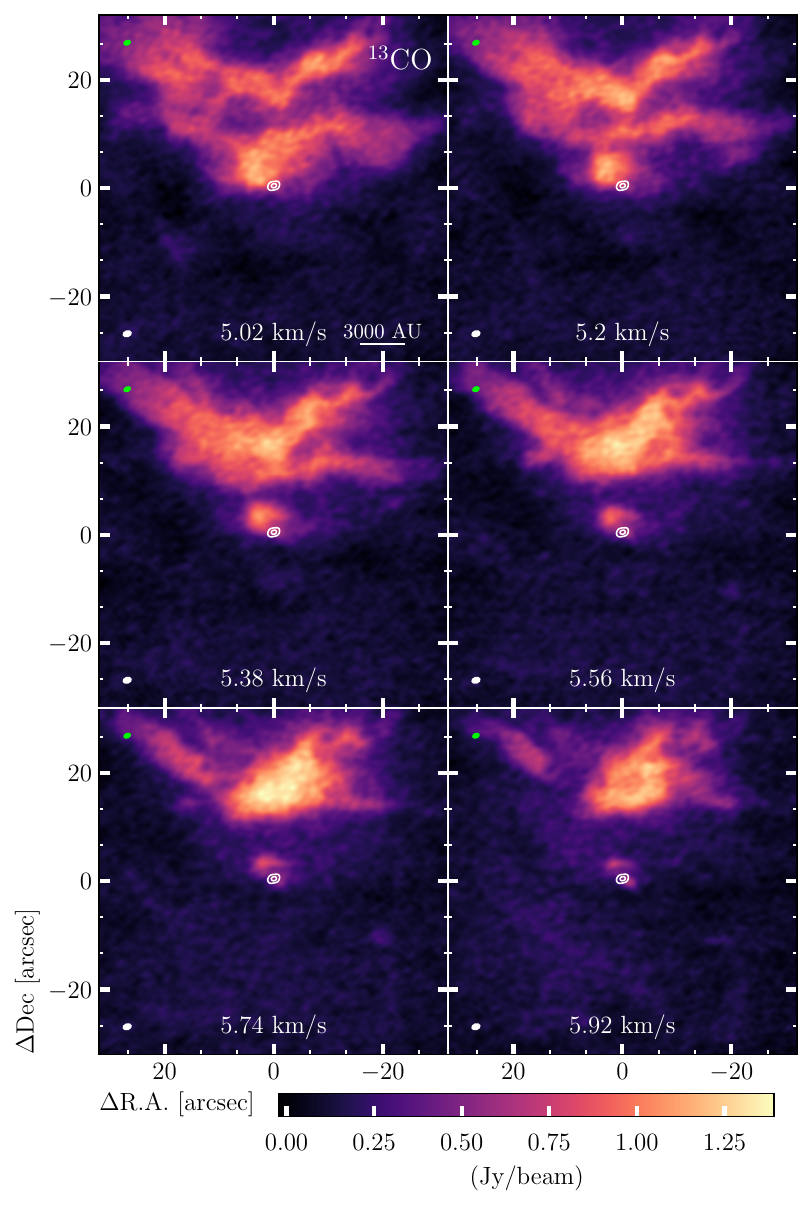}
    \caption{Two individual channel maps of C$^{18}$O (left) and $^{13}$CO (right) emission between channel velocities of 5.02 and 5.92 $\si{{km}.s^{-1}}$. The white contours at the centre of each frame represent continuum emission centred at 232.5 GHz at levels of 0.25 and 0.75 times the maximum detected intensity of 0.11 Jy/beam. The green ellipse in the top left of each frame represents the continuum beam, which has an angular size of 1.50 $\times$ 1.07 arcseconds and a position angle of -66.8$\degr$. The white ellipses visible in the bottom left of each frame in the left and right images represent the C$^{18}$O and $^{13}$CO synthesised beams, respectively. The former has an angular size of 1.78 $\times$ 1.28 arcseconds and a position angle of -74.6$\degr$, while the latter maintains an angular size of 1.77 $\times$ 1.30 arcseconds and a position angle of -72.3$\degr$.}
    \label{fig:1813Comp}
\end{figure*}

Interactions between high-density envelope regions and the primary northern outflow are traced at 5.02 -- 6.28 $\si{{km}.s^{-1}}$. Material present at 5.02 $\si{{km}.s^{-1}}$ forms into bands that appear on either side of HBC 494 and propagate upwards with increased intensity as channel velocity increases to 5.92 $\si{{km}.s^{-1}}$. At this velocity, the bands coalesce into a clump of material which is visible up to 6.28 $\si{{km}.s^{-1}}$. This clump of material is also seen in the same velocity range in $^{12}$CO and $^{13}$CO images, and is likely due to the mixing of two different outflows with differing position angles. This phenomenon is discussed further in section \ref{sec:SecondaryCavity}. This clump, as well as the bands of material, are also seen in $^{13}$CO observations in the same velocity range, although with greater overall extension and higher intensity compared to C$^{18}$O (Fig. \ref{fig:1813Comp}). 
These bands are also detected with nearly identical shapes from 9.34 to 9.88 $\si{{km}.s^{-1}}$, although with much lower intensity (Fig. \ref{fig:18Shadow}).

\subsection{SO observations}
\label{sec:SO}
SO traces shocked material \citep[e.g.][]{2004A&A...415.1021J}, with detections present from 1.6 up to 10.6 $\si{{km}.s^{-1}}$. To map these detections, we created a data cube with the same velocity range as the three prior isotopologues (-20 -- 33.82 $\si{{km}.s^{-1}}$) with a channel width of 0.18 $\si{{km}.s^{-1}}$. SO is able to trace the central object, as well as certain portions of the secondary northern outflow cavity.\par Beginning with the former, emission is present in two distinct velocity ranges: 1.6 to 5.02 $\si{{km}.s^{-1}}$, and 6.1 to 9.52 $\si{{km}.s^{-1}}$ (Figs. \ref{fig:SOfirst} $\&$ \ref{fig:SOsecond}). At the front end first range SO appears as a small lobe at the southwest of HBC 494, which grows in both size and intensity as channel velocity approaches the systemic value. Additionally, SO shifts from the southwest to the northeast, similar to the motion of $^{13}$CO and C$^{18}$O in this velocity range.\par
\begin{figure*}
    \centering
    \includegraphics[width=\textwidth]{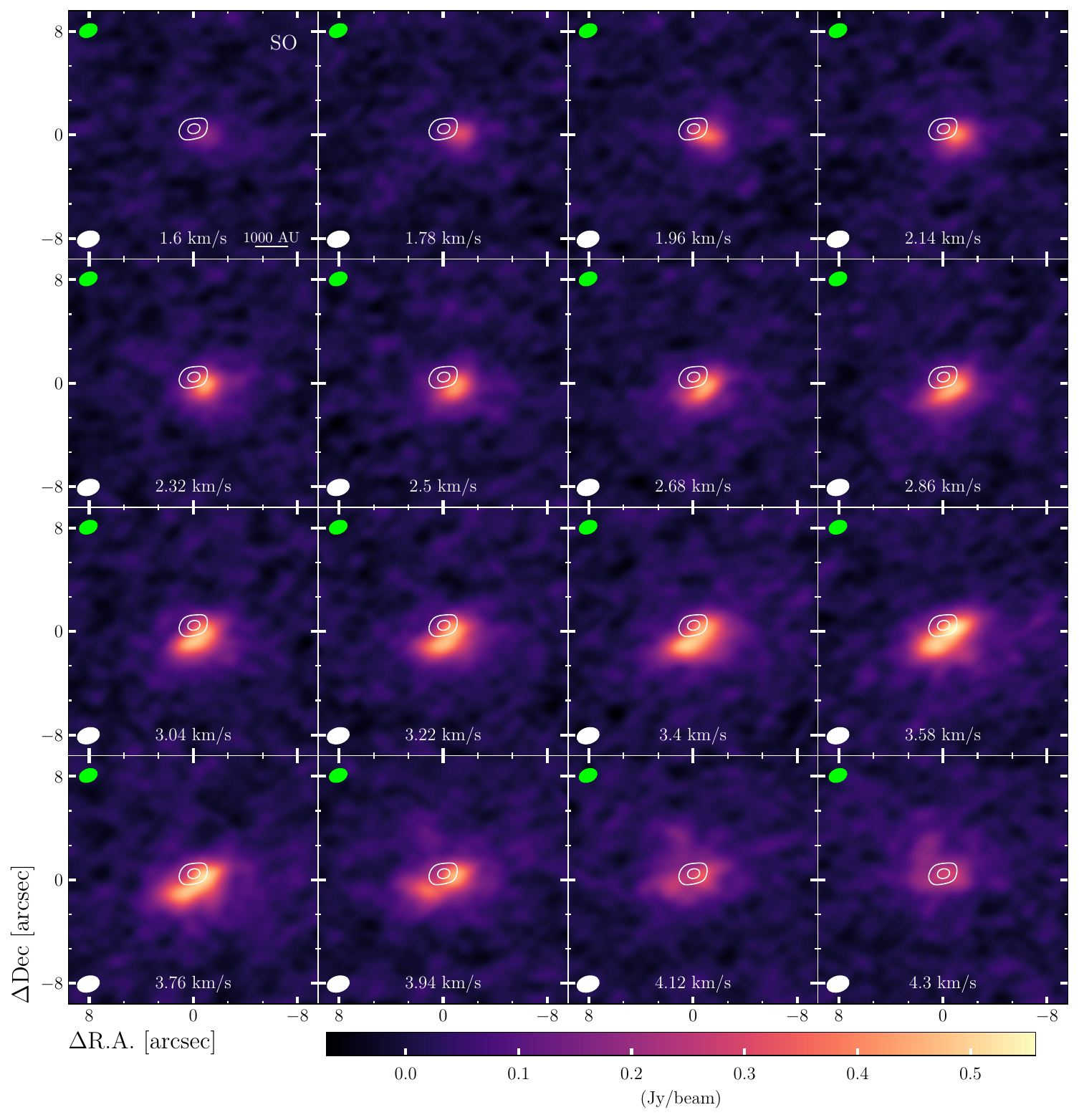}
    \caption{An individual channel map of SO emission between velocities of 1.6 and 4.3 $\si{{km}.s^{-1}}$. The white contours at the centre of each frame represent continuum emission centred at 232.5 GHz, and are shown at levels of 0.25 and 0.75 times the maximum detected intensity of 0.11 Jy/beam. The SO synthesised beam and continuum emission beam are shown in the bottom and top left of each frame, respectively. The former maintains a size and position angle of 1.81 $\times$ 1.28 arcseconds and -72.3$\degr$, while the latter has a size of 1.50 $\times$ 1.07 arcseconds and a position angle of -66.8$\degr$.}
    \label{fig:SOLobes}
\end{figure*}

At 3.58 $\si{{km}.s^{-1}}$ two distinct lobes form with a position angle near that of HBC 494's constituent objects, and are present up to 4.3 $\si{{km}.s^{-1}}$ \citep{2023MNRAS.tmp.1574N}. However, these lobes do not align with continuum observations of the circumstellar disc, either in their position or separation. HBC 494a and HBC 494b maintain a separation of approximately 75 au \citep{2023MNRAS.tmp.1574N}, while the observed SO lobes are separated on average by 830 au. The lobes are also displaced to the south of HBC 494 in all velocity channels where they are intense enough to be seen despite their northeastward motion. Figure \ref{fig:SOLobes} shows a channel map of SO emission between 1.6 and 4.3 $\si{{km}.s^{-1}}$, and illustrates the velocity evolution of the aforementioned lobes as well as their displacement from continuum observations.
SO emission in this velocity range is also spatially coincident with intense $^{13}$CO regions, suggesting that envelope interactions could be responsible for the asymmetric distribution of SO near the disc (Fig. \ref{fig:13SOoverlap}). The SO emission shapes shown in the prior two figures are also present from 6.1 to 9.7 $\si{{km}.s^{-1}}$, although with lower intensity and extension. In addition to mimicking emission seen in the prior velocity range, SO emission in these channels also closely resembles $^{13}$CO surrounding the central binary, again with lower intensity and extension.\par
\begin{figure*}
    \centering
    \includegraphics[width=\columnwidth]{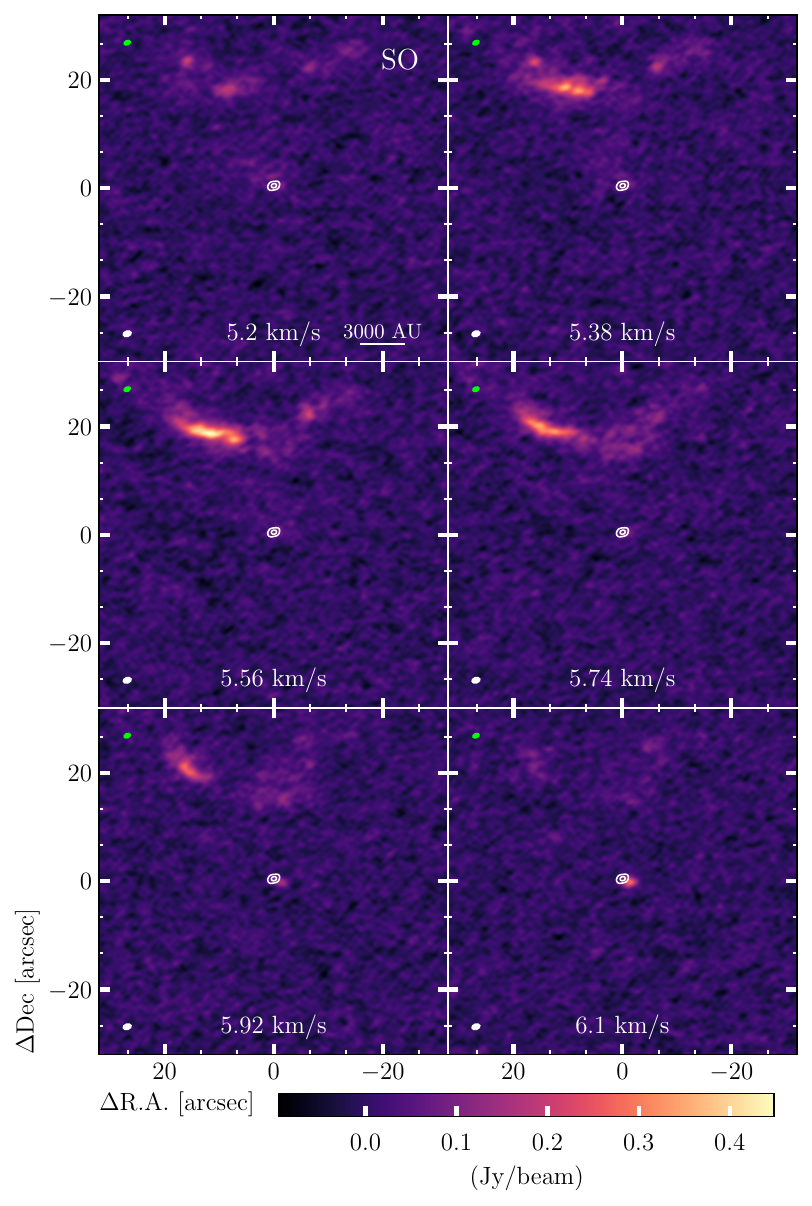}
    \includegraphics[width=\columnwidth]{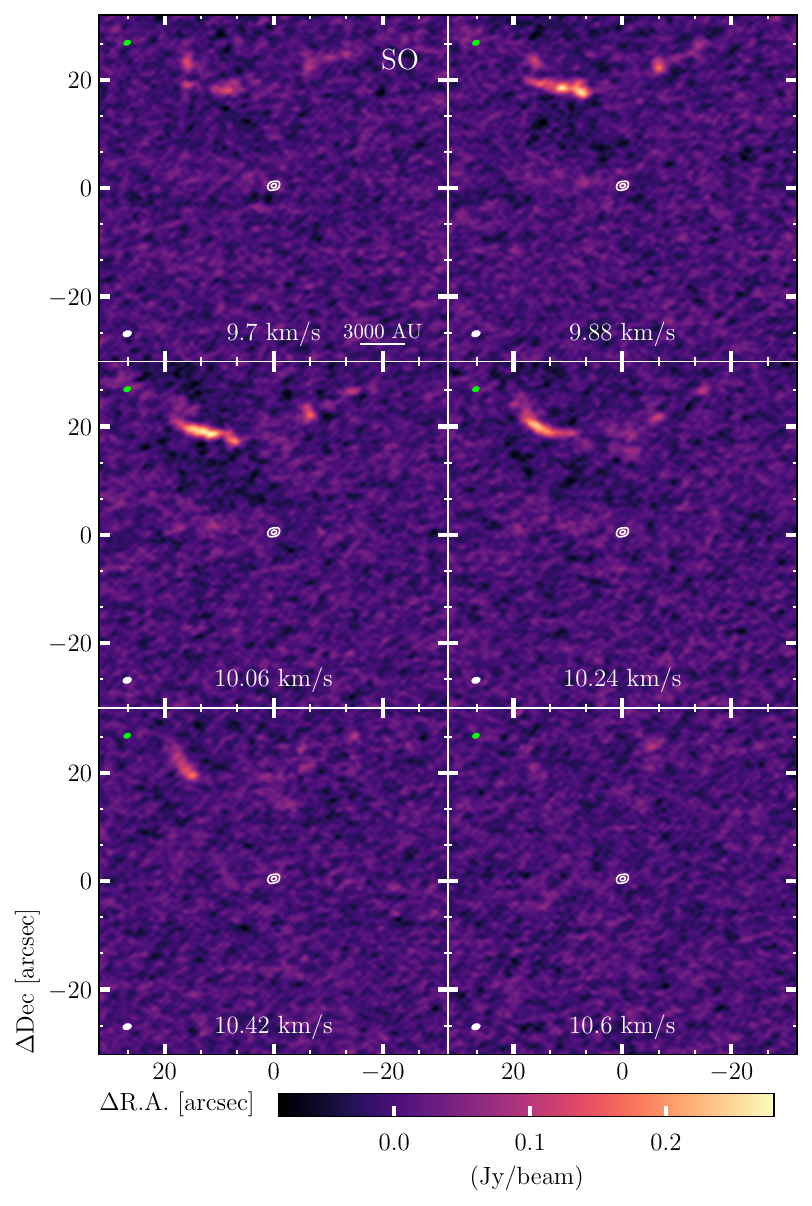}
    \caption{Two channel maps of SO emission between 5.2 and 6.1 $\si{{km}.s^{-1}}$ (left) and 9.7 and 10.6 $\si{{km}.s^{-1}}$ (right). Shown in the centre of each frame are white contours which represent continuum emission, and are at levels of 0.25 and 0.75 times the maximum detected intensity of 0.11 Jy/beam. The bottom and top left of each frame also show the synthesized and continuum beams, respectively. The former has an angular size of 1.81 $\times$ 1.28 arcseconds and a position angle of -72.3$\degr$, while the latter has an angular size of 1.50 $\times$ 1.07 arcseconds and a position angle of -66.8$\degr$.}
    \label{fig:SOCavityComp}
\end{figure*}

As for the secondary northern outflow cavity, detections are present in two distinct velocity ranges: 5.02 to 6.28 $\si{{km}.s^{-1}}$ and 9.34 to 10.6 $\si{{km}.s^{-1}}$. In both of these ranges SO primarily traces the eastern portion of the cavity, with very limited emission present on the western side (Fig. \ref{fig:SOCavityComp}).
The outflow regions traced by SO are also highlighted by $^{13}$CO and C$^{18}$O in the same velocity channels, suggesting that SO surrounding the circumstellar disc may have been entrained in the wind of a previous accretion outburst.

\subsection{P-V Diagrams}
\label{sec:PV}

\begin{figure}
    \centering
    \includegraphics[width=\columnwidth]{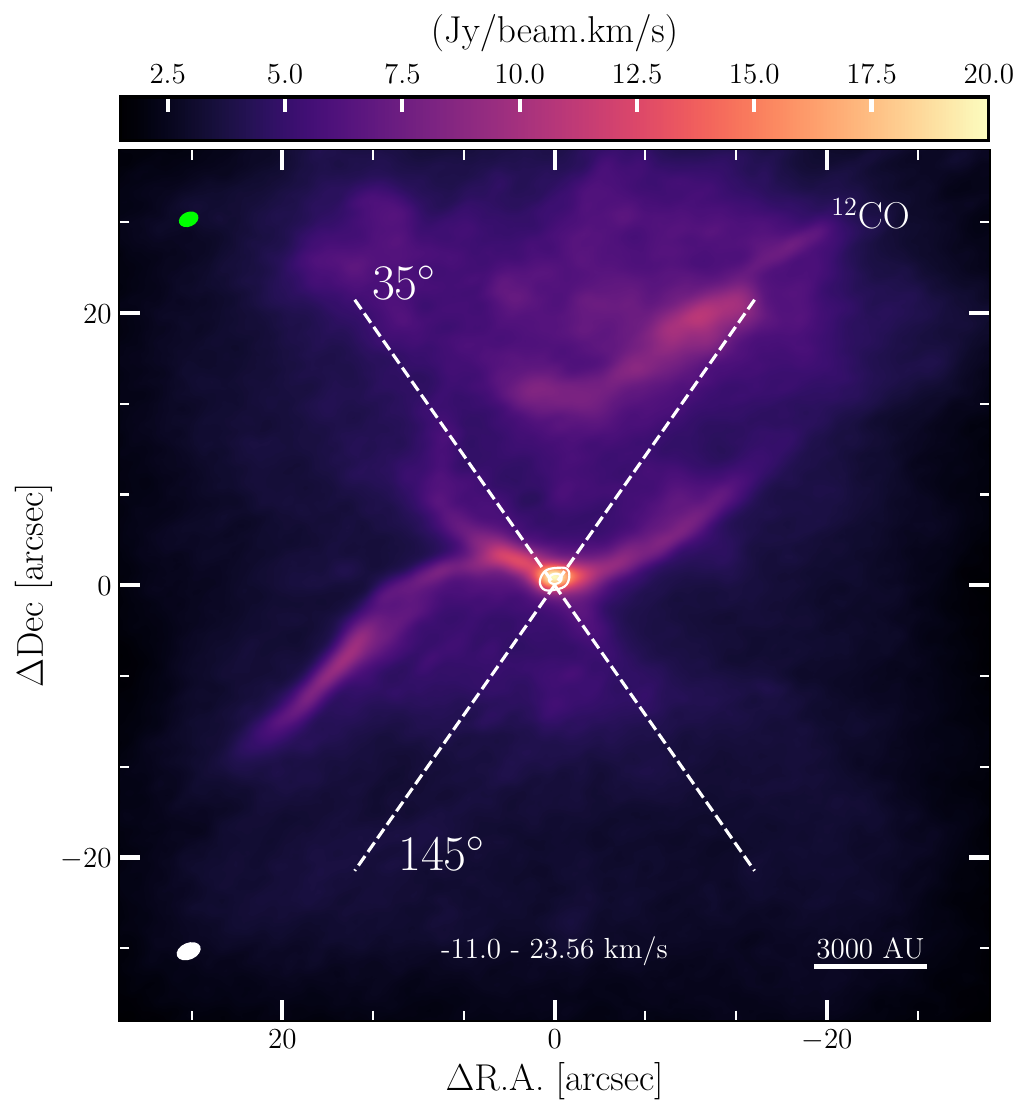}
    \caption{The moment-0 map displayed in Figure \ref{fig:12OutflowAB}c with two overlaid P.A. lines. These lines represent the cuts used to generate the P-V diagrams shown in Figures \ref{fig:PV} and \ref{fig:PV2}.}
    \label{fig:PVLines}
\end{figure}

\begin{figure*}
    \centering
    \includegraphics[width=0.7\columnwidth]{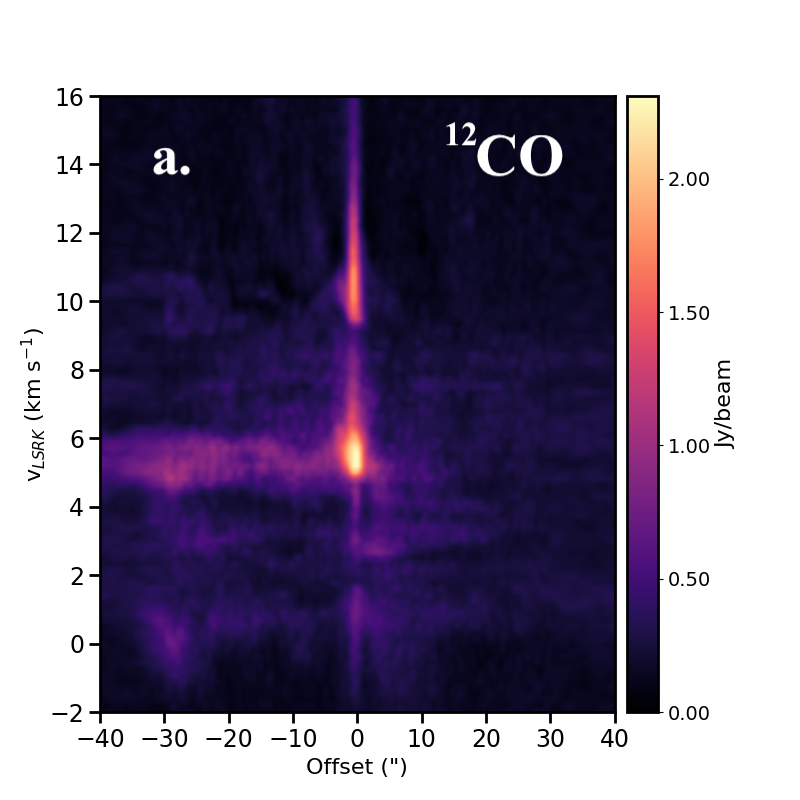}
    \includegraphics[width=0.7\columnwidth]{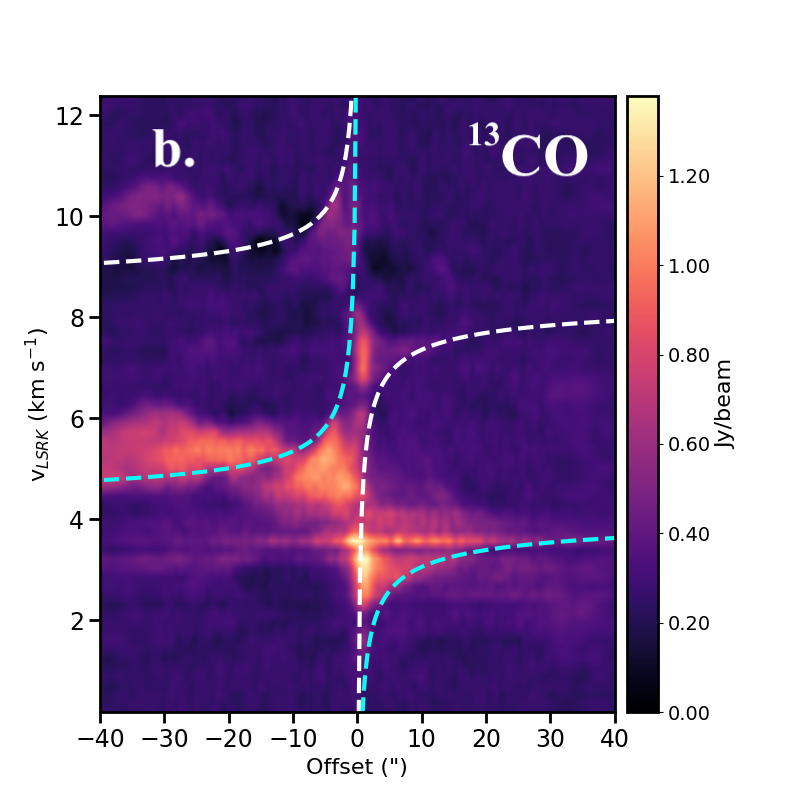} \\
    \includegraphics[width=0.7\columnwidth]{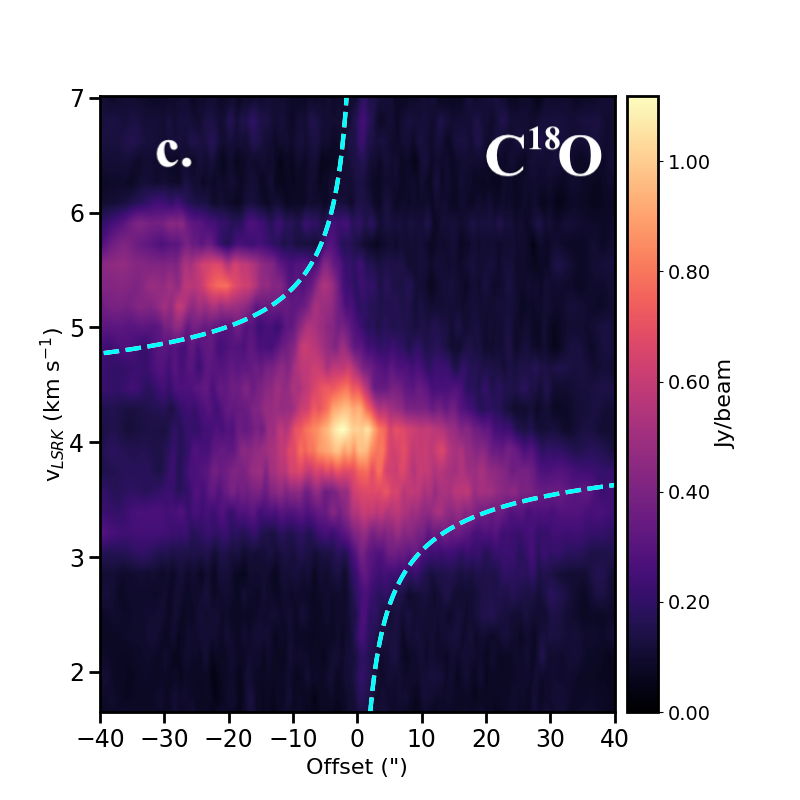} 
    \includegraphics[width=0.7\columnwidth]{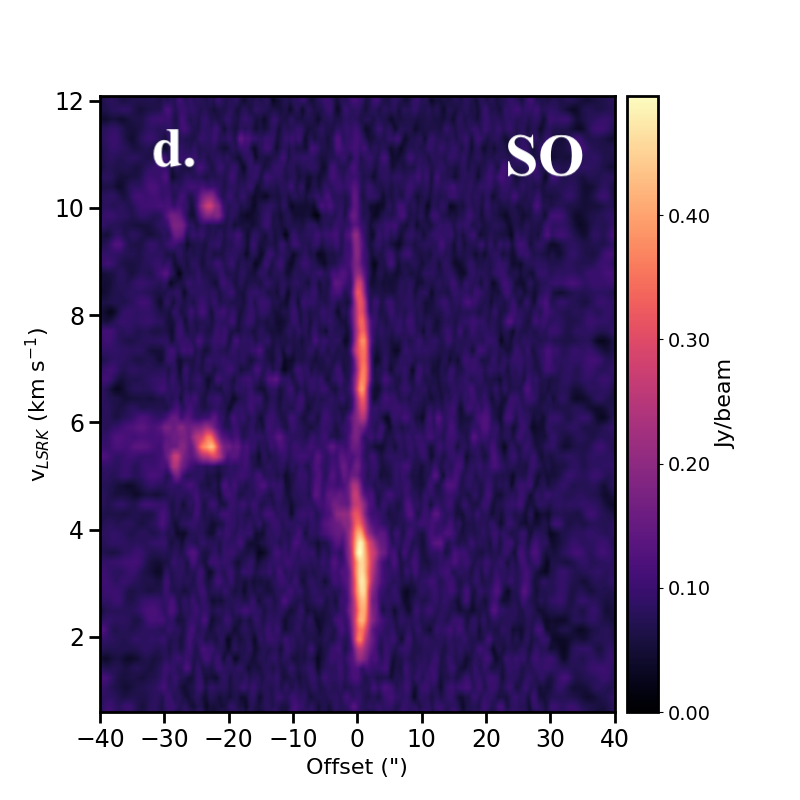}
    \caption{P-V diagrams of the $^{12}$CO, $^{13}$CO, C$^{18}$O, and SO emission obtained perpendicular to the binary system, i.e., P.A. $\sim35\degr$.}
    \label{fig:PV}
\end{figure*}

In order to scrutinise the kinematics and spatial distribution of the detected molecules tracing the bipolar lobes, we build Position-Velocity (P-V) diagrams perpendicular to the position of the binary system, and along the outflow axis traced by the blueshifted $^{12}$CO emission, i.e., position angles (P.A.) of $\sim35\degr$, and $\sim145\degr$, respectively. The position angles used for the cuts are shown in Figure \ref{fig:PVLines}.
Figure \ref{fig:PV} displays the resulting P-V diagram cuts perpendicular to the position of the binary system. $^{12}$CO clearly traces material characteristic of outflow/infall motions near the central binary system \cite[e.g.][]{Tobin2012} with strong blueshifted emission self-absorbed, and an extended emission detected in the velocity range 4.3 $\leq$ V$\rm _{sys}$ $\leq$ 6.0 km s$^{-1}$. The latter feature is associated with emission from the secondary northern cavity. On the other hand, the $^{13}$CO P-V diagram displays mainly Keplerian velocities that may be associated with winds or material close to and interacting with the binary star. In Fig. \ref{fig:PV}, the dashed white lines represent the expected Keplerian velocities for a system of 8 M$_{\sun}$ at an inclination of $\sim65\degr$, a distance to the object of 414 pc, and a V$\rm _{sys}$ of $\sim$ 4.3 km s$^{-1}$ (\citetalias{2017MNRAS.466.3519R}). These Keplerian curves place an upper limit of 8 M$_{\sun}$ for the central stellar masses, HBC 494a and HBC 494b. In addition, and similar to $^{12}$CO, an extended $^{13}$CO feature is extracted along a position angle of $\sim35\degr$. In a similar way to $^{13}$CO, and with expected higher densities, C$^{18}$O emission traces a combination of rotating and envelope/outflow material as seen in Fig. \ref{fig:PV}c. With a more compacted extension, SO emission displays outflowing velocities tracing material towards the central object together with an extra redshifted component that belongs to ejected material.\par

The most striking feature of the P-V diagram analysis is the extent of the $^{13}$CO and SO emission detected as a positive spatial oﬀset ($\sim$ 4.3 km s$^{-1}$ -- redshifted emission). The blue dashed lines in Fig. \ref{fig:PV}b display Keplerian velocities generated with similar parameters as described above except for a positive oﬀset of $\sim$ 4.3 km s$^{-1}$ added in the final outcome. The velocity components of $^{13}$CO and SO trace a similar region where the expelled material possibly -- after a massive and powerful outburst -- preserved rotational motion while expanding. The lack of a symmetric infall signature at a negative offset may come from asymmetrical infall and non-uniformity of the molecular cloud.

\begin{figure*}
    \centering
    \includegraphics[width=0.65\columnwidth]{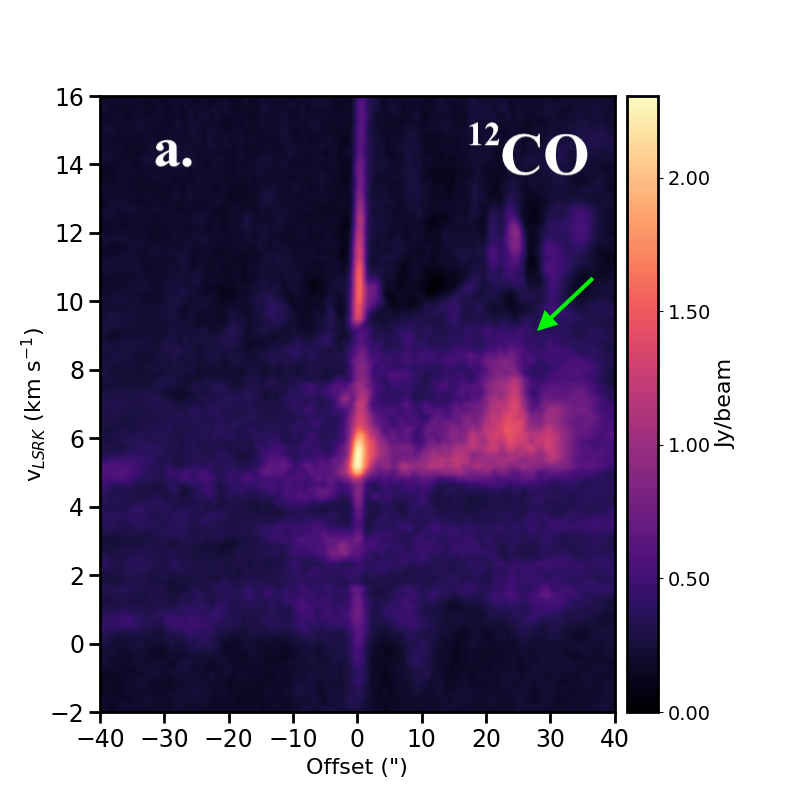} 
    \includegraphics[width=0.65\columnwidth]{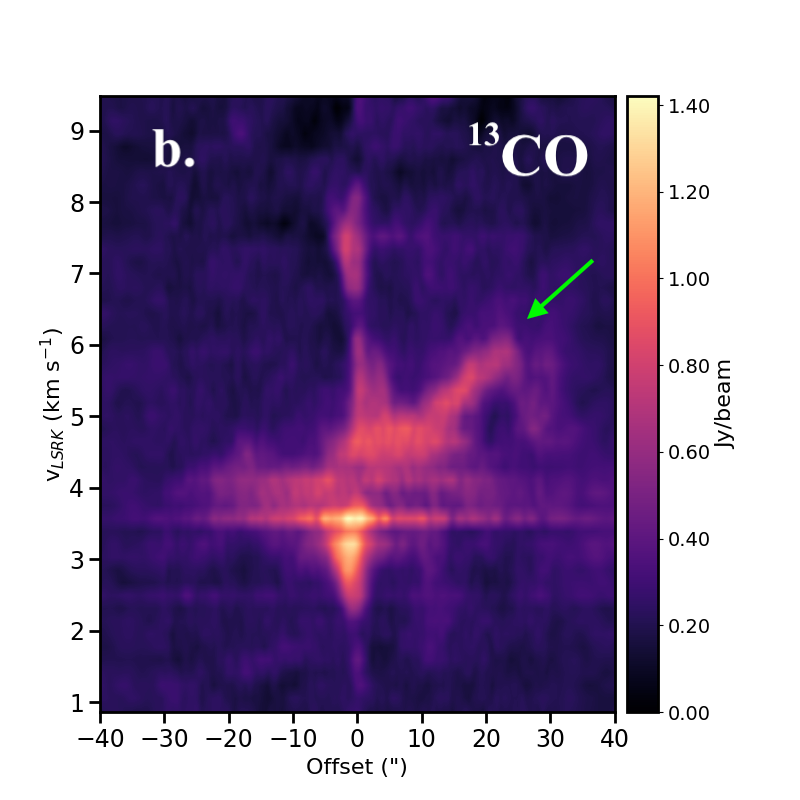} 
    \includegraphics[width=0.65\columnwidth]{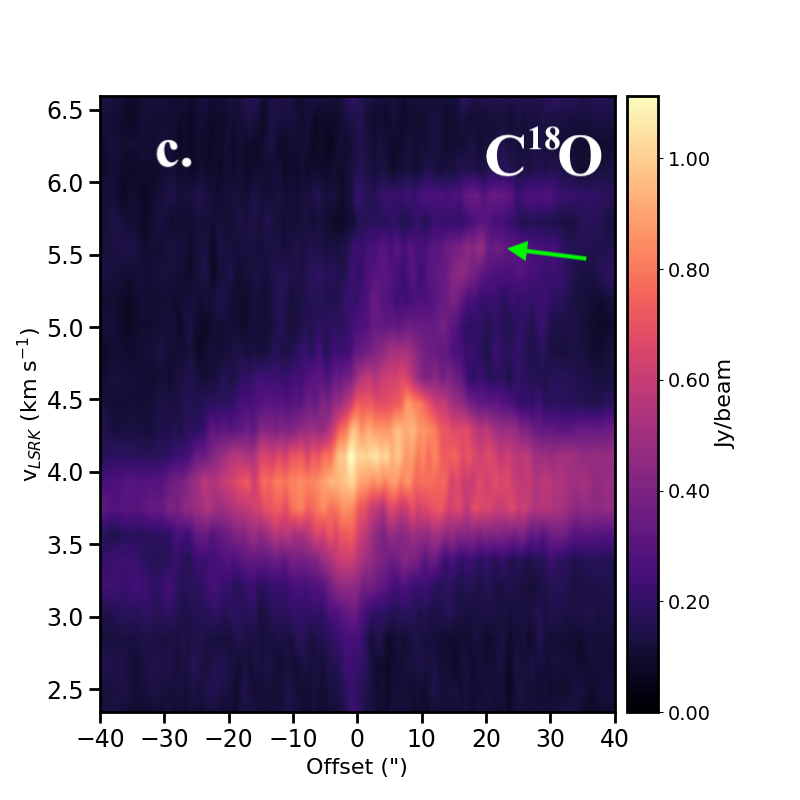} 
    \caption{P-V diagrams of the $^{12}$CO, $^{13}$CO, and C$^{18}$O emission obtained along the outflow rotational axis of P.A. $\sim145\degr$}
    \label{fig:PV2}
\end{figure*}

The P-V cuts also reveal a systematic velocity difference between symmetric positions from the central binary system axis and the secondary northern cavity. Figure \ref{fig:PV2} displays P-V diagrams of the $^{12}$CO, $^{13}$CO, and C$^{18}$O emission extracted along a position angle of $\sim145\degr$, i.e., along the outflow axis traced by blueshifted emission (\citetalias{2017MNRAS.466.3519R}). It is worth noting that there is a redshifted feature that extends toward the northwestern direction from an offset of $\sim$ 10$\arcsec$ to $\sim$ 20$\arcsec$ with a velocity range between $\sim$ 4.5 and 5.5 km s$^{-1}$. This elongated feature, which is indicated in Fig. \ref{fig:PV2} with a magenta arrow, is associated with emission from the secondary northern cavity. This suggests that either the molecular outflows exhibit intermittency or a molecular flow that erupted in two distinct epochs. 

Also, the position and velocity gradients suggest that the redshifted $^{12}$CO, $^{13}$CO, and C$^{18}$O emission traces an overlapping of two northern lobes (the redshifted lobe) with different position angles. It means that it is possible that HBC 494a and HBC 494b are associated with molecular outflows with different rotational axes and opening angles. Inspection of individual channels in the data cubes shows that $^{12}$CO and $^{13}$CO lines likely trace misaligned outflows with rotational axes of 35$\degr$ and 145$\degr$. Nevertheless, it is difficult to evaluate and separate the amount of molecular emission corresponding to each lobe of the bipolar outflows.

\section{Analysis}
\label{sec:CR}
\subsection{Outflow kinematics}
\label{sec:OK}
Because observations recorded with the combination array are able to detect large-scale emission more effectively than main array observations alone, we are able to more accurately determine the kinematic and dynamical properties of HBC 494's outflows. In order to determine the mass, momentum, kinetic energy, and luminosity of each primary outflow, as well as the secondary northern cavity, we make use of techniques detailed in \cite{2014ApJ...783...29D} and references therein.
Our calculations incorporate the H$_{2}$ column density $N_{\si{H_{2}}}$, mean molecular weight per hydrogen molecule $\mu_{\si{H_2}}$ (assumed to be 2.8 for a gas composed of 71$\%$ hydrogen, 27$\%$ helium, and 2$\%$ metals by mass), the mass of a hydrogen atom $m_{\si{H}}$, and the area per pixel $A_{\si{px}}$, to calculate the mass per pixel per velocity channel $M_{\si{v, px}}$:
\begin{equation}
    M_{\si{v, px}} = N_{\si{H_2}}m_{\si{H}}\mu_{\si{H_2}}A_{\si{px}}.
\end{equation}
Column density values are calculated using the methods described in \cite{2014ApJ...783...29D} appendix C. Specifically, we use the CO column density $N_{\si{CO}}$, and an abundance ratio of CO relative to H$_{2}$ of $X_{\si{CO}}=10^{-4}$ to infer $N_{\si{H_{2}}}$. The CO/H$_{2}$ abundance ratio is based on the $^{13}$CO/H$_{2}$ ratio of $2\times10^{-6}$ provided in \cite{2013MNRAS.431.1296R}, and then scaled to CO using the $^{12}$CO/$^{13}$CO ratio of 62 from \cite{1993ApJ...408..539L}. This CO/H$_{2}$ abundance is consistent with other publications using the same analysis procedures \citep[e.g.][]{2023ApJ...947...25H, 2015ApJ...803...22P}. The $N_{\si{CO}}$ calculation is also dependent on the excitation temperature of CO, $T_{\si{ex}}$. We performed one set of calculations using $T_{\si{ex}}=20$ K, and another using $T_{\si{ex}}=50$ K, as these temperatures best represent HBC 494's outflows and allow for straightforward comparison between this work and \citetalias{2017MNRAS.466.3519R}. The total mass of each outflow is then calculated by integrating $M_{\si{v, px}}$ across all pixels and velocities in which outflow emission is detected above a 3$\sigma$ threshold. 

Analogous to their respective classical calculations, the momentum and kinetic energy of each pixel in a given velocity channel are given by
\begin{equation}
    P_{\si{v, px}} = M_{\si{v, px}}|v-v_{\si{sys}}|
\end{equation}
and
\begin{equation}
    K_{\si{v, px}} = \frac{1}{2}M_{\si{v, px}}|v-v_{\si{sys}}|^2
\end{equation}
where again integrating over all relevant pixels and channels produces the total momentum and kinetic energy of the outflow \citep{2015ApJ...803...22P}. In addition, we make estimates of the characteristic extension and velocity for the northern and southern outflows. To determine the characteristic extension for an outflow, we average the length of multiple prominent features, namely the emission wings, cavity filling material, and other emission swept away from the ambient cloud. After accounting for all of these features, we find that the blueshifted outflow has a characteristic extension of 17$\arcsec$, while the redshifted outflow has an extension of 20$\arcsec$. The characteristic velocity is equal to the momentum of the outflow divided by its mass ($V_{\si{char}}=P/M$). With these values in place, the dynamical age of each outflow can be found with $T_{\si{dyn}}={R_{\si{char}}}/{V_{\si{char}}}$. 
 Finally, we provide estimates for outflow mass-loss rate ($\dot{M}=M/T_{\si{dyn}}$), luminosity ($L=K/T_{\si{dyn}}$), and force ($F=\dot{M}/V_{\si{char}}$).
 
To isolate all relevant outflow emission, kinematic and dynamical variable calculations made use of two regions: one encompassing the southern outflow at blueshifted velocities, and one encompassing the northern outflow at redshifted velocities. These regions also serve to exclude emission from HBC 494's disc. Furthermore, velocity channels which showed no emission from either the blueshifted or redshifted outflow, or which showed significant ambient cloud contamination, were excluded from calculations. Velocity channels between -8.3 and 2.32 $\si{{km}.s^{-1}}$ were integrated for blueshifted outflow calculations, and channels between 7.0 and 23.56 $\si{{km}.s^{-1}}$ were integrated for redshifted outflow calculations.

\subsection{Corrections}
\subsubsection{Opacity}
Due to $^{12}$CO and $^{13}$CO being optically thick near systemic velocity, outflow masses and other dynamical properties can be underestimated if optical depth effects are not properly accounted for. Using methods detailed in \cite{2014ApJ...783...29D}, \cite{2016ApJ...832..158Z}, and \cite{2023ApJ...947...25H}, we first correct the optical depth of the $^{13}$CO line using C$^{18}$O, and then use the corrected $^{13}$CO data to correct $^{12}$CO. The calculations necessary for these corrections are prefaced by the assumptions that excitation temperature and beam-filling factors are equivalent for each of the three lines, and that all emission is in local thermal equilibrium (LTE). With a dipole moment of 0.11D, CO is easily thermalized, and thus follows a Boltzmann distribution \citep{10.1063/1.460293}. Due to this easy thermalization, it is valid to assume that $^{12}$CO, $^{13}$CO, and C$^{18}$O have identical excitation temperatures and are in LTE, which greatly simplifies column density calculations. 
The $^{13}$CO correction process begins by determining the ratio of the mean brightness temperature between $^{13}$CO and C$^{18}$O, which can be stated as the following:
\begin{equation}
    \frac{T_{\si{mb,13}}(v)}{T_{\si{mb,18}}(v)}=\frac{1-\exp(-\tau_{v,13})}{1-\exp(-\tau_{v,18})}.
\end{equation}
$T_{\si{mb,13}}$ and $T_{\si{mb,18}}$ are the brightness temperatures of $^{13}$CO and C$^{18}$O respectively, while $\tau_{v,13}$ and $\tau_{v,18}$ represent the respective opacities of the two isotopologues. As opposed to $^{12}$CO and $^{13}$CO, C$^{18}$O is generally found to be optically thin \citep{2018ApJS..236...25K}, and thus the brightness temperature ratio can be approximated to 
\begin{equation} \label{eq:approx}
    \frac{T_{\si{{mb,13}}}(v)}{T_{\si{{mb,18}}}(v)}=X_{13,18}\frac{1-\exp(-\tau_{v,13})}{\tau_{v,13}}
\end{equation}
in which X$_{13,18}$ is the abundance ratio of $^{13}$CO to C$^{18}$O, and is taken to be 8.7 in accordance with \cite{1992A&ARv...4....1W}. Equation \ref{eq:approx} can then be rearranged to produce the $^{13}$CO correction factor, $C(v)_{13}$:
\begin{equation}
    C(v)_{13}=\frac{\tau_{v,13}}{1-\exp(\tau_{v,13})}\approx X_{13,18}\frac{T_{\si{mb,18}}(v)}{T_{\si{mb,13}}(v)}.
\end{equation}
This factor stems from the formula for calculating column density as a function of brightness temperature
\begin{equation}
    N_{2} = \frac{1}{\beta} \frac{8 \pi k {\nu ^{2}}}{A_{21}hc^{3}} \int{T_{\si{B}}dv}
\end{equation}
in which the correction factor is given by $\beta$ \citep{williams_2021}:
\begin{equation}
    \beta = \frac{1-\exp(-\tau_{\nu})}{\tau_{\nu}}.
\end{equation}
The corrected $^{13}$CO line is generated by multiplying the original line by the correction factor across corresponding velocity channels such that  $T_{\si{mb,13}}(v)'\equiv{C(v)_{13}T_{\si{mb,13}}(v)}$. The new $^{13}$CO data cube is now representative of optically thin emission, and can then be used to correct the $^{12}$CO line. Specifically, this method can be replicated with $^{12}$CO in place of the uncorrected $^{13}$CO line, the corrected $^{13}$CO line in place of C$^{18}$O line, and an abundance ratio of $X_{\si{12,13}}=62$ to produce the final opacity-corrected $^{12}$CO data cube.

\begin{figure}
    \centering
    \includegraphics[width=\columnwidth]{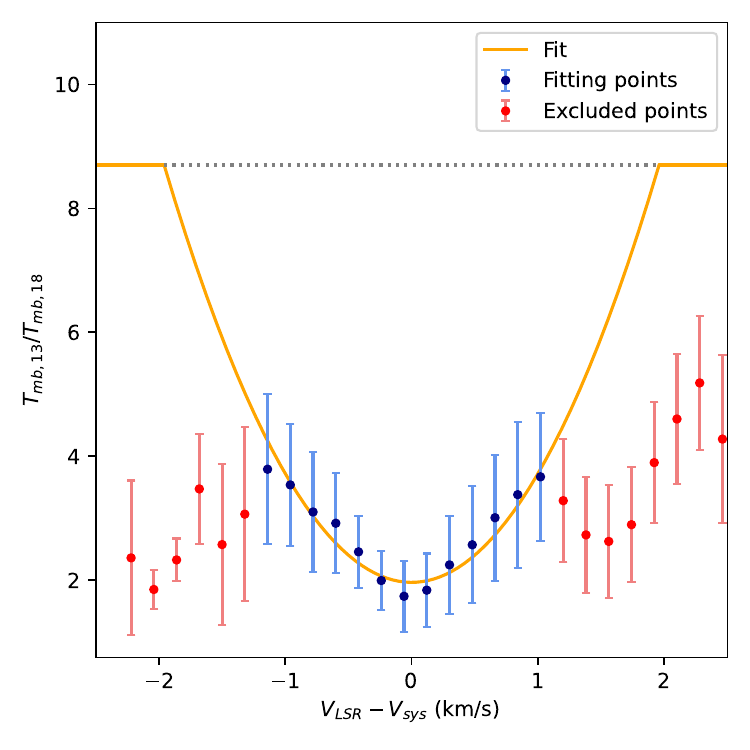}
    \caption{The weighted mean brightness temperature ratio between $^{13}$CO and C$^{18}$O as a function of $v-v_{\si{sys}}$. Each blue point represents those considered in the fitting process, while those in red were not included due to improper optical depth in $^{13}$CO. The data points used to generate the orange parabola encompass velocities between -1.14 and 1.02 $\si{{km}.s^{-1}}$.}
    \label{fig:13COFit}
\end{figure}

\begin{figure}
    \centering
    \includegraphics[width=\columnwidth]{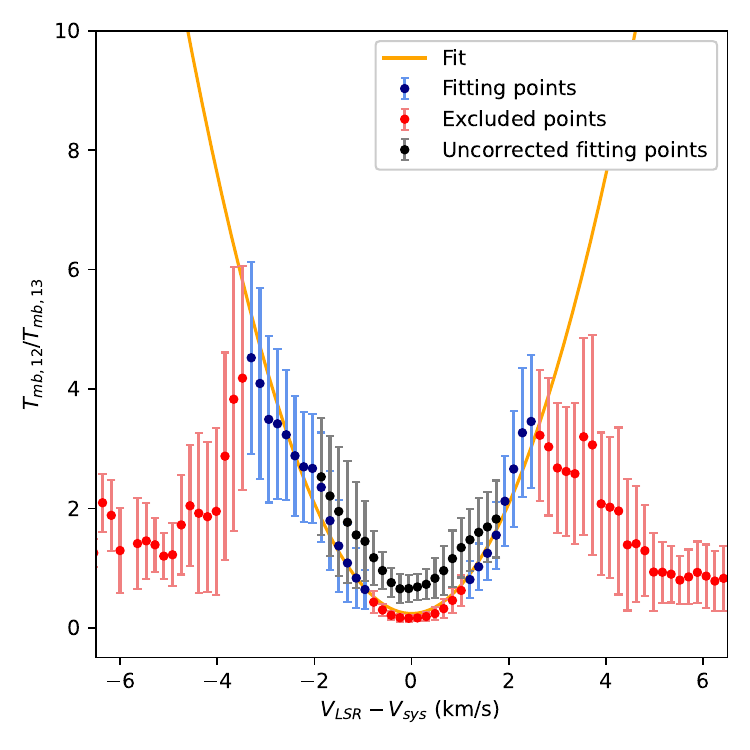}
    \caption{The weighted mean brightness temperature ratio between $^{12}$CO and opacity-corrected C$^{13}$O as a function of $v-v_{\si{sys}}$. Again, points in blue represent those used in the fitting process, while those in red were excluded. Points used to generate the fit encompass velocity channels between -3.48 and 2.82 $\si{{km}.s^{-1}}$.}
    \label{fig:12COFit}
\end{figure}

In order to calculate the brightness temperature ratio for a given velocity channel, all emission was first masked above 3$\sigma$ in order to increase the channel's signal-to-noise ratio and prevent division by small values. Once masked, pixel-wise division was performed in corresponding velocity channels to determine the brightness temperature ratio for two given spectral lines. In order to calculate the weighted mean and standard deviation of the brightness temperature ratios in each channel we applied the weighting scheme described in \cite{2016ApJ...832..158Z}. To properly incorporate high signal-to-noise pixels into the correction process, weights followed an inverse-squared proportionality to the uncertainty in the brightness temperature ratio for each pixel within a channel. Once weighting was applied to each pixel, the average brightness temperature ratio for each channel was determined through the weighted mean, with uncertainty given by the weighted standard deviation. To determine the brightness temperature ratio, and thus the optical depth, for $^{13}$CO in channels where C$^{18}$O was not prominent, we fit a second order polynomial to channels near systemic velocity in which $^{13}$CO is optically thick (Fig. \ref{fig:13COFit}).
The fit used is
\begin{equation}
    \frac{T_{13}}{T_{18}}=(1.96\pm0.27)+(1.76\pm0.61)(v-v_{\si{sys}})^{2} 
\end{equation}
which has a reduced $\chi^{2}$ value of 0.08. The same process was performed with $^{12}$CO and the opacity-corrected $^{13}$CO line. A strong self absorption feature present in low velocity $^{12}$CO channels makes $\frac{T_{\si{mb,12}}}{T_{\si{mb,13}}}$ a poor indicator of optical depth. To avoid the impacts of these self-absorbed channels we exclude emission at velocities between 3.22 and 5.02 $\si{{km}.s^{-1}}$ from the $\frac{T_{12}}{T_{13}'}$ fitting process. This exclusion process resulted in the best-fitting polynomial
\begin{equation}
    \frac{T_{12}}{T_{13}'}=(0.24\pm0.22)+(0.45\pm0.07)(v-v_{\si{sys}})^{2}
\end{equation}
with a reduced $\chi^{2}$ value of 0.12 (Fig. \ref{fig:12COFit}). The fits for both sets of spectral lines are truncated at their respective abundance ratios, meaning that the correction factors for $^{13}$CO and $^{12}$CO have a lower limit of 1.

\subsubsection{Inclination}
In addition to opacity corrections, we also take the inclination of HBC 494 into account when calculating variable estimates, as without them such values can be underestimated by up to factors of 2. Ideally, inclination can be determined using the ratio of the disc's major axis to its minor axis; however, this technique assumes that the disc is circular, flat, and thin \citep{2023ApJ...947...25H}. Due to its binary nature, and the fact that the observations used in this analysis do not resolve HBC 494's individual components, we were unable to make use of this technique. However, the high-resolution images presented in \cite{2023MNRAS.tmp.1574N} are able to effectively resolve both objects in the binary, and thus we take HBC 494a's reported inclination ($i=37.16\degr$) to be the inclination of the system. We apply a series of inclination-based corrections using the trigonometric relationships provided in \cite{2014ApJ...783...29D}, which are shown in Table \ref{tab:corrections} alongside the optical depth correction factors. The results of these two correction processes are listed in Table \ref{tab:values}.

\begin{table*}
    \begin{threeparttable}
    \centering
    \caption{Inclination and opacity correction factors.}
    \label{tab:corrections}
    \begin{tabular}{l c c c c c c}
    \toprule
    \multirow{2}{*}{Property} & \multirow{2}{*}{\makecell{Inclination\\Dependence}} & \multirow{2}{*}{\makecell{ Incl. Correction Factor\\$i=37.16\degr$}}&\multicolumn{2}{c}{Opacity Correction Factors}&\multicolumn{2}{c}{Combined Correction Factor}\\
    \cmidrule(lr){4-5}\cmidrule(lr){6-7}
    &&&Blueshifted&Redshifted&Blueshifted&Redshifted\\
    \hline
    Outflow Mass& N/A & N/A&14.3 & 7.6 & 14.3 & 7.6\Tstrut\\
    Mass-loss rate&$\sin{i}/\cos{i}$&0.80&9.5 & 6.0 & 7.6 & 4.5  \\
    Outflow Momentum&$1/\cos{i}$&1.3&10.1 & 6.1 & 12.5 & 7.8  \\
    Outflow Kinetic energy&$1/\cos^{2}{i}$&1.6&6.0 & 4.3 & 9.5 & 6.8  \\
    Outflow Luminosity&$\sin{i}/\cos^{3}{i}$&1.2&3.8 & 3.4 & 4.5 & 4.0   \\
    Outflow Force&$\sin{i}/\cos^{2}{i}$&0.95&7.0 & 4.8 & 6.6 & 4.4   \\
    Characteristic Velocity&$1/\cos{i}$&1.3&0.7 & 0.8 & 0.8 & 1.0 \\
    Characteristic Extension&$1/\sin{i}$&1.7&N/A&N/A&1.7 & 1.7 \\
    Dynamical Time&$\cos{i}/\sin{i}$&1.3&1.4 & 1.3 & 1.9 & 1.6 \\

    \bottomrule
    \end{tabular}
    \end{threeparttable}
\end{table*}

\begin{table*}
    \begin{threeparttable}
    \centering
    \caption{Outflow properties according to excitation temperature using Cycle-5 data.}
    \label{tab:values}
    \begin{tabular}{clccccccl}
        \toprule
        \multirow{2}{*}{\makecell{Excitation\\Temperature}} & \multirow{2}{*}{Property}& \multicolumn{3}{c}{Blueshifted\tnote{a}} & \multicolumn{3}{c}{Redshifted\tnote{b}} & \multirow{2}{*}{Units} \\
        \cmidrule(lr){3-5} \cmidrule(lr){6-8}
        && Uncorr.\tnote{c} & Incl.\tnote{d} & Opacity + Incl.\tnote{e} & Uncorr. & Incl.& Opacity + Incl. &\\
        \hline
             & Mass&2.1 & 2.1 & 30 & 3.7 & 3.7 & 28 &$10^{-3}\,$M$_{\sun}$ \Tstrut\\
            \multirow{7}{*}{\rotatebox[origin=c]{90}{20 K}} & Mass loss rate&2.1 & 1.6 & 16 & 5.5 & 4.1 & 25&$10^{-7}\,$M$_{\sun}\,$yr$^{-1}$ \\
             & Momentum&0.83 & 1.0 & 10 & 1.8 & 2.3 & 14 &$10^{-2}\,$M$_{\sun}\,\si{{km}.s^{-1}}$\\
             & Kinetic energy&4.3 & 6.8 & 41 & 11 & 18 & 77 & $10^{41}\,$ergs\\
             & Luminosity &0.36 & 0.43 & 1.8 & 1.4 & 1.7 & 5.6 &$10^{-4}\,$L$_{\sun}$\\
             & Force &0.84 & 0.80 & 5.3 & 2.7 & 2.6 & 13 &$10^{-6}\,$M$_{\sun}\,\si{{km}.s^{-1}.yr^{-1}}$\\
             & Char. Velocity&4.0 & 5.0 & 3.4 & 5.0 & 6.3 & 4.9 &$\si{{km}.s^{-1}}$\\
             & Char. Extension&0.04 & 0.07 & 0.07 & 0.03 & 0.06 & 0.06 &pc\\
             & Dynamical Time&9.8 & 13 & 19 & 6.7 & 8.9 & 11 &$10^{3}$ yr\\
             & \\
             & Mass&3.7 & 3.7 & 53 & 6.6 & 6.6 & 51 & $10^{-3}\,$M$_{\sun}$\Tstrut\\
                \multirow{7}{*}{\rotatebox[origin=c]{90}{50 K}}    & Mass loss rate&3.8 & 2.8 & 28 & 9.8 & 7.4 & 45 & $10^{-7}\,$M$_{\sun}\,$yr$^{-1}$\\
             & Momentum&1.5 & 1.9 & 18 & 3.3 & 4.1 & 25 &$10^{-2}\,$M$_{\sun}\,\si{{km}.s^{-1}}$ \\
             & Kinetic energy&7.7 & 12 & 74 & 20.2 & 32 & 140 &$10^{41}\,$ergs\\
             & Luminosity&0.65 & 0.77 & 3.2 & 2.5 & 3.0 & 10 &$10^{-4}\,$ L$_{\sun}$\\
             & Force&1.5 & 1.4 & 9.5 & 4.9 & 4.7 & 22 &$10^{-6}\,$M$_{\sun}\,\si{{km}.s^{-1}.yr^{-1}}$\\
             & Char. Velocity&4.0 & 5.0 & 3.4 & 5.0 & 6.3 & 4.9 &$\si{{km}.s^{-1}}$\\
             & Char. Extension&0.04 & 0.07 & 0.07 & 0.03 & 0.06 & 0.06 &pc\\
             & Dynamical Time&9.8 & 13 & 19 & 6.7 & 8.9 & 11 &$10^{3}$ yr\\
        \bottomrule             
    \end{tabular}
\begin{tablenotes}\footnotesize
\item[a] Emission integrated between -8.3 and 2.32 $\si{{km}.s^{-1}}$.
\item[b] Emission integrated between 7.0 and 23.56 $\si{{km}.s^{-1}}$.
\\
\item[c] Variable calculations performed with neither inclination nor opacity corrections.
\item[d] Inclination-corrected properties based on an angle of $i=37.16\degr$.
\item[e] Properties with inclination and opacity corrections applied.
\end{tablenotes}
\end{threeparttable}
\end{table*}
\section{Discussion}
\label{sec:DIS}
\subsection{Comparison of parameter estimations and observations}
When comparing the results presented in this work to those using the Cycle-2 observations, it is crucial to recognize the differences between the observing cycles and analysis techniques used here and in \citetalias{2017MNRAS.466.3519R}. Starting with the differences in analysis techniques, the most notable change is the inclusion of $^{13}$CO opacity corrections using the $^{13}$CO/C$^{18}$O brightness temperature ratio. Along with correcting the $^{13}$CO data cube, we also use a different scheme to isolate the additional $^{12}$CO outflow emission brought on by the Cycle-5 field of view. Prior work used a mask to remove all emission from the central binary, but included all pixels encompassing both outflows. This method relies on a careful choice of velocity channel integration ranges so as to not include emission from the redshifted outflow in calculations involving the blueshifted outflow, and vice-versa. This method does not work in our case, as the larger Cycle-5 field of view introduces material that is unrelated to either of the outflows in multiple velocity channels. This being the case, it was necessary to create masks to remove emission not related to the outflows. To generate the masks used in our analysis, we constructed two $^{12}$CO moment-0 maps using the integration ranges described in section \ref{sec:OK} and created regions which excluded emission from the central binary and only encompassed each individual outflow. Additionally, \citetalias{2017MNRAS.466.3519R} used a $^{12}$CO/H$_{2}$ abundance ratio of $X_{\si{CO}}=5.6\times10^{-6}$, while we adopt the value $X_{\si{CO}}=1\times10^{-4}$ for our analysis. We also make use of the inclination correction process described in \cite{2015ApJ...803...22P}, while the inclination of HBC 494 was not considered in \citetalias{2017MNRAS.466.3519R}.

For simplicity, and to directly compare the Cycle-2 and Cycle-5 observations, we applied the analysis techniques discussed in section \ref{sec:CR} to the Cycle-2 $^{12}$CO (J=2-1), $^{13}$CO (J=2-1), and C$^{18}$O (J=2-1) data cubes. Details of this observation program can be found in \citetalias{2017MNRAS.466.3519R} section 2. We adopt the velocity integration limits from \citetalias{2017MNRAS.466.3519R}, using channels between -5.25 and 1.5 $\si{{km}.s^{-1}}$ for the blueshifted outflow and channels between 7.0 and 16.5 $\si{{km}.s^{-1}}$ for the redshifted outflow. The results of this analysis are shown in Table \ref{tab:old_values}. 

\begin{table*}
    \begin{threeparttable}
    \centering
    \caption{Outflow properties according to excitation temperature using Cycle-2 data.}
    \label{tab:old_values}
    \begin{tabular}{clccccccl}
        \toprule
        \multirow{2}{*}{\makecell{Excitation\\Temperature}} & \multirow{2}{*}{Property}& \multicolumn{3}{c}{Blueshifted\tnote{a}} & \multicolumn{3}{c}{Redshifted\tnote{b}} & \multirow{2}{*}{Units} \\
        \cmidrule(lr){3-5} \cmidrule(lr){6-8}
        && Uncorr.\tnote{c} & Incl.\tnote{d} & Opacity + Incl.\tnote{e} & Uncorr. & Incl.& Opacity + Incl. &\\
        \hline
             & Mass&0.34 & 0.34 & 7.9 & 0.43 & 0.43 & 7.8 &$10^{-3}\,$M$_{\sun}$ \Tstrut\\
            \multirow{7}{*}{\rotatebox[origin=c]{90}{20 K}} & Mass loss rate&0.68 & 0.52 & 10.1 & 1.7 & 1.3 & 18.0 &$10^{-7}\,$M$_{\sun}\,$yr$^{-1}$ \\
             & Momentum&0.17 & 0.22 & 4.3 & 0.27 & 0.34 & 4.7 &$10^{-2}\,$M$_{\sun}\,\si{{km}.s^{-1}}$\\
             & Kinetic energy&1.0 & 1.6 & 25 & 2.0 & 3.2 & 32 & $10^{41}\,$ergs\\
             & Luminosity &0.17 & 0.20 & 2.7 & 0.7 & 0.8 & 6.1 &$10^{-4}\,$L$_{\sun}$\\
             & Force &0.35 & 0.34 & 5.5 & 1.1 & 1.1 & 11 &$10^{-6}\,$M$_{\sun}\,\si{{km}.s^{-1}.yr^{-1}}$\\
             & Char. Velocity&5.2 & 6.5 & 5.4 & 6.3 & 8.0 & 6.0 &$\si{{km}.s^{-1}}$\\
             & Char. Extension&0.03 & 0.04 & 0.04 & 0.02 & 0.03 & 0.03 &pc\\
             & Dynamical Time&5.0 & 6.5 & 7.9 & 2.5 & 3.3 & 4.3 &$10^{3}$ yr\\
             & \\
             & Mass&0.60 & 0.60 & 14 & 0.77 & 0.77 & 14 & $10^{-3}\,$M$_{\sun}$\Tstrut\\
                \multirow{7}{*}{\rotatebox[origin=c]{90}{50 K}}    & Mass loss rate&1.2 & 0.93 & 18.1 & 3.1 & 2.4 & 32.3 & $10^{-7}\,$M$_{\sun}\,$yr$^{-1}$\\
             & Momentum&0.31 & 0.39 & 7.7 & 0.49 & 0.61 & 8.4 &$10^{-2}\,$M$_{\sun}\,\si{{km}.s^{-1}}$ \\
             & Kinetic energy&1.8 & 2.8 & 45 & 3.6 & 5.7 & 57 & $10^{41}\,$ergs\\
             & Luminosity&0.30 & 0.35 & 4.8 & 1.2 & 1.4 & 11 &$10^{-4}\,$ L$_{\sun}$\\
             & Force&0.64 & 0.60 & 9.8 & 2.0 & 1.9 & 19 &$10^{-6}\,$M$_{\sun}\,\si{{km}.s^{-1}.yr^{-1}}$\\
             & Char. Velocity&5.2 & 6.5 & 5.4 & 6.3 & 8.0 & 6.0 &$\si{{km}.s^{-1}}$\\
             & Char. Extension&0.03 & 0.04 & 0.04 & 0.02 & 0.03 & 0.03 &pc\\
             & Dynamical Time&5.0 & 6.5 & 7.9 & 2.5 & 3.3 & 4.3 &$10^{3}$ yr\\
        \bottomrule             
    \end{tabular}
\begin{tablenotes}\footnotesize
\item[a] Emission integrated between -5.25 and 1.5 $\si{{km}.s^{-1}}$.
\item[b] Emission integrated between 7.0 and 16.5 $\si{{km}.s^{-1}}$.
\\
\item[c] Variable calculations performed with neither inclination nor opacity corrections.
\item[d] Inclination-corrected properties based on an angle of $i=37.16\degr$.
\item[e] Properties with inclination and opacity corrections applied.
\end{tablenotes}
\end{threeparttable}
\end{table*}

\begin{figure*}
    \centering
    \includegraphics[width=\textwidth]{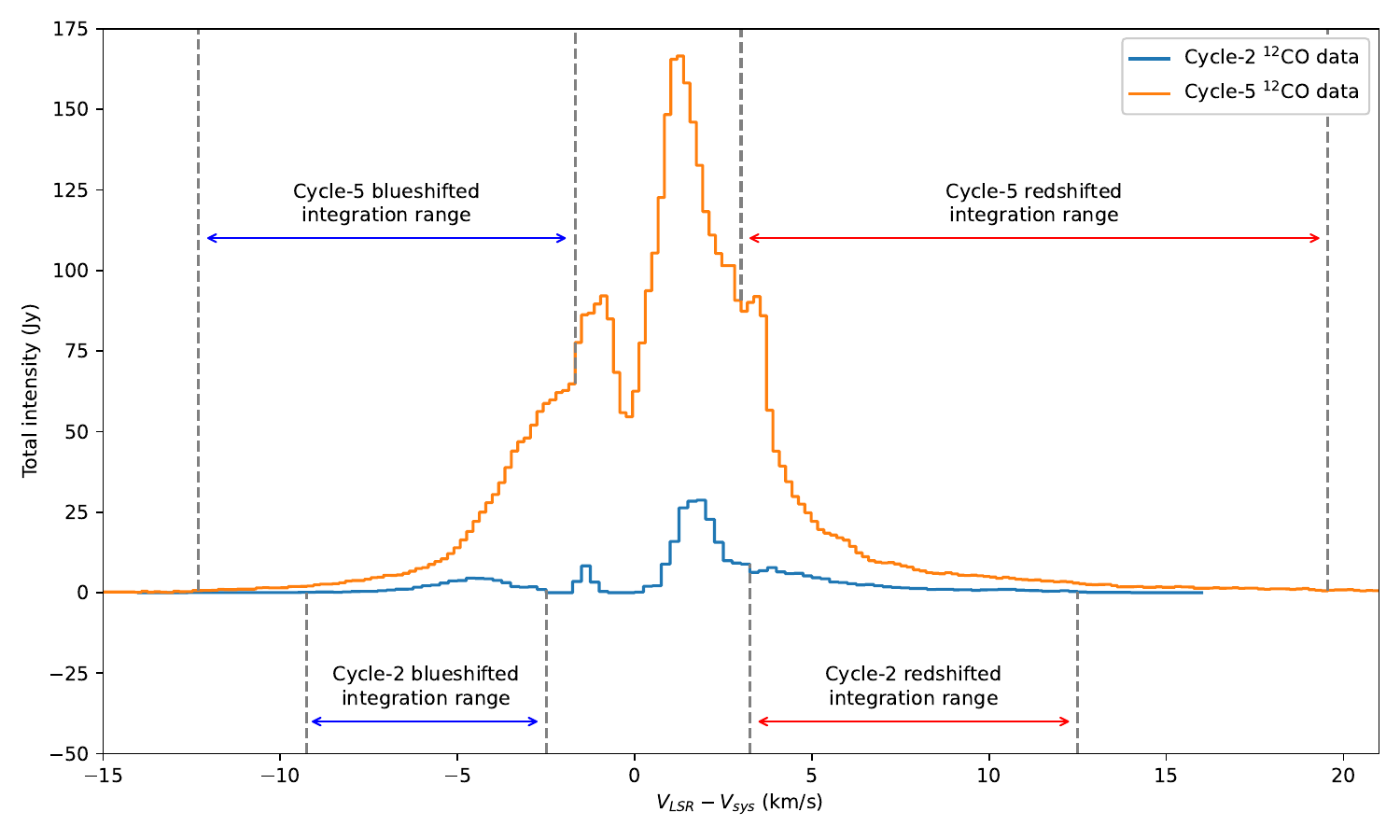}
    \caption{The total intensity of Cycle-2 $^{12}$CO and Cycle-5 $^{12}$CO data at a field of view of $19\arcsec\times19\arcsec$. The velocity integration ranges used in current kinematic calculations, as well as those provided in \protect\citetalias{2017MNRAS.466.3519R}, as shown as dashed grey lines.}
    \label{fig:SpecComp}
\end{figure*}

Observations presented in the prior publication were limited in sensitivity, UV coverage, and FOV, resulting in underestimated kinematic and dynamical variable values. In order to quantify the increases in emission due to the inclusion of the ALMA Total Power and Atacama Compact arrays, we make a per-channel comparison of Cycle-5 $^{12}$CO data cropped to the Cycle-2 field of view ($19\arcsec\times19\arcsec$) against original Cycle-2 $^{12}$CO data (Fig. \ref{fig:SpecComp}). The Cycle-2 data was not corrected for primary beam attenuation, and to account for this, we divided each channel of the $^{12}$CO data cube by a two-dimensional Gaussian with a FWHM=$25\arcsec$ and a maximum of one Jy/beam. We found the sum of all $^{12}$CO emission per channel in each of the two cubes and divided by the synthesised beam area to produce the total Jy in each channel. Comparing the total per-channel emission in these velocity ranges, blueshifted Cycle-5 emission is more intense by factors of 3.11-55.24 with a mean ratio of 8.35. Within the redshifted velocity range, Cycle-5 emission is more intense by factors of 4.30-9.90, with the mean ratio being 4.46. 

\begin{figure*}
    \centering
    \includegraphics[width=\textwidth]{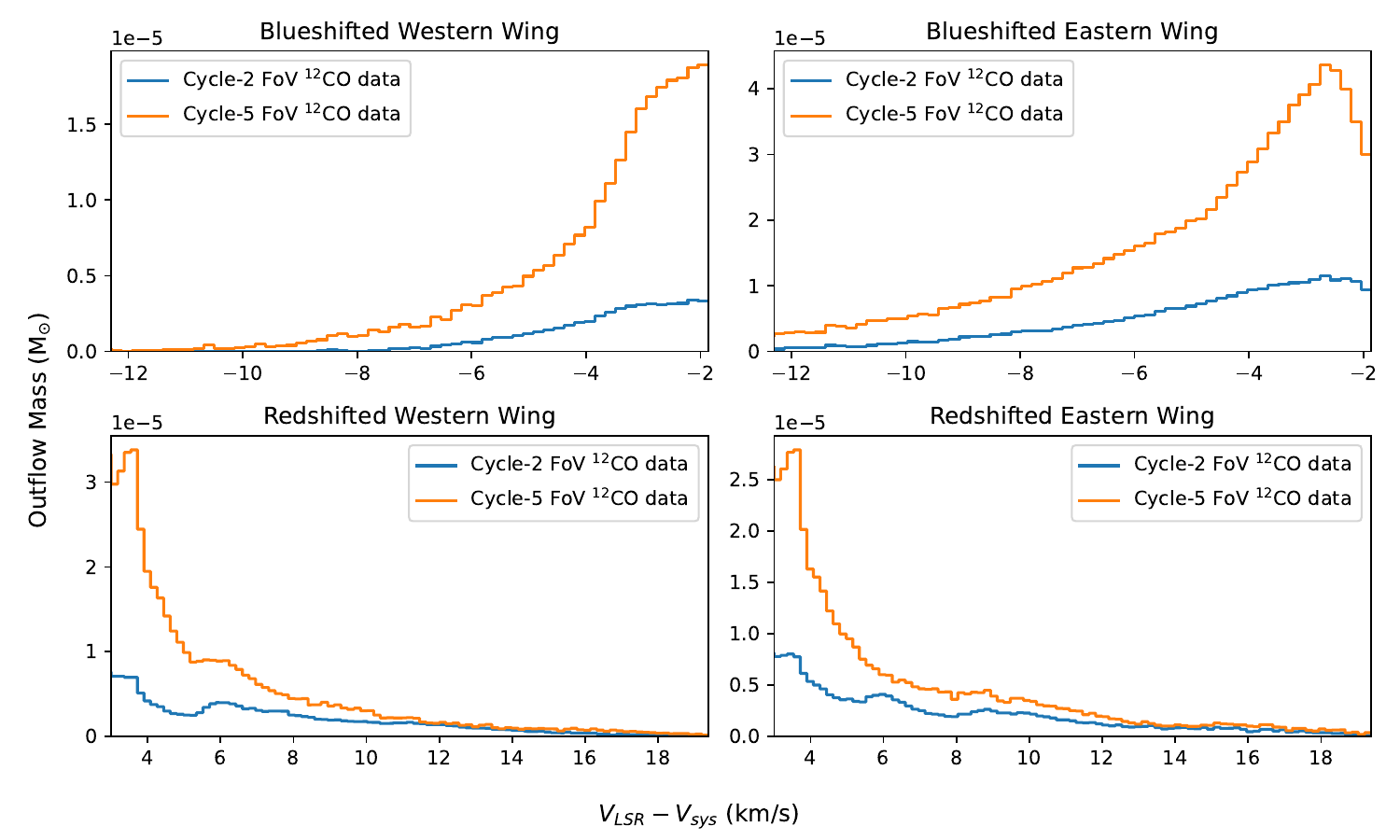}
    \caption{The per-channel mass for each outflow wing over their respective velocity integration ranges.}
    \label{fig:OutflowSpec}
\end{figure*}

In addition to the increases in emission brought by the inclusion of the ALMA Total Power and Atacama Compact arrays, current variable calculations also encompass wider velocity ranges. The blueshifted range is larger by a total of 3.87 $\si{{km}.s^{-1}}$, while the redshifted range is larger by 7.31 $\si{{km}.s^{-1}}$. The combination of increased emission in each channel along with widened velocity ranges for both outflows lead to prominent increases in every estimated variable. Furthermore, the increased field of view of the Cycle-5 observations allows for the full extension of each outflow to be seen. Cycle-2 $^{12}$CO observations display the eastern and western walls of the blueshifted outflow with an extensions of $12\arcsec$ and $5\arcsec$ respectively, while Cycle-5 observations show the same walls with extensions of $27\arcsec$ and $13\arcsec$. Similar increases are seen with the redshifted outflow, as the eastern arm increases in extension from $9\arcsec$ to $11\arcsec$ and the western arm increases from $6\arcsec$ to $18\arcsec$ between observation cycles. To quantify the extra emission provided by the Cycle-5 field of view, we produced a mass spectrum of each outflow wing using Cycle-5 $^{12}$CO data at its original field of view ($60\arcsec\times60\arcsec$) and cropped to the Cycle-2 field of view (Fig. \ref{fig:OutflowSpec}). As with the prior average intensity comparison, the Cycle-5 field of view provides significantly more emission for each of the four outflow wings in comparison to previous observations. The total mass for the blueshifted and redshifted outflow wings increases by factors of 3.6 and 2.6, respectively, due to the increase in field of view.

The changes between observation cycles lead to noticeable differences between the calculations performed with Cycle-2 data and those performed with Cycle-5 data. Firstly, uncorrected and inclination-only corrected Cycle-2 values are lower in comparison to their Cycle-5 counterparts, a trend which aligns with the aforementioned differences between the two cycles. However, the opacity correction process leads to higher force, luminosity, and characteristic velocity values when using the Cycle-2 data. The increases to these values are driven by two factors: the opacity correction process, and the characteristic velocity calculation. The parabolic fits used to produce each Cycle-2 correction factor were flatter than those found in the Cycle-5 process, meaning that every Cycle-2 velocity channel received a significant correction, which then lead to higher variable estimates. Additionally, the use of fewer velocity channels in the Cycle-2 calculations leads an increased momentum/mass ratio, which directly leads to a higher characteristic velocity. This issue propagated throughout the calculation process and led to increases in luminosity and force. Nevertheless, we report the most accurate calculations for HBC 494's outflows through use of the ALMA main, ACA, and TP arrays.

\subsection{Comparison with similar objects}
Along with the comparisons made between current and prior observations regarding HBC 494's outflows, a comparison between HBC 494 and similar objects (i.e., Class 0 and I YSOs, FU Ori objects) can reveal additional information related to HBC 494's outflows and their evolution. We first compare HBC 494 to the numerous objects reported in \cite{2023ApJ...947...25H}, a study which details 21 Class 0, I, and flat spectrum YSOs using ALMA Cycle-6 observations. This study makes use of three spectral lines reported in our work, namely $^{12}$CO (J=2-1), $^{13}$CO (J=2-1), and C$^{18}$O (J=2-1), and also makes use of the same computational methods for determining outflow parameters. Most notably, they report the total outflow mass of the selected Class 0 YSOs to be between 0.5 and 0.9 M$_{\sun}$, and the total outflow mass of the selected Class I YSOs to be between 1.5 and 2.9 M$_{\sun}$. To determine these values, they produce a mass spectrum for each outflow using a combination of $^{12}$CO, $^{13}$CO, and C$^{18}$O. $^{12}$CO emission is used to determine mass in high-velocity channels that lack ambient cloud contamination. As cloud emission becomes more prevalent, $^{13}$CO is used in low-velocity channels, and C$^{18}$O is used in channels closest to the systemic velocity. We apply this technique using a $^{13}$CO/H$_{2}$ ratio of $X_{\si{13,H_{2}}}=2\times10^{-6}$, and a C$^{18}$O/H$_{2}$ ratio of $X_{\si{18,H_{2}}}=2\times10^{-7}$ \citep{2023ApJ...947...25H}. For the blueshifted outflow we integrate $^{12}$CO between -8.3 and 2.32 $\si{{km}.s^{-1}}$, $^{13}$CO between 2.5 and 3.04 $\si{{km}.s^{-1}}$, and C$^{18}$O between 3.22 and 3.4 $\si{{km}.s^{-1}}$. For the redshifted outflow we integrate C$^{18}$O between 4.66 and 4.84 $\si{{km}.s^{-1}}$, $^{13}$CO between 5.02 and 6.82 $\si{{km}.s^{-1}}$, and $^{12}$CO between 7 and 23.56 $\si{{km}.s^{-1}}$. Making use of the opacity corrected $^{12}$CO and $^{13}$CO data cubes, we produce a total outflow mass of 0.63 M$_{\sun}$, which places HBC 494 in the centre of the Class 0 YSO range. \par
It is important to note that the objects selected for this study were chosen specifically for their singular nature, meaning that outflow characteristics stemming from HBC 494's binarity are not present, especially with regard to outflow opening angle. \cite{2023ApJ...947...25H} notes that opening angle generally increases as a YSO evolves, with the largest opening angles coming from the flat-spectrum sources in their study. Specifically, they report opening angles of between $\sim15\degr$ and $\sim130\degr$, with the latter reported for the flat-spectrum source HOPS 200. The aforementioned trend could indicate that HBC 494's binarity may be at least partly responsible for this phenomenon, as it maintains a larger outflow angle than any of the objects documented in this publication.\par
Comparisons of HBC 494's disc to other FUor discs can also provide useful information about the system. FU Ori discs were studied in depth in \cite{2018MNRAS.474.4347C}, specifically those of V883 Ori, V2775 Ori, and V1118 Ori. \cite{2018MNRAS.474.4347C} also includes early observations of HBC 494; however, their observations do not resolve the discs of HBC 494a and HBC 494b individually. They conclude that FUor discs have masses between 80 and 600 M$_{\si{Jup}}$, while EXor discs -- a type of object similar to FUors with less intense outbursting events -- have masses between 0.5 and 40 M$_{\si{Jup}}$. According to \cite{2023MNRAS.tmp.1574N} HBC 494a has a disc mass (gas + dust) of $\sim145$ M$_{\si{Jup}}$, while HBC 494b has a disc mass of $\sim29$ M$_{\si{Jup}}$, a value consistent with EXor type objects. While disc mass alone cannot provide a definitive classification for HBC 494b, it can aid in characterising HBC 494's outflows as a whole. Based on the aforementioned Cycle-2 observations, HBC 494 was thought to present a single set of wide-angle outflows. However, we now believe that HBC 494's outflows are a composite of outflows from HBC 494a and HBC 494b. If possible, untangling the outflows from these two sources may also reveal important properties of each individual object, providing additional information about their specific classifications. Future work can perhaps focus on the orbital evolution of HBC 494, as well as how the individual outflows from its constituent objects impact the system as a whole. \par

\subsection{The secondary cavity}\label{sec:SecondaryCavity}
The existence of HBC 494's secondary cavity could be indicative of multiple phenomena. Firstly, it is possible that this cavity represents a previous outflow from HBC 494 that was expelled outwards due to an accretion outbursting event. The cavity is observed most prevalently at velocities between 6.1 and 8.98 $\si{{km}.s^{-1}}$, and at lower intensity between 10.24 and 13.66 $\si{{km}.s^{-1}}$. Accounting for HBC 494's inclination of $i=37.16\degr$, the cavity is seen at a distance of $24.9\arcsec$ north of the central binary, indicating that it was expelled between 5,300 and 6,200 years ago. These values can provide us with an estimate of when a previous outburst occurred, as \cite{1996ARA&A..34..207H} claims that outburst events must occur multiple times over an interval of $\sim10^{5}$ years, although a prior outburst has not been observed from HBC 494.\par
\begin{figure*}
    \includegraphics[width=\textwidth]{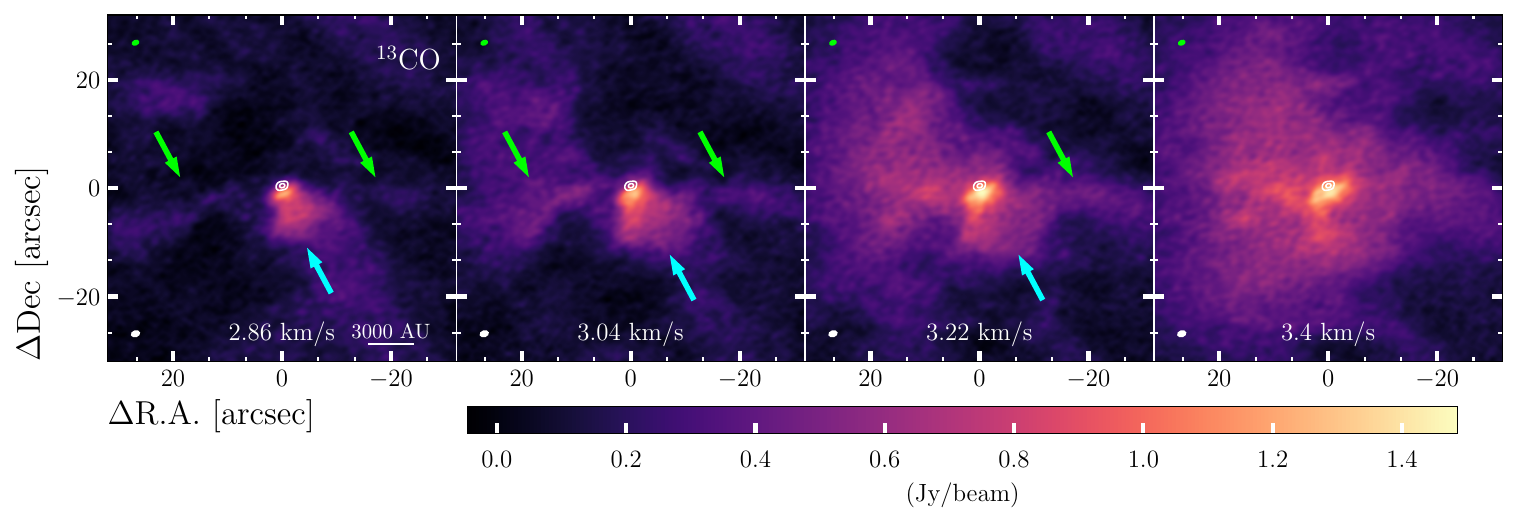}
    \includegraphics[width=\textwidth]{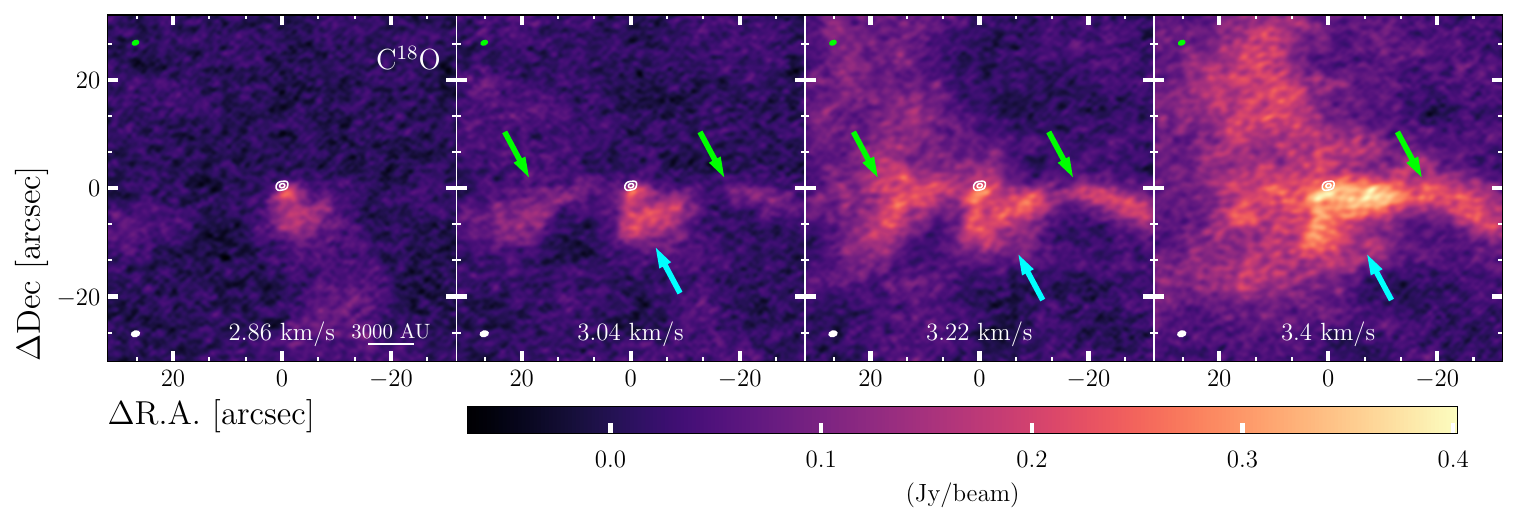}
    \caption{Two channel maps of $^{13}$CO (top) and C$^{18}$O (bottom) which highlight two distinct features with coloured arrows. Specifically, the green arrows highlight a set of outflow wings present on either side of the binary, and the cyan arrows show material representative of an interaction between outflows from HBC 494a and HBC 494b. The white contours seen at the centre of each frame in both channel maps represent continuum emission centred at 232.5 GHz at levels of 0.25 and 0.75 times the maximum intensity of 0.11 Jy/beam. The white ellipse in the bottom left of each frame shows the synthesised beam, which maintains an angular size of 1.77 $\times$ 1.30 arcseconds and a position angle of -72.3$\degr$ for $^{13}$CO, and 1.78 $\times$ 1.28 arcseconds and -74.6$\degr$ for C$^{18}$O. The green ellipse shown in the top left of each represents the continuum beam, which has an angular size of 1.50 $\times$ 1.07 arcseconds and a position angle of -66.8$\degr$.}
    \label{fig:1318Wings}
\end{figure*}

Secondly, the shape of the cavity may serve as additional evidence for overlapping outflows from HBC 494's binary components. The outflows from HBC 494a and HBC 494b, both of which have different position angles, could be the cause of both the primary and secondary cavities' wide opening angle and shape, as well as the wings of emission seen in blueshifted $^{13}$CO and C$^{18}$O observations, which are highlighted in Fig. \ref{fig:1318Wings}. This figure shows two features which evidence outflows with differing position angles: a set of eastern and western outflow wings (green arrows) and a clump of material created by the interaction between outflows from HBC 494a and HBC 494b (cyan arrows). The clump of material is situated in between the position angles for the two suggested outflows. A similar clump of material is seen in $^{12}$CO and corresponds with the northern regions between the two outflow position angles, indicating that a comparable interaction is occurring within the northern outflow cavity. The cavity also shows clear interactions between HBC 494's northern outflow as observed with $^{12}$CO, $^{13}$CO, and C$^{18}$O. These interactions are evidenced by similar emission shapes between the three lines at velocities at which the secondary cavity is detected (8.98 -- 10.6 $\si{{km}.s^{-1}}$). Detections at these velocities, specifically with $^{13}$CO and C$^{18}$O, indicate that the northern side of the system is more embedded in the stellar envelope than the southern side, as $^{13}$CO and C$^{18}$O also trace the envelope.\par
While the specific classification of HBC 494b is currently unknown, it is likely to be a Class 0/I protostar, meaning that outflow interactions may be possible \citep{2008AJ....135.2526C, 2023MNRAS.tmp.1574N}. Additionally, misaligned outflows from binary YSO systems are also noted at length in \cite{2016ApJ...820L...2L}. They report on the outflows of nine binary or higher-n YSOs, and determine outflow position angle differences of between $6\degr$ and $90\degr$ with a median difference of $61\degr$. It is important to note that the observed systems maintain much larger separations than HBC 494, as the minimum separation between any two objects in this study is 1,863 au -- nearly 25 times greater than the separation between HBC 494a and HBC 494b. The gravitational influence of HBC 494's components on one another may be responsible for the large difference in position angle between the two outflow sets.

\section{Conclusions}
\label{sec:CONC}
We present ALMA Cycle-5 observations of $^{12}$CO, $^{13}$CO, C$^{18}$O, and SO, all of which are able to trace different features of HBC 494. To improve upon the results presented in \citetalias{2017MNRAS.466.3519R}, current observations used a combination of the ALMA main, Total Power, and Atacama Compact arrays to increase sensitivity to large-scale emission and reduce spatial filtering. Through image mosiacing, we were also able to greatly improve the field-of-view of our ALMA observations, as those presented in \citetalias{2017MNRAS.466.3519R} were limited to single-pointing images. These improvements in observations led to two important results:
\begin{enumerate}
    \item For the first time in HBC 494's observational history, a secondary outflow cavity was observed above the northern primary cavity. We believe that the cavity was launched upwards from the central binary as the result of a prior accretion outburst. Additionally, it serves as evidence of an overlap between outflows stemming from HBC 494's constituent objects. The cavity has a dynamical age between 5,300 and 6,200 years, which is in line with the time-scales of repeated accretion outburst events \citep{1996ARA&A..34..207H}. P-V diagrams made at angles of $35\degr$ and $145\degr$ reveal that outflows from HBC 494a and HBC 494b are misaligned. This misalignment serves as an explanation as to why the primary outflows present an uncharacteristically wide opening angle. The secondary cavity displays a similarly wide opening angle, suggesting that interactions between the two sets of outflows aided in shaping the secondary cavity. 
    
    \item The full extent of HBC 494's outflows has also been observed for the first time. The southern primary outflow maintains an asymmetric parabolic shape, with its eastern arm showing much greater extension and intensity in comparison to the western arm. The northern primary outflow, however, is much more symmetric in both extension and intensity. The secondary northern cavity mimics the shape of its primary counterpart, although its western arm shows higher intensity than its eastern arm. The asymmetry shown in the primary and secondary outflow cavities, both in shape and emission intensity, could be useful in determining the origins of the secondary cavity and is currently a point of further investigation.
\end{enumerate}
In combination with improved observations, we also present kinematic and dynamical variable estimations using the methods described in \cite{2014ApJ...783...29D}. To perform such calculations, we find the mass of each pixel in a given velocity channel based on a series of assumptions regarding excitation temperature, optical depth, and beam filling factors. Specifically, the CO dipole moment allows for $^{12}$CO and $^{13}$CO to be easily thermalized, and thus can be represented by the Boltzmann distribution. From this, we assume that all emission is in LTE and maintains the same excitation temperature. To find the total outflow mass, we integrate over all pixels and velocity channels in which outflow emission is detected above a $3\sigma$ cutoff, which is determined using the RMS of each data cube. Optical depth effects can lead to underestimations of kinematic and dynamical variables, and thus must be corrected for. To do so, we use the brightness temperature ratio between $^{13}$CO and C$^{18}$O to apply an initial correction to the $^{13}$CO data cube, and then calculate the same ratio between $^{12}$CO and the corrected $^{13}$CO. The second set of corrections are then applied to the $^{12}$CO data cube, and the calculation process is performed again. Without corrections applied, we estimate the mass of the southern primary outflow to be between $2.1\times10^{-3}$ and $3.7\times10^{-3}$ $\si{M_{\sun}}$ based on assumed excitation temperatures of 20 K and 50 K. With corrections, these values increase to $3.0\times10^{-2}$ $\si{M_{\sun}}$ and $5.3\times10^{-2}$ $\si{M_{\sun}}$, respectively. For the northern primary outflow, the uncorrected mass is estimated to be between $3.7\times10^{-3}$ and $6.6\times10^{-3}$ $\si{M_{\sun}}$, while corrected estimates are between $2.8\times10^{-2}$ and $5.1\times10^{-2}$ $\si{M_{\sun}}$. The large increases to each physical parameter created by the correction process are in line with those presented in \cite{2014ApJ...783...29D}. 
The variable estimates made using Cycle-5 data are larger than those using Cycle-2 data, as the use of the ALMA main, Total Power, and Atacama Compact arrays has allowed us to resolve a much greater amount of emission in comparison to prior observations. The use of image mosaicing also contributed to the increased variable estimates, as HBC 494's outflows were observed to their full extent.

\section*{Acknowledgments}
This research was supported through the National Radio Astronomy Observatory’s (NRAO) summer student program in Charlottesville, VA. Additional contributions were made by the Department of Astronomy at the University of Maryland (UMD) in College Park, MD. AF thanks Dr Andrew Harris and Dr Melissa Hayes-Gehrke from UMD for their revisions of this research, and Dr Jim Braatz from the NRAO for his efforts in organizing and running the NRAO summer student program. Lastly, AF thanks Dr John Fourkas and Dr Amy Mullin from UMD for their continued support of this research. This paper makes use of the following ALMA data: ADS/JAO.ALMA$\#$2017.1.00015.S. ALMA is a partnership of ESO (representing its member states), NSF (USA) and NINS (Japan), together with NRC (Canada), MOST and ASIAA (Taiwan), and KASI (Republic of Korea), in cooperation with the Republic of Chile. The Joint ALMA Observatory is operated by ESO, AUI/NRAO and NAOJ. The National Radio Astronomy Observatory is a facility of the National Science Foundation operated under cooperative agreement by Associated Universities, Inc.

%%%%%%%%%%%%%%%%%%%%%%%%%%%%%%%%%%%%%%%%%%%%%%%%%%
\section*{Data Availability}
The data underlying this paper are available in the ALMA archive at https://almascience.nrao.edu under project code 2017.1.00015.S.



\bibliographystyle{mnras}
\bibliography{sources} % if your bibtex file is called example.bib

\begin{thebibliography}{}
\makeatletter
\relax
\def\mn@urlcharsother{\let\do\@makeother \do\$\do\&\do\#\do\^\do\_\do\%\do\~}
\def\mn@doi{\begingroup\mn@urlcharsother \@ifnextchar [ {\mn@doi@} {\mn@doi@[]}}
\def\mn@doi@[#1]#2{\def\@tempa{#1}\ifx\@tempa\@empty \href {http://dx.doi.org/#2} {doi:#2}\else \href {http://dx.doi.org/#2} {#1}\fi \endgroup}
\def\mn@eprint#1#2{\mn@eprint@#1:#2::\@nil}
\def\mn@eprint@arXiv#1{\href {http://arxiv.org/abs/#1} {{\tt arXiv:#1}}}
\def\mn@eprint@dblp#1{\href {http://dblp.uni-trier.de/rec/bibtex/#1.xml} {dblp:#1}}
\def\mn@eprint@#1:#2:#3:#4\@nil{\def\@tempa {#1}\def\@tempb {#2}\def\@tempc {#3}\ifx \@tempc \@empty \let \@tempc \@tempb \let \@tempb \@tempa \fi \ifx \@tempb \@empty \def\@tempb {arXiv}\fi \@ifundefined {mn@eprint@\@tempb}{\@tempb:\@tempc}{\expandafter \expandafter \csname mn@eprint@\@tempb\endcsname \expandafter{\@tempc}}}

\bibitem[\protect\citeauthoryear{{Andrews} \& {Williams}}{{Andrews} \& {Williams}}{2005}]{2005ApJ...631.1134A}
{Andrews} S.~M.,  {Williams} J.~P.,  2005, \mn@doi [\apj] {10.1086/432712}, \href {https://ui.adsabs.harvard.edu/abs/2005ApJ...631.1134A} {631, 1134}

\bibitem[\protect\citeauthoryear{{Arce} \& {Sargent}}{{Arce} \& {Sargent}}{2006}]{2006ApJ...646.1070A}
{Arce} H.~G.,  {Sargent} A.~I.,  2006, \mn@doi [\apj] {10.1086/505104}, \href {https://ui.adsabs.harvard.edu/abs/2006ApJ...646.1070A} {646, 1070}

\bibitem[\protect\citeauthoryear{{Armitage}, {Livio}  \& {Pringle}}{{Armitage} et~al.}{2001}]{2001MNRAS.324..705A}
{Armitage} P.~J.,  {Livio} M.,   {Pringle} J.~E.,  2001, \mn@doi [\mnras] {10.1046/j.1365-8711.2001.04356.x}, \href {https://ui.adsabs.harvard.edu/abs/2001MNRAS.324..705A} {324, 705}

\bibitem[\protect\citeauthoryear{{Audard} et~al.,}{{Audard} et~al.}{2014}]{2014prpl.conf..387A}
{Audard} M.,  et~al., 2014, in {Beuther} H.,  {Klessen} R.~S.,  {Dullemond} C.~P.,   {Henning} T.,  eds, Protostars and Planets VI. p.~387 (\mn@eprint {arXiv} {1401.3368}), \mn@doi{10.2458/azu\_uapress\_9780816531240-ch017}

\bibitem[\protect\citeauthoryear{{Bell} \& {Lin}}{{Bell} \& {Lin}}{1994}]{1994ApJ...427..987B}
{Bell} K.~R.,  {Lin} D.~N.~C.,  1994, \mn@doi [\apj] {10.1086/174206}, \href {https://ui.adsabs.harvard.edu/abs/1994ApJ...427..987B} {427, 987}

\bibitem[\protect\citeauthoryear{{Bonnell} \& {Bastien}}{{Bonnell} \& {Bastien}}{1992}]{1992ApJ...401L..31B}
{Bonnell} I.,  {Bastien} P.,  1992, \mn@doi [\apjl] {10.1086/186663}, \href {https://ui.adsabs.harvard.edu/abs/1992ApJ...401L..31B} {401, L31}

\bibitem[\protect\citeauthoryear{{Cieza} et~al.,}{{Cieza} et~al.}{2018}]{2018MNRAS.474.4347C}
{Cieza} L.~A.,  et~al., 2018, \mn@doi [\mnras] {10.1093/mnras/stx3059}, \href {https://ui.adsabs.harvard.edu/abs/2018MNRAS.474.4347C} {474, 4347}

\bibitem[\protect\citeauthoryear{{Connelley}, {Reipurth}  \& {Tokunaga}}{{Connelley} et~al.}{2008}]{2008AJ....135.2526C}
{Connelley} M.~S.,  {Reipurth} B.,   {Tokunaga} A.~T.,  2008, \mn@doi [\aj] {10.1088/0004-6256/135/6/2526}, \href {https://ui.adsabs.harvard.edu/abs/2008AJ....135.2526C} {135, 2526}

\bibitem[\protect\citeauthoryear{{Dunham}, {Arce}, {Mardones}, {Lee}, {Matthews}, {Stutz}  \& {Williams}}{{Dunham} et~al.}{2014}]{2014ApJ...783...29D}
{Dunham} M.~M.,  {Arce} H.~G.,  {Mardones} D.,  {Lee} J.-E.,  {Matthews} B.~C.,  {Stutz} A.~M.,   {Williams} J.~P.,  2014, \mn@doi [\apj] {10.1088/0004-637X/783/1/29}, \href {https://ui.adsabs.harvard.edu/abs/2014ApJ...783...29D} {783, 29}

\bibitem[\protect\citeauthoryear{{Hartmann} \& {Kenyon}}{{Hartmann} \& {Kenyon}}{1996}]{1996ARA&A..34..207H}
{Hartmann} L.,  {Kenyon} S.~J.,  1996, \mn@doi [\araa] {10.1146/annurev.astro.34.1.207}, \href {https://ui.adsabs.harvard.edu/abs/1996ARA&A..34..207H} {34, 207}

\bibitem[\protect\citeauthoryear{{Herbig}}{{Herbig}}{1966}]{1966VA......8..109H}
{Herbig} G.~H.,  1966, \mn@doi [Vistas in Astronomy] {10.1016/0083-6656(66)90025-0}, \href {https://ui.adsabs.harvard.edu/abs/1966VA......8..109H} {8, 109}

\bibitem[\protect\citeauthoryear{{Hsieh} et~al.,}{{Hsieh} et~al.}{2023}]{2023ApJ...947...25H}
{Hsieh} C.-H.,  et~al., 2023, \mn@doi [\apj] {10.3847/1538-4357/acba13}, \href {https://ui.adsabs.harvard.edu/abs/2023ApJ...947...25H} {947, 25}

\bibitem[\protect\citeauthoryear{{J{\o}rgensen}, {Hogerheijde}, {Blake}, {van Dishoeck}, {Mundy}  \& {Sch{\"o}ier}}{{J{\o}rgensen} et~al.}{2004}]{2004A&A...415.1021J}
{J{\o}rgensen} J.~K.,  {Hogerheijde} M.~R.,  {Blake} G.~A.,  {van Dishoeck} E.~F.,  {Mundy} L.~G.,   {Sch{\"o}ier} F.~L.,  2004, \mn@doi [\aap] {10.1051/0004-6361:20034216}, \href {https://ui.adsabs.harvard.edu/abs/2004A&A...415.1021J} {415, 1021}

\bibitem[\protect\citeauthoryear{{Kenyon}, {Hartmann}, {Strom}  \& {Strom}}{{Kenyon} et~al.}{1990}]{1990AJ.....99..869K}
{Kenyon} S.~J.,  {Hartmann} L.~W.,  {Strom} K.~M.,   {Strom} S.~E.,  1990, \mn@doi [\aj] {10.1086/115380}, \href {https://ui.adsabs.harvard.edu/abs/1990AJ.....99..869K} {99, 869}

\bibitem[\protect\citeauthoryear{{Kepley}, {Tsutsumi}, {Brogan}, {Indebetouw}, {Yoon}, {Mason}  \& {Donovan Meyer}}{{Kepley} et~al.}{2020}]{Kepley2020}
{Kepley} A.~A.,  {Tsutsumi} T.,  {Brogan} C.~L.,  {Indebetouw} R.,  {Yoon} I.,  {Mason} B.,   {Donovan Meyer} J.,  2020, \mn@doi [\pasp] {10.1088/1538-3873/ab5e14}, \href {https://ui.adsabs.harvard.edu/abs/2020PASP..132b4505K} {132, 024505}

\bibitem[\protect\citeauthoryear{{Klaassen}, {Mottram}, {Maud}  \& {Juhasz}}{{Klaassen} et~al.}{2016}]{2016MNRAS.460..627K}
{Klaassen} P.~D.,  {Mottram} J.~C.,  {Maud} L.~T.,   {Juhasz} A.,  2016, \mn@doi [\mnras] {10.1093/mnras/stw989}, \href {https://ui.adsabs.harvard.edu/abs/2016MNRAS.460..627K} {460, 627}

\bibitem[\protect\citeauthoryear{{Koda}, {Teuben}, {Sawada}, {Plunkett}  \& {Fomalont}}{{Koda} et~al.}{2019}]{Koda2019}
{Koda} J.,  {Teuben} P.,  {Sawada} T.,  {Plunkett} A.,   {Fomalont} E.,  2019, \mn@doi [\pasp] {10.1088/1538-3873/ab047e}, \href {https://ui.adsabs.harvard.edu/abs/2019PASP..131e4505K} {131, 054505}

\bibitem[\protect\citeauthoryear{{Kong} et~al.,}{{Kong} et~al.}{2018}]{2018ApJS..236...25K}
{Kong} S.,  et~al., 2018, \mn@doi [\apjs] {10.3847/1538-4365/aabafc}, \href {https://ui.adsabs.harvard.edu/abs/2018ApJS..236...25K} {236, 25}

\bibitem[\protect\citeauthoryear{{Langer} \& {Penzias}}{{Langer} \& {Penzias}}{1993}]{1993ApJ...408..539L}
{Langer} W.~D.,  {Penzias} A.~A.,  1993, \mn@doi [\apj] {10.1086/172611}, \href {https://ui.adsabs.harvard.edu/abs/1993ApJ...408..539L} {408, 539}

\bibitem[\protect\citeauthoryear{{Lee} et~al.,}{{Lee} et~al.}{2016}]{2016ApJ...820L...2L}
{Lee} K.~I.,  et~al., 2016, \mn@doi [\apjl] {10.3847/2041-8205/820/1/L2}, \href {https://ui.adsabs.harvard.edu/abs/2016ApJ...820L...2L} {820, L2}

\bibitem[\protect\citeauthoryear{{Lodato} \& {Clarke}}{{Lodato} \& {Clarke}}{2004}]{2004MNRAS.353..841L}
{Lodato} G.,  {Clarke} C.~J.,  2004, \mn@doi [\mnras] {10.1111/j.1365-2966.2004.08112.x}, \href {https://ui.adsabs.harvard.edu/abs/2004MNRAS.353..841L} {353, 841}

\bibitem[\protect\citeauthoryear{{Menten}, {Reid}, {Forbrich}  \& {Brunthaler}}{{Menten} et~al.}{2007}]{2007A&A...474..515M}
{Menten} K.~M.,  {Reid} M.~J.,  {Forbrich} J.,   {Brunthaler} A.,  2007, \mn@doi [\aap] {10.1051/0004-6361:20078247}, \href {https://ui.adsabs.harvard.edu/abs/2007A&A...474..515M} {474, 515}

\bibitem[\protect\citeauthoryear{{Nogueira} et~al.,}{{Nogueira} et~al.}{2023}]{2023MNRAS.tmp.1574N}
{Nogueira} P.~H.,  et~al., 2023, \mn@doi [\mnras] {10.1093/mnras/stad1614}, \href {https://ui.adsabs.harvard.edu/abs/2023MNRAS.tmp.1574N} {}

\bibitem[\protect\citeauthoryear{{Plunkett}, {Arce}, {Corder}, {Dunham}, {Garay}  \& {Mardones}}{{Plunkett} et~al.}{2015}]{2015ApJ...803...22P}
{Plunkett} A.~L.,  {Arce} H.~G.,  {Corder} S.~A.,  {Dunham} M.~M.,  {Garay} G.,   {Mardones} D.,  2015, \mn@doi [\apj] {10.1088/0004-637X/803/1/22}, \href {https://ui.adsabs.harvard.edu/abs/2015ApJ...803...22P} {803, 22}

\bibitem[\protect\citeauthoryear{{Pudritz} \& {Ray}}{{Pudritz} \& {Ray}}{2019}]{2019FrASS...6...54P}
{Pudritz} R.~E.,  {Ray} T.~P.,  2019, \mn@doi [Frontiers in Astronomy and Space Sciences] {10.3389/fspas.2019.00054}, \href {https://ui.adsabs.harvard.edu/abs/2019FrASS...6...54P} {6, 54}

\bibitem[\protect\citeauthoryear{{Reipurth}}{{Reipurth}}{1985}]{1985A&AS...61..319R}
{Reipurth} B.,  1985, \aaps, \href {https://ui.adsabs.harvard.edu/abs/1985A&AS...61..319R} {61, 319}

\bibitem[\protect\citeauthoryear{{Reipurth} \& {Bally}}{{Reipurth} \& {Bally}}{1986}]{1986Natur.320..336R}
{Reipurth} B.,  {Bally} J.,  1986, \mn@doi [\nat] {10.1038/320336a0}, \href {https://ui.adsabs.harvard.edu/abs/1986Natur.320..336R} {320, 336}

\bibitem[\protect\citeauthoryear{{Ripple}, {Heyer}, {Gutermuth}, {Snell}  \& {Brunt}}{{Ripple} et~al.}{2013}]{2013MNRAS.431.1296R}
{Ripple} F.,  {Heyer} M.~H.,  {Gutermuth} R.,  {Snell} R.~L.,   {Brunt} C.~M.,  2013, \mn@doi [\mnras] {10.1093/mnras/stt247}, \href {https://ui.adsabs.harvard.edu/abs/2013MNRAS.431.1296R} {431, 1296}

\bibitem[\protect\citeauthoryear{{Ru{\'\i}z-Rodr{\'\i}guez} et~al.,}{{Ru{\'\i}z-Rodr{\'\i}guez} et~al.}{2017}]{2017MNRAS.466.3519R}
{Ru{\'\i}z-Rodr{\'\i}guez} D.,  et~al., 2017, \mn@doi [\mnras] {10.1093/mnras/stw3378}, \href {https://ui.adsabs.harvard.edu/abs/2017MNRAS.466.3519R} {466, 3519}

\bibitem[\protect\citeauthoryear{Scuseria, Miller, Jensen  \& Geertsen}{Scuseria et~al.}{1991}]{10.1063/1.460293}
Scuseria G.~E.,  Miller M.~D.,  Jensen F.,   Geertsen J.,  1991, \mn@doi [The Journal of Chemical Physics] {10.1063/1.460293}, 94, 6660

\bibitem[\protect\citeauthoryear{{Tobin}, {Hartmann}, {Bergin}, {Chiang}, {Looney}, {Chandler}, {Maret}  \& {Heitsch}}{{Tobin} et~al.}{2012}]{Tobin2012}
{Tobin} J.~J.,  {Hartmann} L.,  {Bergin} E.,  {Chiang} H.-F.,  {Looney} L.~W.,  {Chandler} C.~J.,  {Maret} S.,   {Heitsch} F.,  2012, \mn@doi [\apj] {10.1088/0004-637X/748/1/16}, \href {https://ui.adsabs.harvard.edu/abs/2012ApJ...748...16T} {748, 16}

\bibitem[\protect\citeauthoryear{Williams}{Williams}{2021}]{williams_2021}
Williams J.~P.,  2021, Introduction to the Interstellar Medium.
Cambridge University Press, \mn@doi{10.1017/9781108691178}

\bibitem[\protect\citeauthoryear{{Wilson} \& {Matteucci}}{{Wilson} \& {Matteucci}}{1992}]{1992A&ARv...4....1W}
{Wilson} T.~L.,  {Matteucci} F.,  1992, \mn@doi [\aapr] {10.1007/BF00873568}, \href {https://ui.adsabs.harvard.edu/abs/1992A&ARv...4....1W} {4, 1}

\bibitem[\protect\citeauthoryear{{Zhang} et~al.,}{{Zhang} et~al.}{2016}]{2016ApJ...832..158Z}
{Zhang} Y.,  et~al., 2016, \mn@doi [\apj] {10.3847/0004-637X/832/2/158}, \href {https://ui.adsabs.harvard.edu/abs/2016ApJ...832..158Z} {832, 158}

\bibitem[\protect\citeauthoryear{{Zhu}, {Hartmann}  \& {Gammie}}{{Zhu} et~al.}{2009}]{2009ApJ...694.1045Z}
{Zhu} Z.,  {Hartmann} L.,   {Gammie} C.,  2009, \mn@doi [\apj] {10.1088/0004-637X/694/2/1045}, \href {https://ui.adsabs.harvard.edu/abs/2009ApJ...694.1045Z} {694, 1045}

\bibitem[\protect\citeauthoryear{{Zhu}, {Hartmann}, {Nelson}  \& {Gammie}}{{Zhu} et~al.}{2012}]{2012ApJ...746..110Z}
{Zhu} Z.,  {Hartmann} L.,  {Nelson} R.~P.,   {Gammie} C.~F.,  2012, \mn@doi [\apj] {10.1088/0004-637X/746/1/110}, \href {https://ui.adsabs.harvard.edu/abs/2012ApJ...746..110Z} {746, 110}

\makeatother
\end{thebibliography}

%\begin{thebibliography}{99}
%\end{thebibliography}

%%%%%%%%%%%%%%%%%%%%%%%%%%%%%%%%%%%%%%%%%%%%%%%%%%

%%%%%%%%%%%%%%%%% APPENDICES %%%%%%%%%%%%%%%%%%%%%

\appendix

\section{Extra channel maps} 
\label{sec:appA}

\begin{figure*}
     \centering
     \includegraphics[width=\textwidth]{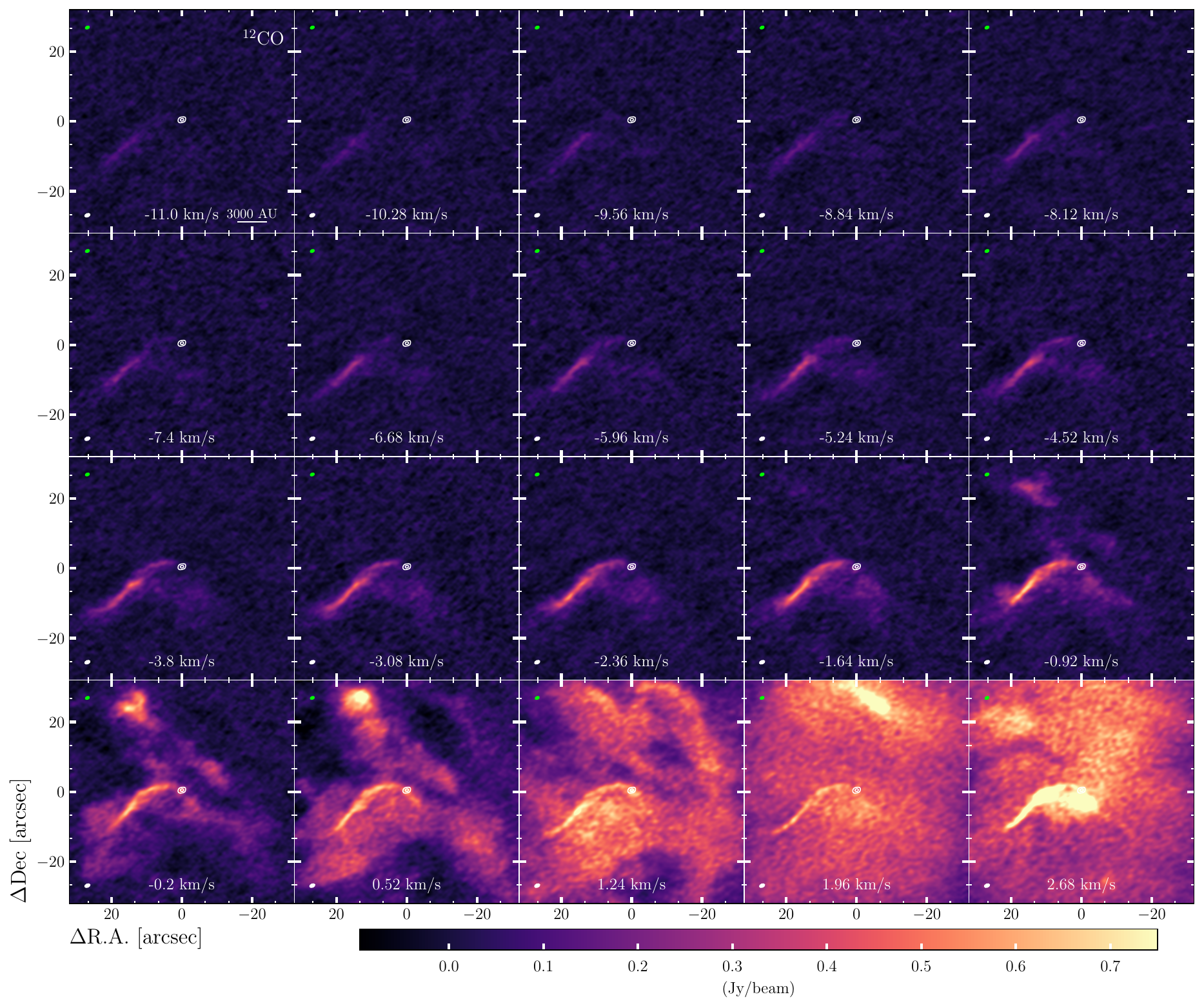}
     \caption{An individual channel map of blueshifted $^{12}$CO emission between -11.0 and 2.68 $\si{{km}.s^{-1}}$. Each frame is spaced by 0.72 $\si{{km}.s^{-1}}$. The white contours in the centre of each panel represent continuum emission centred at 232.5 GHz shown at 0.25 and 0.75 times the maximum recorded intensity of 0.11 Jy/beam. The synthesised beam is shown in the bottom left of each panel by a white ellipse with a size and position angle of 1.85 $\times$ 1.23 arcseconds and -$67.2\degr$, respectively. The continuum beam is represented by a green ellipse in the top left corner of each panel, with a size of 1.50 $\times$ 1.07 arcseconds and a position angle of -$66.8\degr$.}
     \label{fig:12SOutflow4}
\end{figure*}

\begin{figure*}
    \centering
    \includegraphics[width=\textwidth]{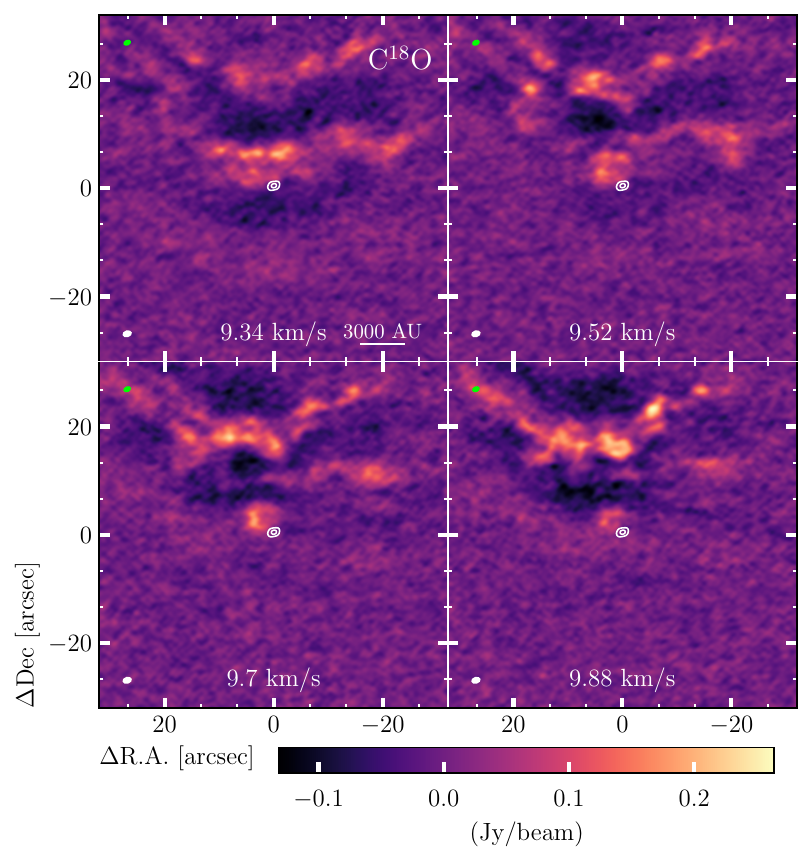}
    \caption{An individual channel map of redshifted C$^{18}$O between 9.34 and 9.88 $\si{{km}.s^{-1}}$. The white contours represent continuum emission centred at 232.5 GHz at levels of 0.25 and 0.75 times the maximum of 0.11 Jy/beam. The C$^{18}$O synthesised beam is shown in the bottom left of each frame in white, while the continuum beam is shown in green at the top left of each frame. The two beam have sizes and positions angles of 1.78 $\times$ 1.28 arcseconds and -74.6$\degr$, and 1.50 $\times$ 1.07 arcseconds and -66.8$\degr$, respectively.}
    \label{fig:18Shadow}
\end{figure*}

\begin{figure*}
    \centering
    \includegraphics[width=\textwidth]{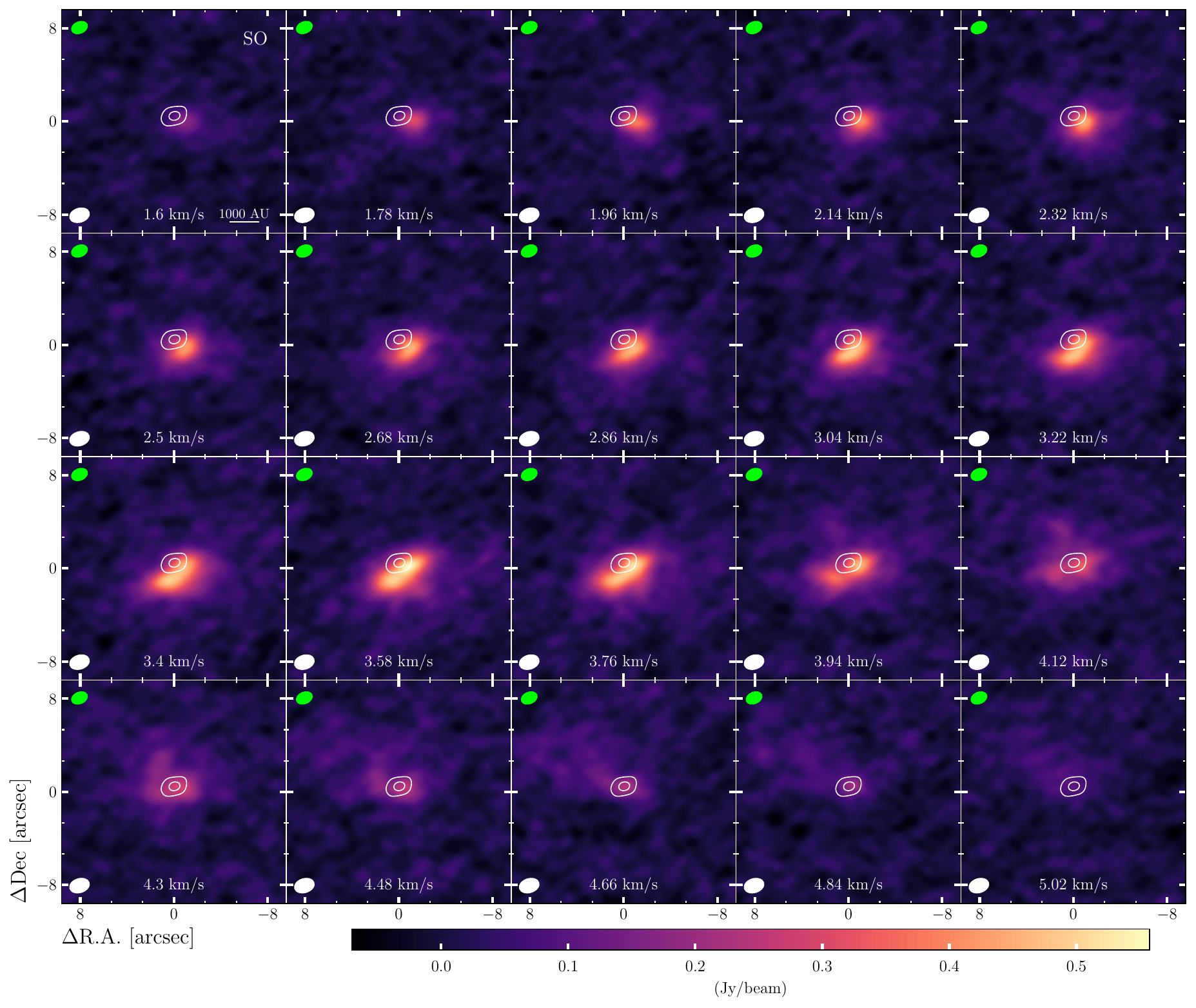}
    \caption{A channel map of SO emission between velocities of 1.6 and 5.02 $\si{{km}.s^{-1}}$. The white contours at the centre of each frame represent continuum emission centred at 232.5 GHz, and are shown at levels of 0.25 and 0.75 times the maximum detected intensity of 0.11 Jy/beam. The SO synthesised beam and continuum emission beam are shown in the bottom left and top left of each frame, respectively. The former maintains a size and position angle of 1.81 $\times$ 1.28 arcseconds and -72.3$\degr$, while the latter has a size and position angle of 1.50 $\times$ 1.07 arcseconds and a position angle of -66.8$\degr$.}
    \label{fig:SOfirst}
\end{figure*}

\begin{figure*}
    \centering
    \includegraphics[width=\textwidth]{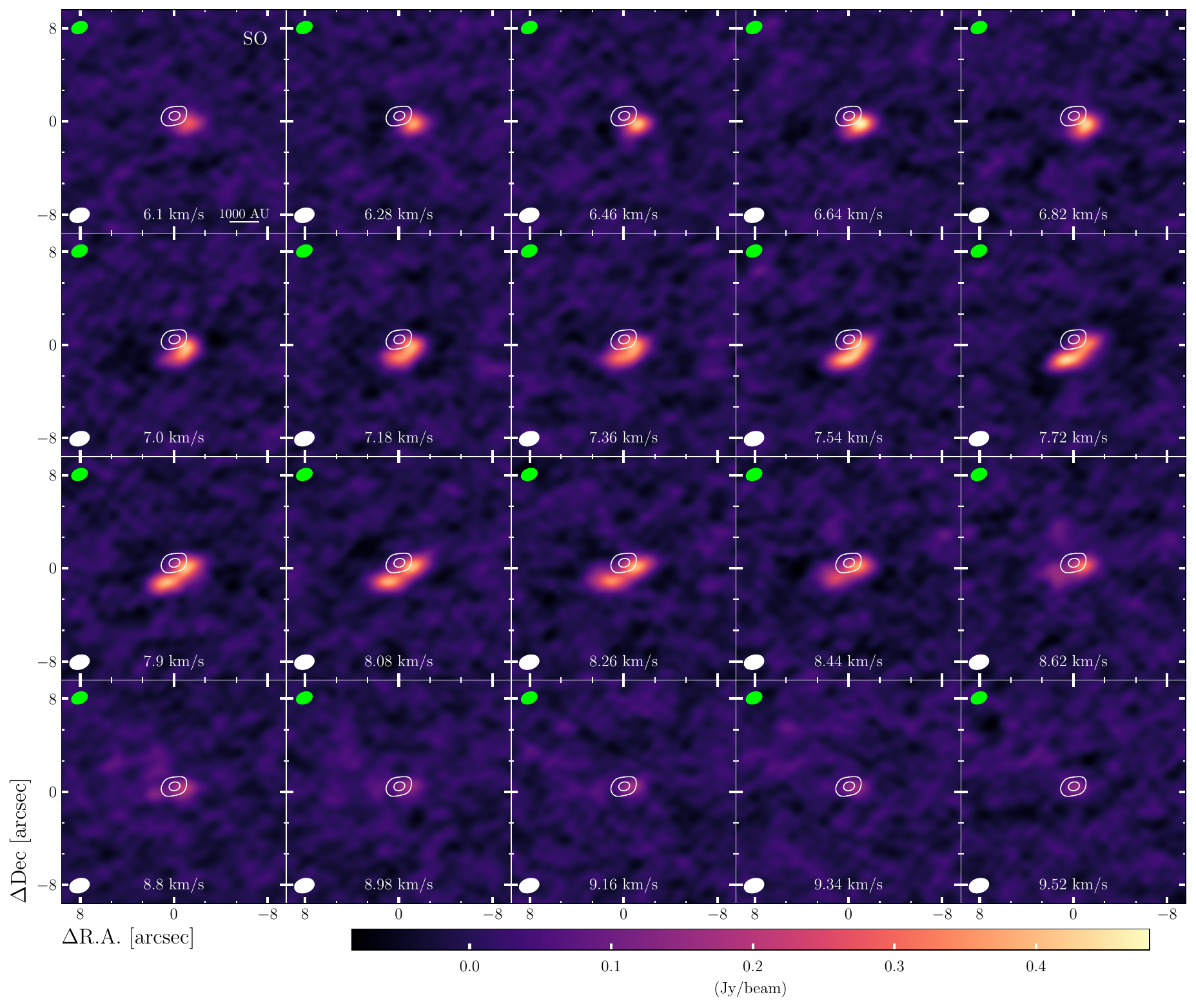}
    \caption{A channel map of SO emission between velocities of 6.1 and 9.52 $\si{{km}.s^{-1}}$. The white contours at the centre of each frame represent continuum emission centred at 232.5 GHz, and are shown at levels of 0.25 and 0.75 times the maximum detected intensity of 0.11 Jy/beam. The SO synthesised beam and continuum emission beam are shown in the bottom left and top left of each frame, respectively. The former maintains a size and position angle of 1.81 $\times$ 1.28 arcseconds and -72.3$\degr$, while the latter has a size and position angle of 1.50 $\times$ 1.07 arcseconds and a position angle of -66.8$\degr$.}
    \label{fig:SOsecond}
\end{figure*}

\begin{figure*}
    \centering
    \includegraphics[width=\textwidth]{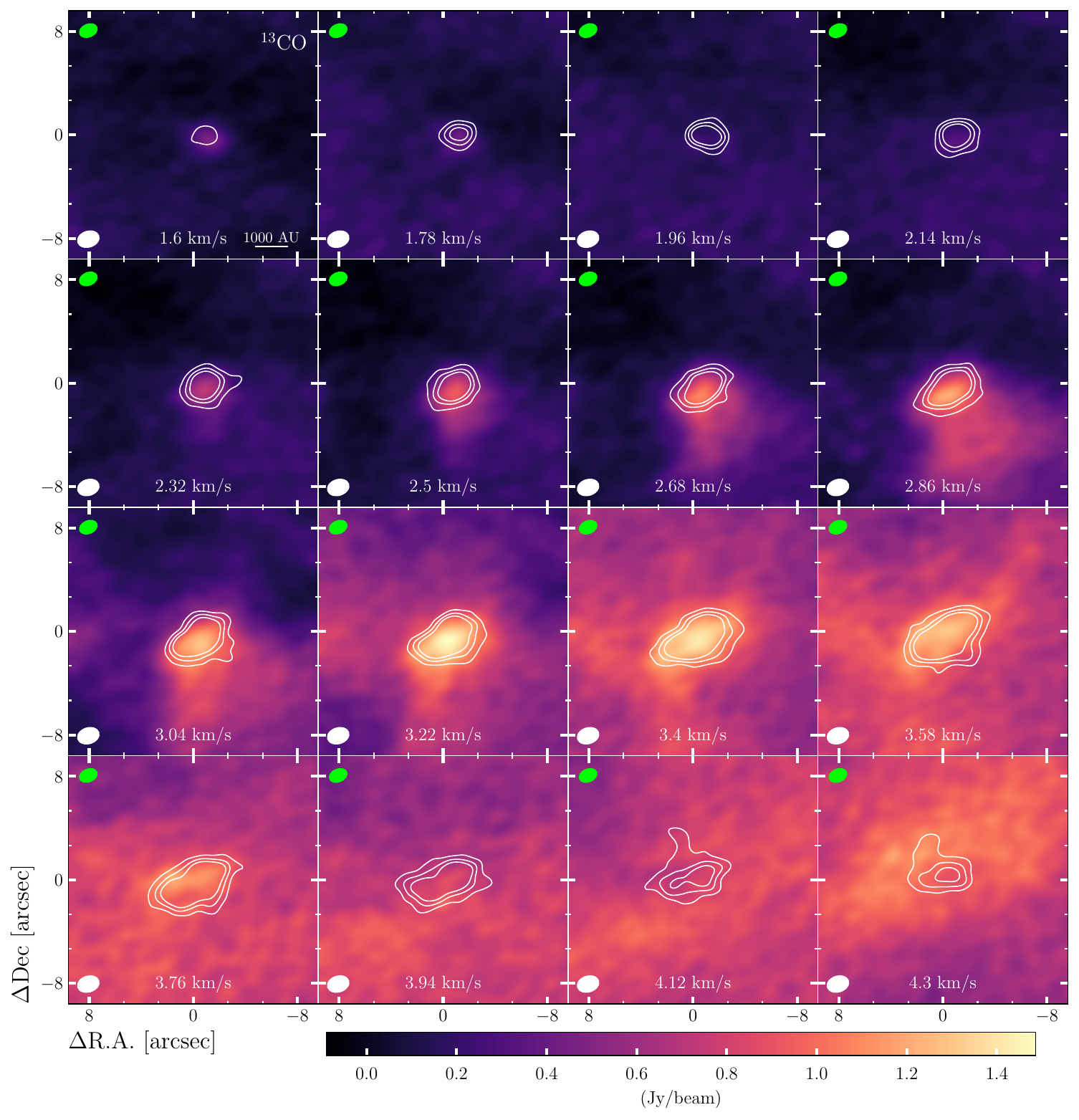}
    \caption{An individual channel map of $^{13}$CO emission between 1.6 and 4.3 $\si{{km}.s^{-1}}$ with SO emission overlaid as a white contour at levels of 5, 7,5, and 10 times the maximum detected intensity of 0.11 Jy/beam. The $^{13}$CO synthesised beam is shown as a white ellipse in the bottom left of each frame, and has an angular size and position angle of 1.77 $\times$ 1.30 arcseconds and -72.3$\degr$, respectively. Likewise, the SO synthesised beam is shown as a green ellipse in the top left of each frame with an angular size of 1.81 $\times$ 1.28 arcseconds and a position angle of -72.3$\degr$.}
    \label{fig:13SOoverlap}
\end{figure*}

%%%%%%%%%%%%%%%%%%%%%%%%%%%%%%%%%%%%%%%%%%%%%%%%%%

% Don't change these lines
\bsp	% typesetting comment
\label{lastpage}
\end{document}